\documentclass[12pt]{iopart}
\usepackage{iopams}
\usepackage{graphicx,float}
\usepackage{color}
\newcommand{\RR}{{R}}
\newcommand{\be}{\begin{equation}}
\newcommand{\ee}{\end{equation}}
\newcommand{\bea}{\begin{eqnarray}}
\newcommand{\eea}{\end{eqnarray}}

\eqnobysec
\def\1.2{\frac{1}{2}}

\begin{document}
\title{Quantum Impurity Entanglement}
\author{Erik S. S{\o}rensen$^1$, Ming-Shyang Chang$^2$, Nicolas Laflorencie$^{2,3}$ and Ian Affleck$^2$}
\address{$^1$Department of Physics and Astronomy, McMaster University, Hamilton, ON, L8S 4M1 Canada}
\address{$^2$Department of Physics \& Astronomy, University of
    British Columbia, Vancouver, B.C., Canada, V6T 1Z1}
\address{$^3$Institute of Theoretical Physics, \'Ecole Polytechnique F\'ed\'erale de Lausanne, Switzerland}
\ead{sorensen@mcmaster.ca, mschang@phas.ubc.ca,\\ nicolas.laflorencie@epfl.ch, iaffleck@physics.ubc.ca}

\date{\today}
\begin{abstract}
Entanglement in $J_1-J_2$, $S=1/2$ quantum spin chains with an impurity is studied using analytic methods
as well as large scale numerical density matrix renormalization group methods. The entanglement is investigated
in terms of the von Neumann  entropy, $S=-\Tr \rho_A\log\rho_A$, for a sub-system $A$ of size $r$
of the chain. The impurity contribution
to the uniform part of the entanglement entropy, $S_{imp}$, is defined and analyzed in detail in both the gapless, $J_2\leq J_2^c$, as well
as the dimerized phase, $J_2>J_2^c$, of the model.
This quantum impurity model is in the universality class of the single channel Kondo model and
it is shown that in a quite universal way the presence of the
impurity in the gapless phase, $J_2\leq J_2^c$, gives rise to a large length scale, $\xi_K$, associated
with the screening of the impurity, the size of
the Kondo screening cloud. The universality of Kondo physics then implies scaling of the form $S_{imp}(r/\xi_K,r/R)$ for
a system of size $R$. Numerical results are presented clearly demonstrating this scaling. At the critical point, $J_2^c$,
an analytic approach based on a Fermi liquid picture, valid at distances $r\gg\xi_K$ and energy scales $T\ll T_K$, is
developed and analytic results at $T=0$ are obtained showing
$S_{imp}=\pi \xi _{K}[1+\pi (1-{r}/{\RR})\cot ({\pi r}/{\RR})]/(12 R)$ for finite $R$.
For $T>0$, in the thermodynamic limit, we find $S_{imp}=[\pi ^{2}\xi _{K}T/(6v)]\coth (2\pi rT/v)$.
In the dimerized phase an appealing picture of the entanglement is developed in terms of a {\it thin soliton} (TS) ansatz
and the notions of impurity valence bonds (IVB) and single particle entanglement (SPE) are introduced. The TS-ansatz
permits a variational calculation of the complete entanglement in the dimerized phase that appears to be
exact in the thermodynamic limit
at the Majumdar-Ghosh point, $J_2=J_1/2$, and surprisingly precise even close to the critical point $J_2^c$.
In appendices the TS-ansatz is further used to calculate $\langle S^z_r\rangle$ and $\langle \vec S_r\cdot \vec S_{r+1}\rangle$
with high precision at the Majumdar-Ghosh point and the relation between the finite temperature entanglement
entropy, $S(T)$, and the thermal entropy, $S_{th}(T)$, is discussed.
Finally, the alternating part of $S_{imp}$ is discussed, together with its relation to the boundary induced dimerization.
\end{abstract}

 \maketitle

\section{Introduction}
Much of the mystery and power of quantum mechanics arises from entanglement,
which leads to Einstein's ``spooky action at a distance'' but is
 now recognized as a resource by the
quantum information community, being essential for quantum  teleportation
or quantum computing~\cite{Bennett00}. Ground states of quantum field theories
and many body theories exhibit fascinating entanglement
properties which are beginning to be understood~\cite{Wilczek94,Osborne02a,Osterloh02,Vidal03,Wei05,Kopp06a}.
A useful measure of many body
entanglement when the total system is in a {\it pure} state is the von Neumann entanglement entropy.
This is obtained
by focusing on bipartite system where space can be divided into 2 regions, $A$ and $B$.  Beginning
with the ground state pure density matrix, region $B$ is traced over
to define the reduced density matrix $\rho_A$. From this the von Neumann
entanglement entropy~\cite{Neumann27,Wehrl78},
\begin{equation}
S(r)\equiv -\Tr [\rho_A\ln \rho_A]\label{eq:vNS}
\end{equation}
is obtained for subsystem of size $r$.
Several other measures of entanglement are in current use such as the concurrence~\cite{Hill97,Wooters98} which is
monotonically related to the entanglement of formation~\cite{Bennett96b}, the distillable entanglement~\cite{Bennett96b,Bennett96prl}
and the relative entropy of entanglement~\cite{Vedral97,Vedral98,Vedral02}. See also~\cite{MHorodecki01}.
Some of these measures have been developed to describe entanglement as it occurs in systems in a mixed state where it is {\it much harder}
to quantify entanglement, see for instance
the seminal paper by Bennett et al.~\cite{Bennett96b}.
Here we focus mainly on bipartite systems in pure states for which the von Neumann entropy, $S$, is an essentially unique
measure of the entanglement.
The rate at which $S$ grows
with the spatial extent, $r$, of region $A$  is not only a fundamental
measure of entanglement, it is also crucial~\cite{Osborne02b,Vidal04,Verstraete04,MPS} to the functioning
of the Density Matrix Renormalization Group (DMRG) a powerful numerical method
for solving many body problems~\cite{White92,Schol05}.
For systems with finite
correlation lengths, it is generally expected that $S_A$
grows with the area of the boundary of region $A$~\cite{Bombelli86,Srednicki93}.
In the one-dimensional case, conformally invariant systems (with infinite
correlation length) have $S(r)\to (c/3)\ln r$~\cite{Wilczek94,Cardy04} where $c$ is the
``central charge'' characterizing the conformal field theory (CFT).
Entanglement entropy has recently been shown to be a useful
way of characterizing topological phases of many body theories~\cite{Kitaev06,Levin06,Fendley06,Furukawa06}, which cannot
be characterized by any standard order parameter.
Entanglement entropy is also closely related to the thermodynamic
entropy of black holes and to the ``holographic principle'' relating
bulk to boundary field and string theories~\cite{Fendley06,Ryu06,Ryu07}.
Hence, due to these latter developments, even though the von Neumann entanglement entropy may only give
an incomplete description of entanglement in {\it mixed states}, such as would be the case at finite temperature or in
the presence of noise from the environment, an understanding of its behavior in such
states is important from the condensed matter perspective. We discuss the precise connection between
the finite temperature entanglement entropy and the thermal entropy in some detail in~\ref{app:FiniteT}.

Recent experiments~\cite{Ghosh03} on the
magnetic salt LiHo$_x$Y$_{1-x}$F$_4$ have been interpreted as evidence for
quantum entanglement in the magnetic susceptibility and electronic
specific heat at temperatures approaching 1K. 
Theoretical work
have shown that macroscopic entanglement is in principle
observable at much higher temperatures~\cite{Vedral04} and should
also be observable using other probes such as neutron
scattering~\cite{Brukner06}. Experimental measures of the
concurrence have thus been obtained~\cite{Ghosh03,Brukner06}. Some
of the
theoretical~\cite{Horodecki96,Terhal00,Lewenstein00,Jordan04,Toth05,Wu05,Wiesniak05,Anders06}
and experimental~\cite{Rappoport06,Vertesi06} have focused on
establishing {\it entanglement witnesses} (EW) for detecting
entanglement. An EW is an hermitian operator $W$ with
$\langle\rho\rangle_W\equiv \tr(W\rho)\geq 0$ for all separable
$\rho$. Thus, if $\langle\rho\rangle_W<0$, then $\rho$ is
non-separable. Energy fluctuations~\cite{Jordan04}, the persistent
current~\cite{Jordan04}, the temperature~\cite{Anders06} and the
magnetic susceptibility~\cite{Bose05,Wiesniak05,Brukner06} have
been proposed as EW's. While an EW can detect entanglement it does
not provide a quantitative measure of the strength of the
entanglement. Analysis based on the magnetic susceptibility as an
EW have been interpreted as a signature of entanglement at
temperatures as high as 365K in the nanotubular system
Na$_2$Va$_3$O$_7$~\cite{Vertesi06}, close to 100K in the
warwickite MgTiOBO$_3$~\cite{Rappoport06} and around 20K in the
pyroborate MgMnB$_2$O$_6$~\cite{Rappoport06}.

Comparatively few results~\cite{
Cardy04,
Zhou05,Levine04,Wang04,Peschel05,Fan06,Zhao06,Cho06} have been
obtained on entanglement in systems with impurities.
For CFT's impurity interactions will, under the renormalization group (RG), flow
to fixed points which can be represented by conformally invariant boundary conditions.
These conformal boundary conditions can be characterized by
the zero temperature thermodynamic impurity entropy~\cite{Affleckg},
$\ln g$, a length independent term in the thermodynamic entropy of
a semi-infinite system.  Consequently, it was argued~\cite{Cardy04} that the entanglement
entropy of a semi-infinite CFT is $(c/6)\ln (r/a)+\ln g$,
where $a$ is a constant which is non-universal but independent
of the boundary condition. At the fixed point, the $\ln g$ term is then part
of the impurity contribution to the entanglement entropy. The RG flow between
different fixed points was numerically studied in terms of $\ln g$ for the quantum Ising and $XXZ$ chain in~\cite{Zhou05}.
It is then of considerable interest  to quantitatively define what we
call the ``impurity entanglement entropy":
\begin{equation}
S_{imp}=S({\rm with\ impurity})-S({\rm no\ impurity}),
\label{eq:simp}
\end{equation}
the additional entanglement entropy that arises from adding an impurity in region $A$.
A specific implementation of Eq.~(\ref{eq:simp}) suitable for our numerical work will
be discussed in section~\ref{sec:simp}.

\begin{figure}[!ht]
\begin{center}
\includegraphics[height=2.5cm,clip]{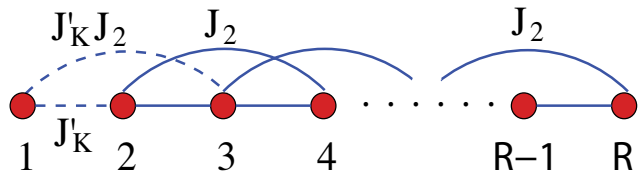}
\end{center}
\caption{Schematic picture for the $J_1-J_2$ spin chain model (\ref{eq:spinch}) with an impurity located at the left boundary and coupled with $J_{K}^{\prime }$.}
\label{fig:chain}
\end{figure}

\begin{figure}[!ht]
\begin{center}
\includegraphics[width=10cm,clip]{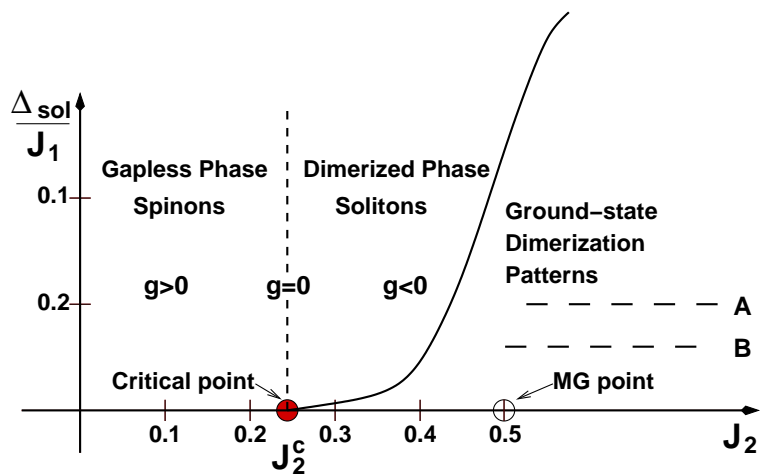}
\end{center}
\caption{Phase diagram of the $J_1-J_2$ model as a function of $J_2$. The critical
point $J_2^c\approx.2412$ separates a gapless Heisenberg phase from the dimerized phase.
  At the Majumdar-Ghosh (MG) point, $J_2=J_1/2$, the two exact ground-states, $A$ and $B$,  for a system
  with an even number of spins are shown. For an odd number of spins a single spin (soliton) will separate
  regions with these two dimerization patterns.
}
\label{fig:phasediag}
\end{figure}

Here we focus on the impurity entanglement entropy in the Kondo model and 
a related spin-chain model.  The 3-dimensional (3D) Kondo model Hamiltonian is:
\begin{equation}
H=\int d^{3}r[\psi ^{\dagger }(-\nabla ^{2}/2m)\psi +J _{K}\delta^3(\vec r)\psi
^{\dagger }(\vec{\sigma}/2)\psi \cdot \vec{S}],  \label{H3D}
\end{equation}
where  $\psi (\vec{r})$ is the electron annihilation operator (with
spin-index suppressed) and the $S^{a}$ are $S=1/2$ spin operators. 
Most of our numerical work is performed on the related $J_1-J_{2}$ family of antiferromagnetic 
$S=1/2$ Heisenberg spin chain models.  For $J_{2}\leq J_{2}^{c}\simeq 0.2412$ the $J_1-J_2$ spin
chain is gapless and the low energy behavior of an impurity is {\it equivalent} to that of the Kondo model.
In \ref{app:field_theory} we review the connection between these 
models.  (This connection is pursued further in \cite{KEAC}.) 
In the $J_1-J_2$ spin chain model the equivalence to the Kondo model is achieved by
modeling the impurity as a weakened coupling, $J_K'$, 
at the end of an open chain. (see Fig.~\ref{fig:chain}.)
We then write the hamiltonian for the spin chains as:
\begin{equation}
H =J_{K}^{\prime }\left( \vec{S}_{1}\cdot \vec{S}_{2}+J_{2}\vec{S}%
_{1}\cdot \vec{S}_{3}\right) +
\sum_{r=2}^{\RR-1}\vec{S}_{r}\cdot \vec{S}_{r+1}+J_{2}\sum_{r=2}^{\RR-2}\vec{%
S}_{r}\cdot \vec{S}_{r+2}.  \label{eq:spinch}
\end{equation}
While unimportant for our
analytic work, the spin chain representation, Eq.~(\ref{eq:spinch}),
dramatically alleviates the numerical work compared to using the Hamiltonian, Eq.~(\ref{H3D}).
As discussed in \ref{app:field_theory}, the finite size corrections 
are much smaller when $J_2$ is fine-tuned to the critical point $J_c\approx .2412$~\cite{Eggert96},
a fact that, from a numerical perspective, presents a considerable advantage.
For $J_2>J_2^c$ the spin chain enters a dimerized
phase~\cite{Haldane82} with a gap and the relation between
Eq.~(\ref{eq:spinch}) and Kondo physics no longer holds. The two-fold degenerate dimerized
ground-state is exactly known at the Majumdar-Ghosh~\cite{MG69} (MG) point, $J_2=J/2$, with
the two ground-states corresponding to the two possible dimerization patterns. In the dimerized
phase the fundamental excitations can be viewed as single $S=1/2$ {\it solitons} separating regions with
these two distinct dimerization patterns. A sketch of the phase-diagram of the $J_1-J_2$ model summarizing
these points is shown in Fig.~\ref{fig:phasediag}.

At low energies, the spin chain model is equivalent to the Kondo 
model in any dimension, $D$. This is also true for the entanglement entropy as
we show in \ref{app:3D-1D}.
A physical motivation  for studying the Kondo model in the $D=2$-dimensional case
is provided by the possibility of using the spins of
gated semi-conductor quantum dots as  qubits~\cite{Loss98,Simon06,Eriksson04}. The  various
quantum dot spins would ideally be entangled only with each other.
However, in practice, they would also be entangled with the
surrounding conduction electrons which would cause limitations
on the functioning of such a quantum computer~\cite{Costi03,Cho06}. More generally,
any physical realization of a quantum computer has
a dissipative environment which is entangled, to some extent
with the qubits. The Caldeira-Leggett (spin-boson) model~\cite{Caldeira} of a spin
interacting with an ohmic dissipative environment is equivalent,
at low energies to the Kondo model and entanglement in the Caldeira-Leggett model
has been considered in recent work~\cite{Costi03,Kopp06b}.
The understanding of entanglement as it occurs in these quantum impurity models
is therefore of considerable interest.

Spin chains have also been proposed
for the purpose of creating quantum communication
channels~\cite{Lloyd03,Bose03,Plenio05} and subsequent work has showed
that entanglement in $S=1/2$ spin chain models can be used to
establish perfect state
transfer~\cite{Christandl04,Burgarth05a,Burgarth05b}. In most cases
the couplings in the bulk of the chain are not uniform but vary
with $r$, however, a model closely related to
Eq.~(\ref{eq:spinch}), where only the couplings at the end of the
chain are modified have also been shown to accommodate perfect
state transfer~\cite{Wojcik05}. The algorithm for perfect state
transfer proposed in~\cite{Christandl04} has been experimentally
realized using a three-qubit liquid NMR quantum computer
simulating a Heisenberg $XY$ interaction~\cite{Zhang05}. The
possibility for using quantum spin chains as perfect quantum state
mirrors has also recently been emphasized~\cite{Karbach05,Fitzsimons06}. 
From this perspective, entanglement in the spin chain model, Eq~(\ref{eq:spinch}),
is clearly of interest even aside from the relation to Kondo physics.

In the ground state of the Kondo model, the spin of the impurity
is ``screened'' meaning that it effectively forms a singlet
with the conduction electrons. This screening is expected to
take place at an exponentially large length scale
\begin{equation}
\xi_K=v/T_K\propto e^{1/\lambda_K}\label{eq:xiK}.
\end{equation}
Here $v$ is the velocity of the low energy excitations (fermions
or spin-waves) and $T_K$, the Kondo temperature, is the energy
scale at which the renormalized Kondo coupling becomes large. One
of the interesting features of entanglement entropy in many body
systems is its universality, which makes it useful for studying
quantum phase transitions~\cite{Osborne02a,Osterloh02,Vidal03,Cardy04}. A major conclusion of our work is that
this universality extends to $S_{imp}$ in the following sense.
When $\xi_K$ is large compared to all microscopic length scales in
the system, and the system size $R$ is $\infty$, we find that
$S_{imp}(r/\xi_K)$ is a universal scaling function depending only
on the ratio  $r/\xi_K$. For finite systems of extent $R$ we find
when $r,R-r,\xi_K\gg 1$:
\begin{equation}
S_{imp}\equiv S_{imp}(r/R,r/\xi_K)
\label{eq:simpscal}
\end{equation}
In this case, there are actually
two different scaling functions depending on whether the total spin of
the finite system ground state is $0$ or $1/2$. (For the spin chain
these two cases correspond to $R$ even or odd respectively.)
We emphasize that this holds for entanglement in both models Eq.~(\ref{eq:spinch}) and (\ref{H3D}).
This extends to entanglement entropy
the well-known universality of other properties
of the Kondo model.  We expect this universal scaling property
to hold generally for quantum impurity models as was
remarked on in~\cite{Zhou05}. An immediate consequence of
this universality is that the same impurity entanglement
entropy occurs, for large $r$, in many different microscopic models.
It is insensitive, for example, to the dimensionality of space and the range
of the Kondo interaction (as long as it is finite).
In \cite{Sorensen06} preliminary density matrix group (DMRG) results obtained for the spin chain model, Eq.~(\ref{eq:spinch}), at  $J_2=J_2^c$
showed that
the $r$-dependence of  $S_{imp}$ confirms this picture and the presence of
the length scale $\xi_K$ was demonstrated. In section~\ref{sec:simp} we provide additional evidence supporting this scaling
at $J_2=J_2^c$ as well as results for for $J_2<J_2^c$.

The universal scaling functions are generally not amenable to
analytic calculation except in certain limiting cases. The
most straightforward of these is
when $\xi_K\ll r$, in which case it is possible to perform calculations using
Nozi\`eres local Fermi liquid theory (FLT) approach~\cite{Nozieres},
as developed in~\cite{Affleck90,Affleck91a,Affleck91b}.
In \cite{Sorensen06} initial analytical results using this approach were reported and in section~\ref{sec:FLT} we present
a detailed derivation and additional results notably at finite temperature.

When the Kondo coupling approaches either the weak coupling fixed point, $J_K'=0$, or the strong coupling
fixed point, $J_K'=1$, one might have assumed that the impurity entanglement entropy would vanish. This turns
out to be the case at the strong coupling fixed point, $J_K'=1$. However, as we initially reported in~\cite{Sorensen06},
the impurity entanglement at the weak coupling fixed point is {\it non-zero}. In 
sections~\ref{sec:simp},\ref{sec:numres} we discuss in detail this fixed point entanglement
entropy and further develop the intuitive picture for understanding it.

For $J_2>J_2^c$ the spin chain enters a dimerized
phase~\cite{Haldane82} with a gap and the relation between
Eq.~(\ref{eq:spinch}) and Kondo physics no longer holds. 
However, entanglement as it occurs in spin chains without impurities,
viewed as model systems for entanglement, is currently the focus of intense studies and is 
relatively
well established both from a static~\cite{Arnesen01,Gunlycke01,Osborne02a,Latorre04,
Verstraete04b,Cardy04,Fan04,Peschel04,Subrahmanyam04,Laflorencie06} and dynamic perspective~\cite{Amico04,Cardy05,dechiara06}.
Here we show that it is possible to obtain almost
exact analytical results for the fixed
point entanglement for the $J_1-J_2$ model, Eq.~(\ref{eq:spinch}), in the dimerized phase. 
This approach is based
on describing the lowest lying excitations in the dimerized phase
as simple domain walls or single site {\it solitons}~\cite{SS81} which we refer to as thin solitons (TS)
and is described in detail in section~\ref{sec:ssansatz}.
While the use of gapped spin chains for the purpose of quantum communication
and quantum computing is less evident than for gapless chains, the relative simplicity
of the entanglement as it occurs in the dimerized phase allows for the development
of an appealing intuitive picture of how the entanglement arises in terms of {\it single particle
entanglement} (SPE) and {\it impurity valence bonds} (IVB). These concepts can be rigorously established
using the TS approach in the dimerized phase and, more importantly, appear to be quite general concepts applicable
to other models even in the absence of a gap. (See section~\ref{sec:simp}).

The outline of the paper is as follows: In
Section~\ref{sec:simp} we discuss in detail our definition of $S_{imp}$ and show additional evidence for the scaling behavior of
Eq.~(\ref{eq:simpscal}) at $J_2=J_2^c$. The intuitive picture for understanding the impurity entanglement in terms of SPE and IVB is developed in Section~\ref{sec:SPEIVB} along with
the fixed point entanglement.  Weak scaling
violations, related to another single-site measure for the impurity entanglement, $s_{imp}$, are discussed in Section~\ref{sec:smallimp}. Section~\ref{sec:FLT} contains a detailed account of the FLT approach
for calculating $S_{imp}$ and Section~\ref{sec:ssansatz} describe the thin soliton  approach to
performing variational calculations for the entanglement in the dimerized phase for $J_2 > J_2^c$.
Numerical results for the fixed point entanglement are presented in section~\ref{sec:numres}.
Most of the numerical results presented are density matrix renormalization group (DMRG) calculations
performed on parallel SHARCnet computers keeping $m=256$ states.
As we discuss in section~\ref{sec:simp} our working definition of $S_{imp}$
only focus on the uniform part of the entanglement.
In section~\ref{sec:sa} we therefore present results for the alternating part of the entanglement as well
as the dimerization.
Finally we briefly summarize our main results in section~\ref{sec:conclusion}.  

This paper contains a number of appendices, some of 
which may be of quite general interest. In \ref{app:field_theory} we review 
field theory results on the $D$-dimensional Kondo model and the spin chain, 
and their relationship to each other. 
In \ref{app:3D-1D} we extend these results to prove that the impurity entanglement 
entropy is the same for the $D$-dimensional Kondo model and 
the spin chain model.  This appendix also contains 
a new derivation of the free fermion entanglement entropy in $D$-dimensions. 
In \ref{app:7p} we derive the 7-point formula used in the numerical work for extracting
the uniform and alternating part of the entanglement entropy.
In \ref{app:SU2} 
we prove that the entanglement entropy is the 
same for all linear combinations of 
spin up and spin down elements of a doublet state.  
The connections between finite temperature entanglement entropy and thermal entropy are discussed in \ref{app:FiniteT}.
In \ref{app:TSMG}
we present new results on the Majumdar-Ghosh model based 
on  the thin soliton ansatz, including $\langle S^z_r\rangle$ in the 
ground state with open boundary conditions and an odd number of sites, and 
the dimerization, $\langle\vec S_r\cdot \vec S_{r+1}\rangle$. 

\section{The impurity Entanglement Entropy~\label{sec:simp}}
We begin by discussing our definition of the impurity entanglement
entropy, Eq.~(\ref{eq:simp}). For impurity problems such as the
Kondo model it is quite standard in experimental situations to
define the impurity contribution to, for example, the
susceptibility or the specific heat by subtracting reference
values with the impurity absent. We have therefore defined the
impurity entanglement analogously. For the $S=1/2$ spin chain 
the entanglement entropy has not only a uniform but
also a staggered part~\cite{Laflorencie06}. For $r\gg 1$ we can
write:
\begin{equation}
S(r,\RR)=S_U(J'_K,r,{\RR})+(-1)^rS_A(J'_K,r,{\RR}),
\end{equation}
where $S_U(J'_K,r,{\RR})$ and $S_A(J'_K,r,{\RR})$ are slowly varying functions
or $r$. The entanglement entropy is also strongly dependent on whether the total spin of
the ground-state of the system is $0$ or $1/2$, or, equivalently, whether $R$ is even or odd.

[In general, for $\RR$ odd, the ground state is a spin doublet.  The entanglement 
entropy is the same for the spin up or down element of the doublet 
or for any linear combination of these states.  We give a formal 
proof of this in \ref{app:SU2}.  The result follows from 
the fact that {\it any} linear combination of spin up and down 
can be obtained by a spin rotation from the spin up state.  It is 
intuitively obvious that the entanglement entropy doesn't 
depend on the direction of the spin quantization axis. For 
higher spin states the situation is more complex and in general 
different elements of the spin multiplet have different entanglement entropy.] 

Theoretical work~\cite{Wilczek94,Cardy04}
has established that for a uniform $S=1/2$ spin chain with periodic boundary conditions, or in general
for gapless 1-D models, the entanglement entropy
is given by:
\begin{equation}
S^{PBC}(r,{\RR})=\frac{c}{3}\ln \left[\frac{\RR}{\pi} \sin \left(\frac{\pi r}{\RR}\right)\right]+{s_1},
\label{eq:spbc}
\end{equation}
with $c=1$ and $s_1\simeq 0.726$~\cite{Jin04}. In this case there is no alternating term in the entanglement
entropy.
For the purpose of studying quantum impurity models,
our focus is here
exclusively on the case of an {\it open} $S=1/2$ spin chain where 
the alternating term, $S_A$, is non-zero~\cite{Laflorencie06} and it
is known~\cite{Cardy04} 
that for a gapless system of linear extent $R$, the uniform part has a bulk part,
\begin{equation}
S_{U0}(r,{\RR})\approx \frac{1}{6}\ln \left[\frac{2\RR}{\pi} \sin \left(\frac{\pi r}{\RR}\right)\right]+\ln g +s_1/2,
\label{eq:sobc}
\end{equation}
independent of $J_K^\prime$, {\it as well as} an impurity contribution that will depend on $J_K'$ and that
is our main focus.
Here $\ln g$ is the zero temperature thermodynamic impurity entropy~\cite{Affleckg}.
We see that the presence of the boundary, in addition to generating an alternating term in $S$, have modified the uniform
part of the entanglement entropy with
respect to the result for periodic boundary conditions, Eq.~(\ref{eq:spbc}). Note that both Eqs.~(\ref{eq:sobc}) and Eq.~(\ref{eq:spbc})
only are valid for critical models.

The impurity models we consider are defined using systems with open boundary conditions and when considering
the impurity contribution to the entanglement care has to be taken with respect to the alternating
term generated by the open boundary conditions. We do this by initially focusing only on the uniform part
of the entanglement entropy and the contribution arising from the impurity to this part.
Returning to our fundamental definition of $S_{imp}$, Eq.~(\ref{eq:simp}),
we view $S$(no impurity)
as the entanglement
in a system without the impurity site, i.e. with one less site, $R-1$ and all couplings equal to unity. We {\it do not} define $S$(no impurity)
as $S$ with $J_K'=0$ since, as we shall
discuss in detail later, for $J_K'=0$  the impurity spin
can have a non-trivial entanglement
with the rest of the chain. It then follows that the complete impurity contribution to the entanglement entropy cannot be obtained by subtracting
results for a system with $J_K'=0$.
We thus define the uniform part of the impurity entanglement entropy precisely as:
\begin{equation}
S_{imp}(J'_K,r,{\RR})\equiv S_U(J'_K,r,{\RR})-S_{U}(1,r-1,{\RR}-1),\ r>1.\label{Simpdef}
  \end{equation}
For the subtracted part all couplings have unit strength. In a
completely analogous manner one can also define the alternating
part of the impurity entanglement entropy, $S_{imp}^A$, which we
shall discuss in section~\ref{sec:sa} where it is shown that
for $S^A_{imp}$ the scaling form is modified from that of Eq.~(\ref{eq:simpscal}).
Our focus is therefore on the uniform part. For our
numerical DMRG results a procedure for extracting the uniform and
staggered part of the entanglement entropy is needed. We have
found it sufficient to extract these functions by assuming that
the uniform and staggered parts locally can be fitted by
polynomials. If 7 sites surrounding the point of interest are used
for the fitting a 7 point formula can easily be derived as
outlined in \ref{app:7p}. The numerical work has been done using density matrix renormalization
group~\cite{White92} (DMRG) techniques in a fully parallelized version, keeping $m=256$ states.
When performing calculations with $J_K'=0$ and $R$ even we have found it necessary to use spin-inversion symmetry~\cite{Sorensen98b}
under the DMRG iterations in order to select the desired singlet ground-state.

Other measures of the impurity entanglement entropy
could have been defined and in section~\ref{sec:smallimp} we will discuss
one of these, the single site entanglement of the impurity spin with the rest
of the chain, $s_{imp}$.  As we shall see, this quantity gives an incomplete picture of
the impurity entanglement. Another possibility would be to define a {\it relative entropy}~\cite{Vedral02}
between the state with the impurity and a reference state without the impurity. It would be interesting
to explore this latter possibility. We expect that similar scaling would be found as we show here with our
definition of $S_{imp}$.

\begin{figure}[!ht]
\hfill\includegraphics[width=14cm,clip]{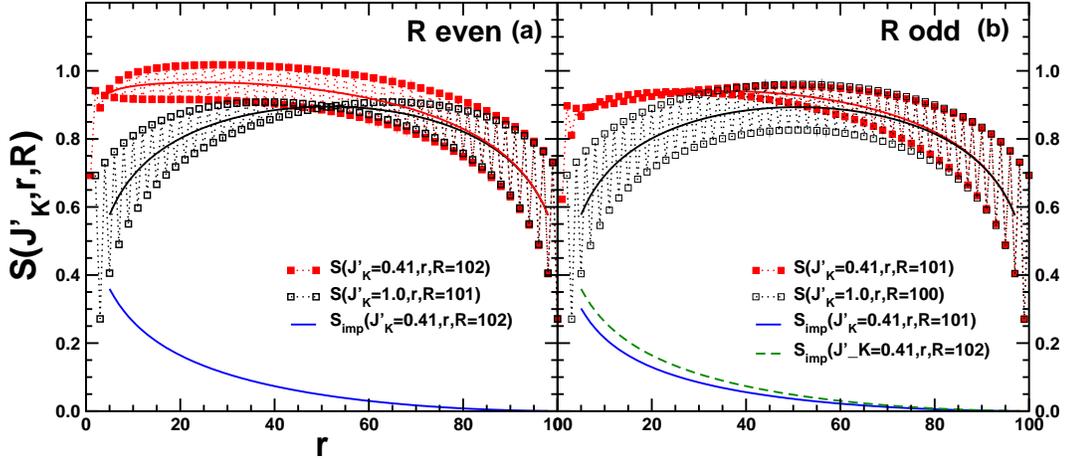}
\caption{(a) DMRG results for the total entanglement entropy, $S(J'_K,r,R)$ for a 102 site spin chain at $J^c_2$,
  with a $J'_K=0.41$ Kondo impurity ($\blacksquare$) along with $S(J'_K=1,r-1,R-1)$ ($\square$).
    Uniform parts (solid lines) and the resulting $S_{imp}(J'_K=0.41,r,R=102)$ for $R$ even.
(b) DMRG results for the total entanglement entropy, $S(J'_K,r,R)$ for a 101 site spin chain at $J^c_2$,
  with a $J'_K=0.41$ Kondo impurity ($\blacksquare$) along with $S(J'_K=1,r-1,R-1)$ ($\square$).
    Uniform parts (solid lines) and the resulting $S_{imp}(J'_K=0.41,r,R=101)$ for $R$ odd.
    For comparison we also show $S_{imp}(J'_K=0.41,r,R=102)$ for $R$ even from panel (a) (dashed line).
}
\label{fig:RawSimp}
\end{figure}
To gain some insight into how the impurity influence the entanglement entropy and lead to a non-zero
$S_{imp}$ we show in Fig.~\ref{fig:RawSimp} data for $J_K'=0.41$ for both $R$ even, Fig.~\ref{fig:RawSimp}(a) and $R$
odd, Fig.~\ref{fig:RawSimp}(b). There are significant differences between the results for $R$ even and $R$ odd. It is
also clear that the influence of the impurity is felt more strongly for $R$ even compared to $R$ odd. The resulting
$S_{imp}$ is clearly {\it bigger} for $R$ even for all $r$ as shown by the comparison in Fig.~\ref{fig:RawSimp}(b).
In the limit where $J_K'\to 0$ this difference becomes more pronounced
since $S_{imp}$ for $R$ even {\it increases} with decreasing $J_K'$ while for $R$ odd it {\it tends to zero}
with decreasing $J_K'$.
For the value of $J_K'=0.41$ used in Fig.~\ref{fig:RawSimp} we shall later find that
$\xi_K=25.65$ significantly smaller than $R$ and
it would have been natural to expect features in $S_{imp}$ signaling the presence of
this length scale.
However, $S_{imp}$ is a {\it monotonically decreasing function} of $r$ for both parities of
$R$ and no particular features are observed in $S_{imp}$ for $r$ of the order of $\xi_K$. It is
important to note that this fact does {\it not} imply a violation of scaling of the form Eq.~(\ref{eq:simpscal}).

\subsection{Scaling of $S_{imp}$ at $J_2^c$\label{sec:simpscal}}

The scaling of $S_{imp}(J_K',r,\RR)$ for fixed $r/\RR$ at
$J_2=J_2^c$ was considered in detail in Ref.~\cite{Sorensen06}. It
was shown that $S_{imp}$ with $r/R$ fixed follows the expected
scaling form, Eq.~(\ref{eq:simpscal}), and is a function of a single variable $r/\xi_K$. 
We expect this scaling to hold for $r,R-r,\xi_K \gg 1$ when the results
are not influenced by microscopic parameters such as the lattice spacing. For the
moderate system sizes of $\RR<102$ ($\RR$ even) and $\RR<101$
($\RR$ odd) $S_{imp}$ shows a strong dependence on the parity of
$\RR$ and two different scaling functions for $\RR$ even and odd
were found. For $r/\xi_K\gg 1$ these two scaling functions are
essentially the same but they differ significantly for $r/\xi_K
\ll 1$. In the limit where $\RR\to\infty$ the two scaling
functions eventually coincide. By requiring the data to collapse
according to the scaling form, a naive estimate of $\xi_K$ can be
obtained for both $\RR$ even and odd. In table~\ref{tab:xiKScal}
we list the resulting $\xi_K$ obtained through such an analysis.
%
\begin{table}
  \centering
  \begin{tabular}{|c|cccccccccc|}
    \hline\hline
      $J'_K$ & 
0.8    &  
0.6    &  
0.525  &  
0.45   &  
0.41   &  
0.37   &  
0.30   &  
0.25   &  
0.225  & 
0.20   \\
    \hline
      $\xi_K$ $\RR$ even& 
 1.89      & 
 5.58      & 
 9.30      & 
 17.40     & 
 25.65     & 
 40.5      & 
 111       & 
 299       & 
 $\sim$565     & 
 $\sim$1196    \\
    \hline
      $\xi_K$ $\RR$ odd &
 1.65      &
 5.45      &
 9.30      &
 17.40     &
 25.65     &
 39.2      &
 127       &
 411       &
 $\sim$ 870  & 
 $\sim$ 2200  \\

  \hline
  \end{tabular}
  \caption{The numerically determined values for $\xi_K(J_K')$ using naive rescaling of $S_{imp}(J_K',r,\RR)$ at fixed $r/\RR$ for $J_2=J_2^c$.
  For $\RR$ odd system sizes of $\RR=19\ldots 101$ have been used and for $\RR$ even $\RR=18\ldots 102$. The estimates become unreliable
  once $\xi_K\gg\RR$.}\label{tab:xiKScal}
\end{table}

We expect $S_{imp}$ to be a scaling function of only 2 parameters $S_{imp}(r/\xi_K,r/\RR)$, or, equivalently $S_{imp}(R/\xi_K,r/\RR)$
for $r,R,\xi_K\gg 1$. It should
therefore also be possible to test this scaling by looking directly at $S_{imp}$ for fixed $\RR/\xi_K$ which should be a function of the single
variable $r/\RR$. We use the previously determined $\xi_K$ listed in table~\ref{tab:xiKScal} to test this assumption by selecting 2 sets of data,
$\RR=400, J'_K=0.41, \xi_K=25.65, \RR/\xi_K=15.59$ and $\RR=86, J'_K=0.60, \xi_K=5.58, \RR/\xi_K=15.41$.
\begin{figure}[!ht]
\begin{center}
\includegraphics[width=10cm,clip]{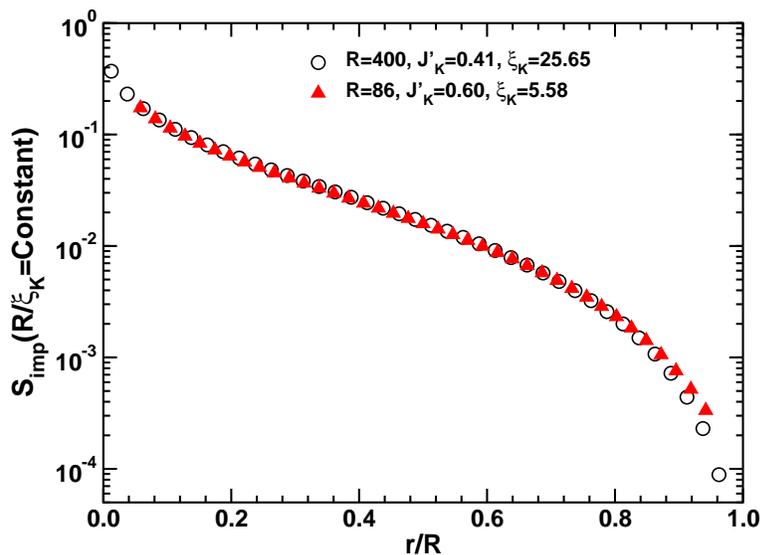}
\end{center}
\caption{DMRG results for $\RR=400, J'_K=0.41, \xi_K=25.65$ yielding $ \RR/\xi_K=15.59$ (open circles) and $\RR=86, J'_K=0.60, \xi_K=5.58$ with the
  ratio $\RR/\xi_K=15.41$ (full triangles) plotted versus $r/\RR$. $m=256$ states was kept in the DMRG calculation.
}
\label{fig:ScalRovxiK}
\end{figure}
Our results are shown in Fig.~\ref{fig:ScalRovxiK} where we observe an excellent data collapse
using the previously determined values of $\xi_K$ confirming the expected scaling form. Combined with the
results presented in \cite{Sorensen06}, this provides rather strong numerical evidence for the presence of
the large length scale $\xi_K$ in the impurity entanglement entropy and demonstrates the scaling picture
arising from the universal aspects of Kondo physics.

\subsection{The Fixed Point Entanglement}
It is instructive to consider what might be termed the fixed point entanglement,
the entanglement occurring at the two fixed points at weak coupling $J_K'=0$ and at strong coupling $J_K'=1$.
When $R$ is odd and $J'_K=0$ the impurity spin is free while the remainder of the system is in a singlet ground-state.
Due to the subtraction in Eq.~(\ref{Simpdef}) $S_{imp}$ as well as the alternating part $S^A_{imp}$ is then
zero. Surprisingly, as was shown in \cite{Sorensen06}, $S_{imp}(J'_K=0,r,R)$ is non-zero when $R$ is even. We
now discuss this in detail.

We consider the case $\RR$ even and the limit $J_K^{\prime}\to 0$.
That is, out of the 4 degenerate states at $J_K'=0$ we focus only
on the singlet state, which is uniquely picked out by this limiting procedure.
This state has a non-zero entanglement of the impurity spin
with the rest of the system.
On the other hand, to calculate the impurity entanglement entropy,
defined in Eq. (\ref{Simpdef}), we must subtract $S$ for
a chain of odd length, $\RR -1$, with $J_K'=1$. This odd
length system has a spin doublet ground state. There is no simple relationship between
$S(r,\RR ,J_K'=0)$ and $S(r-1,\RR -1,J_K'=1)$.

Explicitly, we may write the spin singlet ground state for
a chain of even length $\RR$ in the form:
\begin{equation}
|\psi >= (1/\sqrt{2})[|\uparrow >|\Downarrow >-|\downarrow >|\Uparrow >].
\end{equation}
Here the first arrow refers to the state of the impurity spin and the
second, double arrow, to the total $S^z$ of all the other spins. We may
write the pure state density matrix as:
\begin{eqnarray}
\rho_P \equiv |\psi ><\psi |=\frac{1}{2}\Big[&& |\uparrow ><\uparrow
|\otimes |\Downarrow ><\Downarrow |
+|\downarrow ><\downarrow |\otimes |\Uparrow ><\Uparrow |  \nonumber \\
&&-|\uparrow ><\downarrow |\otimes |\Downarrow ><\Uparrow |
-|\downarrow ><\uparrow |\otimes |\Uparrow ><\Downarrow |\Big] .
\label{rho}
\end{eqnarray}
Now consider doing the partial trace over the spins at sites $j$ with $%
r<j\leq R$. This trace can be done on each term separately in Eq. (\ref{rho}%
). It is convenient to define 4 operators on the Hilbert space of the $R-1$
spins $2,3,\ldots R$:
\begin{equation}
\rho_{m,m^{\prime}}\equiv \Tr_B|m><m^{\prime}|.
\end{equation}
Here $m=\Uparrow$ or $\Downarrow$ (i.e. $\pm 1/2$). $\Tr_B$ means tracing
over the spins at sites $j$ with $r<j\leq R$.
\begin{equation}
\rho_{\Downarrow \Uparrow}^\dagger =\rho_{\Uparrow \Downarrow}.
\end{equation}
While $\rho_{\Downarrow \Downarrow}$ and $\rho_{\Uparrow \Uparrow}$ are
themselves reduced density matrices for the $R-1$ site model, $%
\rho_{\Downarrow \Uparrow}$ is clearly \textit{not} a reduced density matrix
since it is not Hermitian. The two density matrices $\rho_{\Downarrow
\Downarrow}$ and $\rho_{\Uparrow \Uparrow}$ are related by spin-inversion.
The total reduced density matrix, including the impurity spin, can then be
written:
\begin{equation}
\rho = \frac{1}{2}\left(
\begin{array}{cc}
\rho_{\Downarrow \Downarrow} & -\rho_{\Downarrow \Uparrow} \\
-\rho_{\Uparrow \Downarrow} & \rho_{\Uparrow \Uparrow}
\end{array}
\right).  \label{eq:rhoJK0}
\end{equation}
When considering the fixed point impurity entanglement entropy $S_{imp}(J'_k=0,r,R)$
the term we subtract in Eq.~(\ref{Simpdef}), the entanglement entropy
for a chain of odd length $R-1$ with $J_K'=1$,
 is given by $-Tr\rho_{\Uparrow \Uparrow}\log \rho_{\Uparrow \Uparrow}$
(or $-Tr\rho_{\Downarrow \Downarrow}\log \rho_{\Downarrow
\Downarrow}$). Since, $\rho_{\Uparrow \Downarrow}$ is non-zero in
the case at hand we see that $S_{imp}(J'_k=0,r,R)$ is non-zero for
$R$ even, also in the limit $R\to\infty$. The resulting
$S_{imp}(J'_k=0,r/R)$ was numerically calculated at $J_2^c$ in
\cite{Sorensen06} using DMRG methods and was shown to crossover in
an approximately linear manner from $\ln(2)$ at small $r/R$ to
zero at $r/R=1$. We show additional results in
section~\ref{sec:numres}. This cross-over can be
understood~\cite{Sorensen06} in terms of a term corresponding to
the zero temperature thermodynamic impurity entropy $\ln g$ shown
to be present in the entanglement entropy at conformally invariant
boundary fixed points~\cite{Cardy04}.

If we now consider $S_{imp}(J'_K=1,r/R)$ it will, up to a sign change, be the same for both $R$ even and $R$ odd since it
is the difference in the uniform part of the entanglement entropy for an odd and even sized system. Since we expect the
uniform part of the entanglement entropy to be independent of the parity of $R$ in the thermodynamic limit we expect
$S_{imp}(J'_K=1,r/R)$ to approach zero for increasing $R$ as was demonstrated numerically in \cite{Sorensen06}.
We show additional results for $S_{imp}(J'_K=1,r/R)$ at $J_2^c$ in section~\ref{sec:numres}.

\section{Impurity Valence Bonds (IVB) and Single Particle Entanglement (SPE)\label{sec:SPEIVB}}
In this section we define two heuristic quantities, the impurity valence bond (IVB) and the single
particle entanglement (SPE). As we shall see these quantities capture essential parts of the impurity entanglement
as described in the previous two sections.
The IVB was first discussed in \cite{Sorensen06} and closely related ideas were developed by Refael and Moore in \cite{Refael04}
(see also Ref.~\cite{alet07}).
Here we recapitulate some of the results and further develop the ideas. In section~\ref{sec:ssansatz} a much more complete
development valid in the dimerized phase, $J_2>J_2^c$, showing more rigorously the presence of IVB and SPE terms in
the entanglement entropy.

We start by discussing the single particle entanglement. We think in terms of a tight binding model describing
a finite chain with a single particle present in a state where the particle has probability $p$ for being in region $A$
and $(1-p)$ for being in region $B$. The wave-function can quite generally
be written $|\Psi\rangle=\sum_i\psi_i|i\rangle$, with $|i\rangle$ the coordinate space states, 
from which it follows that $p=\sum_{i\in A}|\psi_i|^2$.
The reduced density matrix for region $A$ can then be written
$p|1\rangle\langle 1|+(1-p)|0\rangle\langle 0|$ where $|1\rangle$ is the state with the particle in region $A$ and
$|0\rangle$ the state with the particle in region $B$ (absent from $A$). It immediately follows that the entanglement entropy is given
by:
\begin{equation}
S_{\rm SPE}=-p\ln p-(1-p)\ln(1-p).
\label{eq:SPE}
\end{equation}
We shall refer to this as the single particle entanglement
contribution to the entanglement and we imagine that if a free
spinon (or soliton) is present in the ground-state it should give
rise to such a contribution to the {\it uniform} part of the
entanglement entropy. Such a term would be present in the uniform
part of the entanglement entropy for a uniform chain ($J'_K=1$)
for $R$ odd, where a single unpaired spin is present, but would likely be negligible for $R$ even at
$J_2^c$.  If we assume that the single spinon (soliton) picture is relevant at $J_2^c$ it follows that
$S_{imp}(J'_k=1,r/R)$ for $R$ {\it odd} should be given by the
SPE. In \cite{Sorensen06} this was shown {\it not} to be the case
but as we shall see  it the SPE {\it is}
a very good approximation in the dimerized phase 
and in particular at the Majumdar-Ghosh point.

\begin{figure}[th]
\begin{center}
\includegraphics[height=3.5cm,clip]{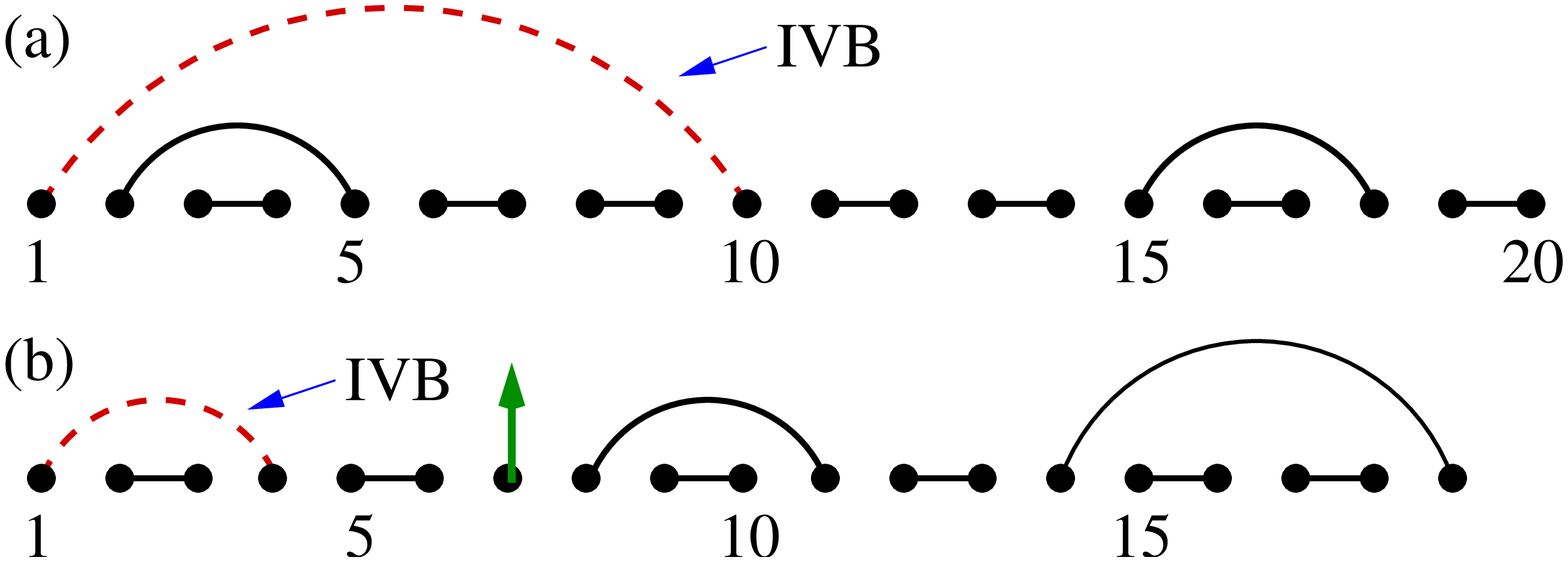}
\end{center}
\caption {
Typical impurity  valence bond configurations for $R$ even (a) and
$R$ odd (b). Note the unpaired spin on the $7^{\rm th}$ site in (b).
}
\label{fig:ivb}
\end{figure}

The other component of the heuristic picture is the formation of
an ``impurity valence bond" (IVB) between the impurity and a site
in the chain.  See Fig.~\ref{fig:ivb}. When the IVB is cut by the
boundary between regions $A$ and $B$ we expect it to give rise to
a contribution of $\ln 2$ to the impurity part of the entanglement
entropy. If the IVB does {\it not} cut this boundary the
contribution to $S_{imp}$ arising from the IVB is zero. We then
see that:
\begin{equation}
S_{imp}=(1-p)\ln 2,
\end{equation}
where $p$ is the probability that the IVB connects the impurity spin to a site in region $A$.
In the limit where $J_k'\to 0$, the probability will, as above, be given by 
$p=\sum_{i\in A}|\psi_i|^2$ where $\psi_i$ now describe the wave-function of a single unpaired spin (soliton).
At $J_2^c$ we expect the fixed point entanglement entropy $S_{imp}(J'_k=0,r/R)$ for $R$ even to follow
this form. In \cite{Sorensen06} this was shown to work well even at $J_2=J_2^c$.
Care has to be given to the fact that the parity of $R$ will influence this picture and
we now discuss the different situations in a preliminary manner, postponing a more complete
discussion to section~\ref{sec:ssansatz}.

{\it $R$ even and $J_K' < 1$:} This situation is shown in Fig.~\ref{fig:ivb}(a). If $J_K'=0$ we would expect the probability of forming an IVB
between the impurity and a given site in the chain to be almost uniform throughout the chain. Hence, $S_{imp}$ should decrease
approximately linearly with $r/R$ from $\ln 2$ to zero as mentioned above. For $J_K'$ small but non-zero we expect the typical size of an IVB to be of
the order of $\xi_K$ when $\xi_K<R$. In that case $S_{imp}$ should decrease monotonically to zero for $r/\xi_K\gg 1$. For $r/\xi_K,r/R\ll 1$ we would expect it
to approach $\ln 2$. This behavior is largely confirmed by the numerical DMRG calculations.

{\it $R$ odd and $J_K' < 1$:} This situation is shown in
Fig.~\ref{fig:ivb}(b). Since $R$ is odd and the total spin of the
system is $1/2$ there is an unpaired spin present in the
ground-state. When $\xi_K\gg R$ the unpaired spin is the impurity
spin and we simply have $S_{imp}=0$. As $\xi_K$ decreases the
probability of creating an IVB {\it increases} due to screening of
the impurity. However, the average length of the IVB when it is
present {\it decreases} with $\xi_K$. In section~\ref{sec:simpscal} we
showed clear numerical evidence for a monotonically decreasing
$S_{imp}$ with $r$ for $\RR /\xi_K$ constant, however, if we instead now imagine keeping $r/R$ fixed
and varying $R/\xi_K$ we see that the two above effects will trade off
and give rise to a maximum in $S_{imp}$ for $\xi_K\approx R$~\cite{Sorensen06} with $R$ odd.

Eventually, in the limit $R\to\infty$, the parity of $R$ no longer plays a role and we obtain the same $S_{imp}(r/\xi_K)$ for
both $R$ even and $R$ odd. However, for mesoscopic systems these effects could be of importance.

\subsection{The Entanglement entropy at the MG point}
While the entanglement entropy at the critical point, $J_2^c$, is less amenable to the
heuristic approach of this section, this intuitive picture sheds considerable light on
the entanglement in the dimerized phase occurring for $J_2>J_2^c$ as we now discuss. 
In fact, almost exact expressions for the entanglement entropy can be obtained for
the Majumdar Ghosh model. In section~\ref{sec:ssansatz} we present a much more detailed
approach based on a variational wave-function. At the MG point, the heuristic expression
developed in this section for the entanglement entropy, follow directly from the
variational approach. 
\begin{figure}[!ht]
\begin{center}
\includegraphics[height=3.5cm,clip]{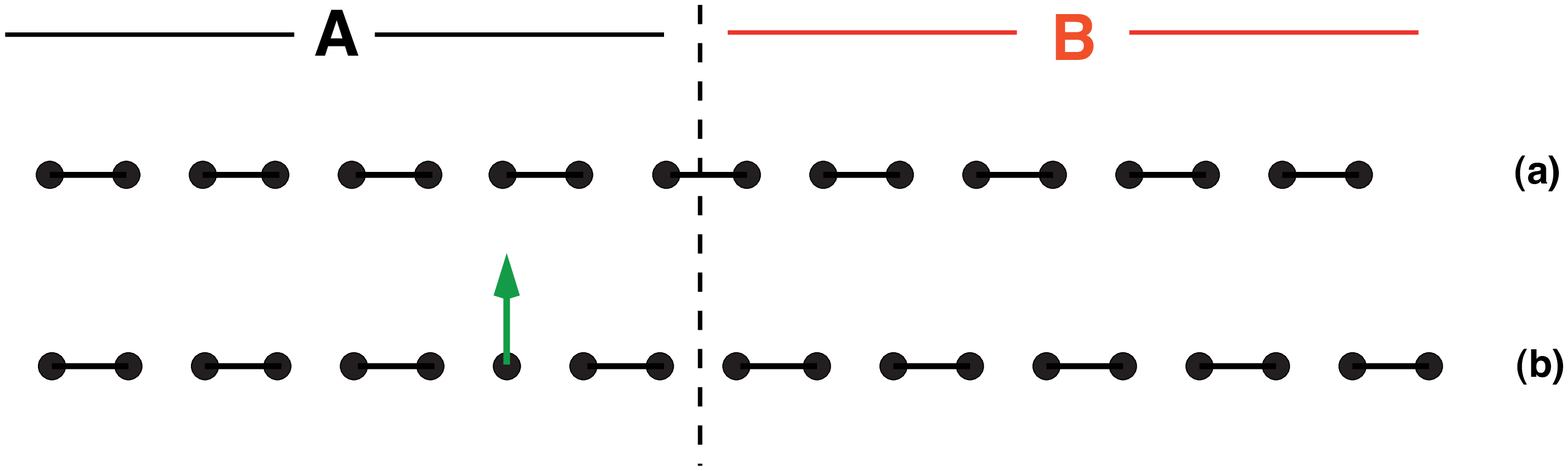}
\end{center}
\caption {MG state divided in two regions $A$ and $B$. (a) The total number of site $R$ is even and the ground-state is a singlet. (b) When $R$ is odd, the ground-state contains a soliton 
${\color{green}\uparrow}$.}
\label{fig:MG}
\end{figure}

We now focus on the Majumdar-Ghosh model, $J_2=J/2$.
Consider first the case $J_K'=1$ and $\RR$ even. 
Then the ground state is a trivial nearest neighbor valence bond state on 
every link between sites $2i-1$ and $2i$, for $i=1,2,3\ldots \RR/2$, as depicted in Fig.~\ref{fig:MG} (a).  When $r$ is even, $S=0$ since the 
ground state is a direct product of singlet states in regions $A$ and $B$. 
On the other hand, for $r$ odd, there is one valence bond, between sites $r$ and $r+1$,
connecting regions $A$ and $B$. In this case $S=\ln 2$, as exemplified in Fig.~\ref{fig:MG} (a). 

Now consider $J_K'=1$ but $\RR$ odd. The ground state 
now contains one soliton (unpaired spin) on one of the {\it odd} sites. This soliton 
can be in region $A$ (see Fig.~\ref{fig:MG} (b)) with probability $p(r)$ or region $B$ with probability 
$1-p$, where:
\be p \equiv \sum_{n \in A} |\psi_n^{sol}|^2 ,\ee
and $\psi_n^{sol}$ describe the amplitude for the soliton to be on site $n$.
The fact that regions $A$ and $B$ are sharing the soliton produces 
a ``single particle entanglement'' precisely as outlined above:
\be S_{SPE}=-p\ln p -(1-p)\ln (1-p).\ee 
However, this is not the whole story.  We must also take into account 
that there may be a nearest neighbor valence bond connecting regions 
$A$ and $B$. Whether or not this is present depends both on 
the parity of $r$ and on whether the soliton is in region $A$ or $B$. 
If $r$ is even, then this valence bond is present when the 
soliton is in $A$.  Conversely, if $r$ is odd, then it is 
present when the soliton is in region $B$. When this valence 
bond is present, it contributes an additional $\ln 2$ to $S$.  
The probability of it being present is $p$ for $r$ even and $1-p$ for $r$ odd. 
Adding this extra term we obtain:
\be 
S(r)=-p\ln p-(1-p)\ln (1-p)+[1/2+(-1)^r(p-1/2)]\ln 2.
\label{eq:HeurSodd}
\ee

As remarked earlier, when $J_K'=0$ and $\RR$ is odd, the impurity site 
is unentangled with the rest of the chain which contains 
only nearest neighbor valence bonds, between sites $2i$ and $2i+1$.
The only source 
of entanglement between $A$ and $B$ is a valence bond 
from site $r$ and $r+1$  when $r$ is even. Thus
\be S=(1/2)[1+(-1)^r]\ln 2.\ee

Now, consider the case $J_K'=0$ and $\RR$ even. Then there is an impurity 
valence bond stretching from site $1$ to some even site. All 
other valence bonds have length $1$.  
The right hand member of the impurity valence bond is again a soliton 
separating the $2$ different nearest neighbor valence bond ground states, 
but it now forms a singlet with the spin at site $1$. The soliton 
again contributes its single particle entanglement.
 Since the soliton forms a singlet with site $1$ this 
contributes $\ln 2$ to $S$ when the IVB terminates in region $B$, 
with probability $[1-p(r)]$ but make no contribution when it 
terminates in region $A$. Furthermore, there may be an 
additional nearest neighbor valence bond entangling regions $A$ and $B$ and 
contributing another $\ln 2$ to $S$. When 
$r$ is even, this occurs when the IVB terminates in region $B$, with 
probability $(1-p)$.  When 
$r$ is odd it occurs when the IVB terminates in region $A$, with probability $p$. 
Combining the contributions of the SPE, the IVB 
and the possible nearest neighbor 
valence bond gives:
\be 
S=-p\ln p - (1-p)\ln (1-p)+[3/2-p+(-1)^r(1/2-p)]\ln 2.
\label{eq:HeurSeven}
\ee

Both Eqs~(\ref{eq:HeurSeven}) and (\ref{eq:HeurSodd}) are shown to agree very
well with numerical results at the MG-point presented in section~\ref{sec:numres}.

\section{$s_{imp}$ and Weak Scaling Violations~\label{sec:smallimp}}

The simplest measure
of how a qubit is entangled with the environment would be to take
system $A$ to be the impurity spin (qubit) itself and
regard the environment as region $B$~\cite{Cho06}. Often this is referred to
as single site entanglement. If system $A$ only contains the single impurity
spin at the boundary of the chain it becomes impossible to extract the uniform
part and we can therefore not use Eq.~(\ref{Simpdef}) to
define the impurity entanglement. Instead we have to use the complete entanglement
entropy including both uniform and alternating parts. To distinguish it from $S_{imp}$ in Eq.~(\ref{Simpdef}) we therefore denote it
by $s_{imp}$ and define it as $S(J_K^{\prime},1,\RR)$ with {\it no} subtraction.
We now analyze this quantity.

First consider the case of $\RR
$ even. Then the ground state is a spin singlet. Since the impurity has
spin-1/2, as does the rest of the system, we can write the spin-zero ground
state in the form:
\begin{equation}
|\psi>=(1/\sqrt{2})[|\uparrow >\otimes |\Downarrow >-|\downarrow >\otimes
|\Uparrow >],
\end{equation}
where the single arrow labels the state of the impurity and the double arrow
the state of the rest of the system. Tracing out the rest of the system
gives a two-dimensional density matrix for the impurity spin which is
diagonal with elements $1/2$ and hence a maximal entanglement entropy of $%
\ln 2$. The case of odd $\RR$ is more interesting. Now the ground
state is a doublet with total spin $S_T=1/2$. Let us focus on the state with
$S^z_T=+1/2$. This must have the form:
\begin{equation}
|\psi > = a|\uparrow >\otimes |0,0>+b|\downarrow >\otimes |0,1>,
\end{equation}
where $|0,0>$ denotes a spin singlet state of the rest of the system and $%
|1,1>$ denotes an $S=1$, $S^z=1$ state of the rest of the system. All states
are normalized to $1$, so it follows that $|a|^2+|b|^2=1$. The density
matrix is again diagonal with matrix elements $|a|^2$ and $|b|^2$ and hence
\begin{equation}
s_{imp}=-|a|^2\ln |a|^2-|b|^2\ln |b|^2.
\end{equation}
On the other hand, the magnetization of the impurity in the ground states is
given by:
\begin{equation}
m_{imp}=(|a|^2-|b|^2)/2=(2|a|^2-1)/2.
\end{equation}
Thus we may write:
\begin{equation}
s_{imp} = -\sum_{\pm}(1/2\pm m_{imp})\ln [(1/2\pm m_{imp})].
\label{eq:single}
\end{equation}
(This formula is also trivially true for the case $\RR$ even in
which case $m_{imp}=0$.) For $\RR$ odd, $m_{imp}$, and hence $s_{imp}$,
shows an interesting dependence on $\RR$ which reflects Kondo
physics. $m_{imp}$ was studied, for the usual fermion Kondo model, in
\cite{Sorensen96} and \cite{Barzykin98,Barzykin99} for example. For weak
Kondo coupling and relatively short chains, $\RR\ll \xi_K$, $%
m_{imp}\approx 1/2$. On the other hand, for stronger Kondo coupling or
larger chains, the magnetization is progressively transferred from the
impurity to the rest of the chain, associated with screening of the
impurity. In the limit of an infinite \textit{bare} Kondo coupling, $%
m_{imp}=0$ since the impurity spin then forms a singlet with one other
electron and all of the magnetization resides in the other electrons, which
have individual magnetization of order $1/\RR$.

Nonetheless, $m_{imp}$ is \textit{not} a scaling function of $\RR%
/\xi_K$. This is associated with the fact that the operator $\vec S_{imp}$
has an anomalous dimension\cite{Barzykin98,Barzykin99}, $\gamma_{imp}(\lambda_K)$. Thus
it obeys the renormalization group equation:
\begin{equation}
\left[\RR\frac{\partial}{\partial \RR} +\beta (\lambda_K)%
\frac{\partial}{\partial \ \lambda_K} +\gamma_{imp}(\lambda_K)\right]
m_{imp}(\RR,\lambda_K)=0.  \label{RGm}
\end{equation}
The weak coupling $\beta$-function, the variation of the effective
Kondo coupling as we vary the length scale is:
\begin{equation}
\frac{d\lambda_K}{d\ln L}=-\beta (\lambda_K)=\lambda_K^2-\frac{1}{2}%
\lambda_K^3+\ldots\label{beta1}
\end{equation}
We then see that if $%
\gamma_{imp}(\lambda_K)$ were zero, (\ref{RGm}) would imply that $m_{imp}$
was a function of $\lambda_{eff}(\RR)$ only, or equivalently a
function of $\xi_K/\RR$ only. However, a non-zero $\gamma_{imp}\neq
0 $, as occurs here, implies scaling violations. At lowest
non-vanishing order \cite{Abrikosov70,Barzykin98,Barzykin99} in $\lambda_K$,
\begin{equation}
\gamma_{imp}(\lambda_K) = \lambda_K^2/2 + \ldots  \label{gamma}
\end{equation}
The general solution of (\ref{RGm}) is:
\begin{equation}
m_{imp} = \exp \{\int_0^{\lambda_K^0} [\gamma (\lambda_K)/\beta (\lambda_K)]
d\lambda_K\} f(\xi_K/\RR),  \label{mimpscale}
\end{equation}
where $f$ is some function of $\xi_K/\RR$ or equivalently of $%
\lambda_{K}(\RR)$. From (\ref{beta1}) and (\ref{gamma}) we see
that
\begin{equation}
\exp\{\int_0^{\lambda_K^0} [\gamma (\lambda_K)/\beta (\lambda_K)]
d\lambda_K\} = 1+\lambda_K^0/2 + O[(\lambda_K^0)^2].
\end{equation}
The fact that $m_{imp}$ has this residual dependence on the bare coupling
which cannot be adsorbed into the renormalized coupling at scale $\RR
$ implies a violation of scaling. However, since this effect vanishes as the
bare coupling, $\lambda_K^0\to 0$, we refer to it as a ``weak scaling
violation''.

\begin{figure}[!ht]
\begin{center}
\includegraphics[height=7cm,clip]{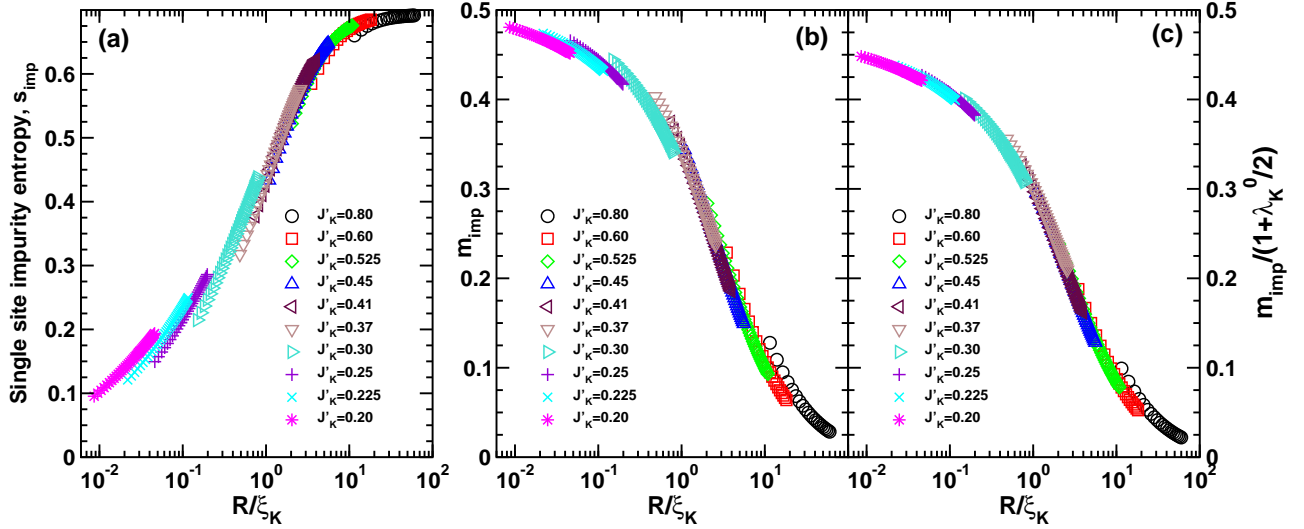}
\end{center}
\caption{
(a) Weak scaling violations for the single site impurity entanglement entropy $s_{imp}$ (a) [Eq.~(\ref{eq:single})] 
and the local impurity magnetization $m_{imp}=\langle S_{1}^{z}\rangle$ (b). 
In panel (c) is shown an improved scaling plot of 
$m_{imp}/(1+\lambda_K^0/2)$.
The $\xi_K$ used in all panels are from table~\ref{tab:xiKScal}.
All data are for odd length chains between $R=19\ldots 101$.
}
\label{fig:violation1}
\end{figure}
This form of $m_{imp}$, exhibiting weak scaling violation, can be confirmed
by an explicit perturbative calculation. This calculation was actually
presented earlier in \cite{Sorensen96} for the 1D lattice version of
the fermionic Kondo Hamiltonian:
\begin{equation}
H=-t\sum_{j=1}^{\RR-2}(\psi_j^\dagger \psi_{j+1}+h.c.) +J_K\vec
S_{imp}\cdot \psi^\dagger_1\frac{\vec \sigma}{2}\psi_1.
\end{equation}
In \cite{Sorensen96} $\RR-1$ was called $L$. The result was:
\begin{equation}
m_{imp}=\frac{1}{2}-\left[\frac{J_K}{(L+1)}\right]^2\sum_{k,k^{\prime}}\theta
( \epsilon_{k^{\prime}})\theta (-\epsilon_k)\left[ \frac{\sin k \sin
k^{\prime}}{ \epsilon_k-\epsilon_{k^{\prime}}}\right]^2.  \label{mimppert}
\end{equation}
Here $\theta (x)$ is the Heavyside or step function, and $\epsilon_k=-2t\cos
k$ is the dispersion relation. The allowed values of $k$ and $k^{\prime}$
occurring in the sum in (\ref{mimppert}) are $k=\pi n/\RR$ for $%
n=1,2,3,\ldots$. Assuming $\RR\gg 1$, we replace the sums by
integrals. The integrals diverge logarithmically at $k=k^{\prime}=k_F=\pi /2$%
. Letting $k=\pi /2-q$ and $k^{\prime}=\pi /2-q^{\prime}$, the integrals
near $q, q^{\prime}=0$ take the form:
\begin{eqnarray}
m_{imp}&\approx& \frac{1}{2}-\left[ \frac{J_K}{\pi v}\right]^2 \int_0^\infty
dq\int_0^\infty dq^{\prime}\frac{1}{(q+q^{\prime})^2}  \nonumber \\
&\approx & \frac{1}{2}-\left[ \frac{J_K}{\pi v}\right]^2 \int_0^\infty \frac{%
dq}{q}  \nonumber \\
&\approx & \frac{1}{2}-\left[ \frac{J_K}{\pi v}\right]^2\ln (\RR/a).
\end{eqnarray}
Here we have replaced the small $q$ limit of the integral by $1/\RR$
and the large $q$ limit by a short distance cut off, $a$, of order a lattice
spacing. The continuum limit of this tight binding model gives the
Hamiltonian of (\ref{H1D})
with $\lambda_K=J_K/\pi t$ and $v=2t$, so we may
write this as:
\begin{equation}
m_{imp} \approx \frac{1}{2}-\left[ \frac{\lambda_K^0}{2}\right]^2\ln (
\RR/a).
\end{equation}
The lowest order correction to the effective coupling as determined by the
first (quadratic) term in the $\beta$-function is:
\begin{equation}
\lambda_K(\RR) \approx \lambda_K^0+(\lambda_K^0)^2\ln (\RR%
/a^{\prime}) +\ldots ,
\end{equation}
where $a^{\prime}$ is another short distance cut off. Thus we see that, to $%
O[(\lambda_K^0)^2]$ we can write:
\begin{eqnarray}
m_{imp}&\approx& \left[ 1+\frac{\lambda_K^0}{2}\right]\left[ \frac{1}{2} -%
\frac{\lambda_K^0}{4}-\frac{(\lambda_K^0)^2}{4} \ln (\RR%
/a^{\prime})\right]  \nonumber \\
&& \left[ 1+\frac{\lambda_K^0}{2}\right]\left[ \frac{1}{2}- \frac{\lambda_K(%
\RR)}{4}\right] .
\end{eqnarray}
This has the form of (\ref{mimpscale}) with the scaling function:
\begin{equation}
f[\lambda_K(\RR)] \approx \frac{1}{2}-\frac{\lambda_K(\RR)}{4%
} +\ldots \approx \frac{1}{2}-\frac{1}{4\ln (\RR/\xi_K)}+\ldots .
\end{equation}
The $\ldots$ represents terms of $O(\lambda_K(R)^2$ and higher in the
effective coupling at scale $\RR$.

The fact that $m_{imp}$ exhibits weak scaling violations implies that $s_{imp}$
does also, as displayed in Figs.~\ref{fig:violation1}(a).
In Fig.~\ref{fig:violation1} we show DMRG results obtained at the critical
point $J_2=0.2412$ for various odd lengths $19\le \RR\le 101$ and Kondo
couplings $0.2\le J'_K\le 0.8$.  
The local magnetization at the impurity site
$m_{imp}=\langle S_{1}^{z}\rangle$ is shown in Fig.~\ref{fig:violation1}(b)
as a function of $\RR/\xi_K$, using the values of $\xi_K(J'_K)$
determined previously (table~\ref{tab:xiKScal}). Clearly both $s_{imp}$ and  $m_{imp}$ 
violate scaling since the various curves \textit{do not} fall on top of each other. 
As outlined above we expect $m_{imp}/(1+\lambda_K^0/2)$ to scale much better and
this is shown in Fig~\ref{fig:violation1}(c). Some deviations from scaling are still
visible but clearly the scaling has improved. We expect the remaining discrepancies
could be improved upon by using more optimal values for $\xi_K$ instead of the ones determined
from other scaling plots which have significant uncertainties associated with them
for either very large or very small $\xi_K$.
%
%

\begin{figure}[!ht]
\begin{center}
\includegraphics[width=10cm,clip]{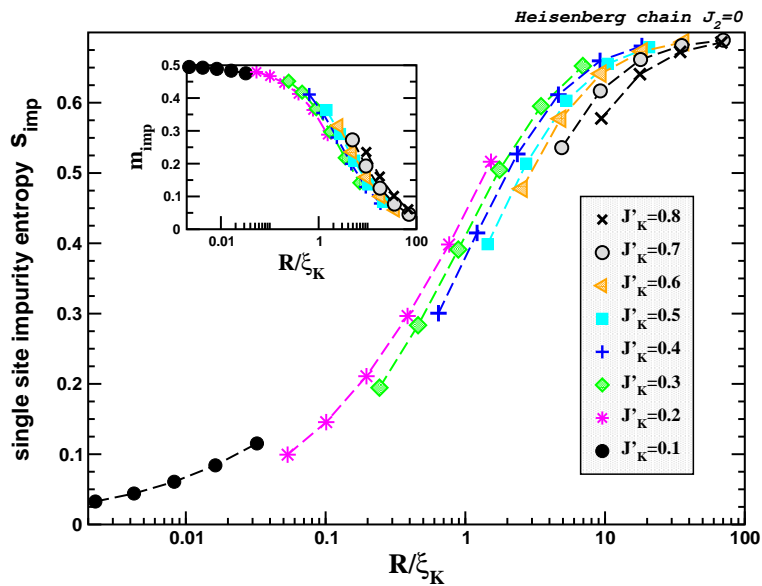}
\end{center}
\caption{Weak scaling violations for the local impurity magnetization 
  $m_{imp}=\langle S_{1}^{z}\rangle$ (inset) and the single site impurity
    entanglement entropy $s_{imp}$ [Eq.~(\ref{eq:single})] (main panel). 
    QMC results obtained for the critical spin chain [model (\ref{eq:spinch})] at $J_2=0$ with 
    various odd lengths $\RR$ and Kondo couplings $J'_K$ indicated on the plot. 
    The scaling violations are shown as a clear non-universality as a function of  $\RR/\xi_K(J'_K)$}
\label{fig:violation2}
\end{figure}

The Kondo physics also is expected for $J_2\leq J_2^c$
and therefore also for the unfrustrated chain at $J_2=0$~\cite{KEAC}.
The unfrustrated spin chain model can be investigated using Quantum
Monte Carlo (QMC) methods as an alternative to the DMRG
computations. Since QMC works for any temperature, it allows to
investigate finite temperature scaling properties like the spin
susceptibility (see Ref.~\cite{KEAC}) from where we can
get precise estimates for the Kondo temperature $T_K=v_s/\xi_K$.
These estimates, given in table~\ref{tab:NN} (along with estimates
obtained in Ref.~\cite{KEAC} by solving the Bethe Ansatz
equations for this model~\cite{Frahm97}), are then used to check
the scaling violations at $J_2=0$ of the local magnetization at
the impurity site as well as the single site impurity entanglement
entropy. As in the frustrated case discussed above, also here the
weak scaling violations are clearly present. In
Fig.~\ref{fig:violation2}, we show the QMC results obtained for
various odd lengths $9\le\RR\le 257$ and Kondo couplings $0.1\le
J'_K\le 0.8$ where $m_{imp}$ violates scaling as well as
$s_{imp}$.
\begin{table}
  \centering
  \begin{tabular}{|c|cccccccc|}
    \hline
      $J'_K$ & 
$0.1$&
$0.2$&
$0.3$&
$0.4$&
$0.5$&
$0.6$&
$0.7$&
$0.8 $\\
  \hline
      $\xi_K$&
$\sim 4\times 10^3$&
$ 168$&
$ 37$&
$ 14$&
$ 6.24$&
$ 3.5$&
$ 1.8$&
$0.95$\\
  \hline
  \end{tabular}
  \caption{$\xi _{K}(J_{K}^{\prime })$ for $J_2=0$, as determined from scaling of the finite temperature 
    impurity susceptibility and Bethe Ansatz calculations (reported in Ref.~\cite{KEAC}).}
  \label{tab:NN}
\end{table}

\subsection{Discussion}
In \cite{Barzykin98,Barzykin99} a rather general discussion was given of which
physical quantities are given by pure scaling functions and which ones
exhibit weak scaling violations. In general quantities defined far from the
impurity, such as the local susceptibility, are pure scaling functions
whereas those that depend explicitly on the impurity spin operator exhibit
weak scaling violations. In particular, the impurity susceptibility and
specific heat or (thermodynamic) entropy are given by pure scaling
functions. This is because they only involve conserved quantities, the total
spin and Hamiltonian operators which have zero anomalous dimension. (Roughly speaking, for the
case of the total spin operator, the anomalous dimension of
the impurity spin operator is canceled by the anomalous dimension of the
electron spin density operator near the impurity.) We expect 
 the impurity entanglement entropy, $S_{imp}(J_K',r,\RR)$ to be
a pure scaling function when $r\gg 1$, since it is a quantity defined far
from the impurity.
Indeed, in the approach of Cardy and Calabrese, reviewed 
in Sec.~\ref{sec:FLT}  the entanglement 
entropy is obtained from Green's functions of 
an operator $\Phi_n$, which is inserted at the positions $\pm r$. Since 
this operator is inserted at a location far from the boundary, 
we expect that it will not have any anomalous dimension associated 
with the Kondo interaction. Therefore the one-point function 
obeys the RG equation of Eq. (\ref{RGm}) with zero anomalous 
dimension, implying that it is a pure scaling function. 
Indeed our DMRG data is
consistent with such scaling behavior as we have shown in Fig. (\ref{fig:ScalRovxiK}).

\section{Fermi Liquid Theory for $S_{imp}$~\label{sec:FLT}}

In the limit $\xi _{K}\ll r$, within the FLT as outlined in \ref{app:field_theory}, we can
also calculate $S_{imp}$ by treating $H_{int}$ in Eq.~(\ref{Hintcon}) in
lowest order perturbation theory. The entanglement entropy is obtained by
means of the replica trick. If Tr$\rho (r)^{n}$ is known,
\begin{equation}
S=-\lim_{n\to 1}\frac{d}{dn}[{\rm Tr}\rho (r)^{n}],  \label{rep}
\end{equation}
where $\rho (r)$ is the reduced density matrix for the subsystem $(0,r)$. In
the path integral representation of Euclidean space-time,
\begin{equation}
{\rm Tr}\rho (r)^{n}=\frac{Z_{n}(r)}{Z^{n}},
\end{equation}
where $Z_{n}(r)$ is the partition function on an $n$-sheeted
Riemann surface ${\cal R}_{n}$, with the sheets joined at the cut
extending from $r$ to ${R}$ \cite{Wilczek94, Cardy04}. Now the
original problem has been transformed into the calculation of the
partition function with a nontrivial geometry. We use 
the approach where the Hamiltonian is written in terms 
of left movers only, obeying PBC on an interval of length $2\RR$. 
(See \ref{app:field_theory}). 
 In the critical
region, the system is conformally invariant and CFT methods are
applicable. Starting with zero temperature, the $n$-sheeted
Riemann surface ${\cal R}_{n}$, can be mapped to the usual complex
plane ${\cal C}$ \cite{Cardy04}. Then the expectation value of the
energy momentum tensor $T$ on ${\cal R}_{n}$ is simply given by
the Schwartzian derivative:
\begin{equation}                                                                                                                                                          
\langle T(w)\rangle _{\mathcal{R}_{n}}=\frac{\Delta                                                                                                                       
  _{n}(u_{1}-u_{2})^{2}}{(w-u_{1})^{2}(w-u_{2})^{2}},  \label{ExpectationT}                                                                                                 
  \end{equation} 
where $\Delta _{n}=(c/24)(1-(1/n)^{2})$ and we have $u_{1}=ir$ and                                                                                                        
$u_{2}=-ir$ here.                                                                                                                                                         
Then Calabrese and Cardy~\cite{Cardy04} observed its important connection to the
correlators on ${\cal C}$ through the Ward identity:
\begin{eqnarray}
\left\langle T(w)\right\rangle _{{\cal R}_{n}} &\equiv &\frac{\int [d\phi
]T(w)e^{-S_{E}({\cal R}_{n})}}{\int [d\phi ]e^{-S_{E}({\cal R}_{n})}}
\nonumber \\
&=&\frac{\left\langle T(w)\Phi _{n}(r)\Phi _{-n}(-r)\right\rangle _{{\cal C}}%
}{\left\langle \Phi _{n}(r)\Phi _{-n}(-r)\right\rangle _{{\cal C}}}.
\label{T}
\end{eqnarray}
The fictitious primary operators $\Phi _{\pm n}$ on the branch
points have the left scaling dimensions $\Delta _{n}$. They
concluded that Tr$\rho ^{n}=$ $Z_{n}/Z^{n}$ behave identically to
the $n$-th power of $\left\langle \Phi _{n}(r)\Phi
_{-n}(-r)\right\rangle _{{\cal C}}$ under the conformal mappings
or explicitly,
\begin{equation}
{\rm Tr}\rho ^{n}\cong \tilde{c}_{n}(2r/a)^{(c/12)(n-1/n)}.
\end{equation}
Applying the replica trick, the entanglement entropy is $S\sim (c/6)\ln (2r)$%
. Then they extended the result to finite system size ${R}$ or infinite
system size and finite temperature $T=1/\beta $ by applying the
corresponding conformal mapping to $\left\langle \Phi _{n}(r)\Phi
_{-n}(-r)\right\rangle _{{\cal C}}$ \cite{Cardy04}.

Now with the presence of the local irrelevant interaction Eq.~(\ref{Hintcon}%
), we should calculate perturbatively the correction to the partition
function $Z_{n}$ in order to get the impurity entanglement entropy $S_{imp}$%
. Luckily, the irrelevant interaction is just the energy momentum tensor
itself and its expectation value on the $n$-sheeted Riemann surface is just
Eq.~(\ref{ExpectationT}). The correction to $Z_{n}$ of first order in $\xi
_{K}$ is:
\begin{equation}
-\delta Z_{n}=-(\xi _{K}\pi )n\int_{-\infty }^{\infty }d\tau \langle {\cal H}%
_{s,L}(\tau ,0)\rangle _{{\cal R}_{n}}.
\end{equation}
${\cal H}_{s,L}=T/(2\pi )$ where $T(\tau ,x)$ is the conventionally
normalized energy-momentum tensor for the $c=1$, free boson conformal field
theory corresponding to the spin excitations of the original free fermion
model. After doing the simple integral and taking the replica limit, for ${R}%
\rightarrow \infty $, we get
\begin{equation}
S_{imp}=\pi \xi _{K}/(12r).  \label{FLT_ISE}
\end{equation}

In principle, in order to extend the FLT calculation to finite ${R}$, we
will need the conformal mapping from a infinite $n$-sheeted Riemann surface
to a finite one. On the other hand, we can also try to exploit Eq.~(\ref{T})
following ideas similar to Ref.~\cite{Cardy04} by applying the standard
finite size conformal mapping to both $\left\langle T\Phi _{n}\Phi
_{-n}\right\rangle _{{\cal C}}$ and $\left\langle \Phi _{n}\Phi
_{-n}\right\rangle _{{\cal C}}$. Then, in the first order perturbation,
\begin{equation}
-\delta (\frac{Z_{n}}{Z^{n}}) =-(\frac{\xi _{K}}{2})n
\int_{-\infty }^{\infty }d\tau [(\frac{\pi }{2{R}})\frac{\sinh [i\pi r/({R}%
)]}{\sinh [\frac{\pi (v\tau +ir)}{2{R}}]\sinh [\frac{\pi (v\tau -ir)}{2{R}}]}%
]^{2}.
\label{deltaZn} 
\end{equation}
We use the integral
\begin{equation}
\int_{0}^{\infty }\frac{dx}{\cosh ax-\cos t}=\frac{t}{a}\csc t,
\label{integral}
\end{equation}
from Ref.~\cite{IntegralTable} and differentiate Eq.~(\ref{integral}) with
respect to $t$ on the both sides. Applying this result and the
product-to-sum hyperbolic identity to Eq.~(\ref{deltaZn}), we can complete
the integral and after the replica limit we get:
\begin{equation}
S_{imp}=\frac{\pi \xi _{K}}{12{R}}[1+\pi (1-\frac{r}{{R}})\cot (\frac{\pi r}{%
{R}})].  \label{FLT_FSE}
\end{equation}
Of course, Eq.~(\ref{FLT_FSE}) reduces to Eq.~(\ref{FLT_ISE}) for ${R}$ $\gg
r$ and both of them agree with the scaling form of $S_{imp}$. Interestingly,
Eq.~(\ref{FLT_FSE}) can be regarded as the first order Taylor expansion in $%
\xi _{K}/r$ and $\xi _{K}/{R}$ of $S_{U}=(1/6)\ln [{R}\sin \pi r/{R}]$ with $%
r$ and ${R}$ both shifted by $\pi \xi _{K}/2$. Consistently, Eq.~(\ref
{FLT_ISE}) can also be obtained from expanding $(1/6)\ln (r+\pi \xi _{K}/2)$. 
In fact, many other quantities such as impurity susceptibility, specific
heat and ground state energy correction can be also obtained in this fashion
by shifting the size of the total system, ${R}$, to ${R}+\pi \xi _{K}/2.$

Within CFT methods, we can also calculate $S_{imp}$ for infinite
${R}$ but at finite temperature $\beta $. We apply the standard
finite temperature
conformal mapping to $\left\langle T\Phi _{n}\Phi _{-n}\right\rangle _{{\cal %
C}}$ and $\left\langle \Phi _{n}\Phi _{-n}\right\rangle _{{\cal C}}$. The
result for first order perturbation is just to replace $2{R}$ by $\beta $
and sinh by sin in Eq.~(\ref{deltaZn}) with the integral from $-\beta /2$ to
$\beta /2$. Completing the straightforward integral yields
\begin{equation}
S_{imp}=[\pi ^{2}\xi _{K}T/(6v)]\coth (2\pi rT/v),  \label{FLT_FT}
\end{equation}
valid for $T,v/r\ll T_K$.
In the intermediate temperature regime, $v/r\ll T\ll T_K$ and hence $rT\gg v$, Eq.~(\ref{FLT_FT}) approaches the
thermodynamic impurity entropy, $S_{imp}\to \pi ^{2}\xi _{K}T/(6v)=\pi
^{2}T/(6T_{K})=c_{imp}(T)$, the well-known impurity specific heat. ($%
T_{K}\equiv v/\xi _{K}$ is the Kondo temperature.) 
This is  consistent with the observation that in this limit the entanglement
entropy approaches the thermodynamic entropy as noted in Ref.~\cite{Cardy04}.
In \ref{app:FiniteT} this connection is explored in more detail and it is
argued that quite generally $S$ will approach the thermodynamic entropy
for $T\gg v/r$.

\begin{figure}[!ht]
\begin{center}
\includegraphics[width=10cm,clip]{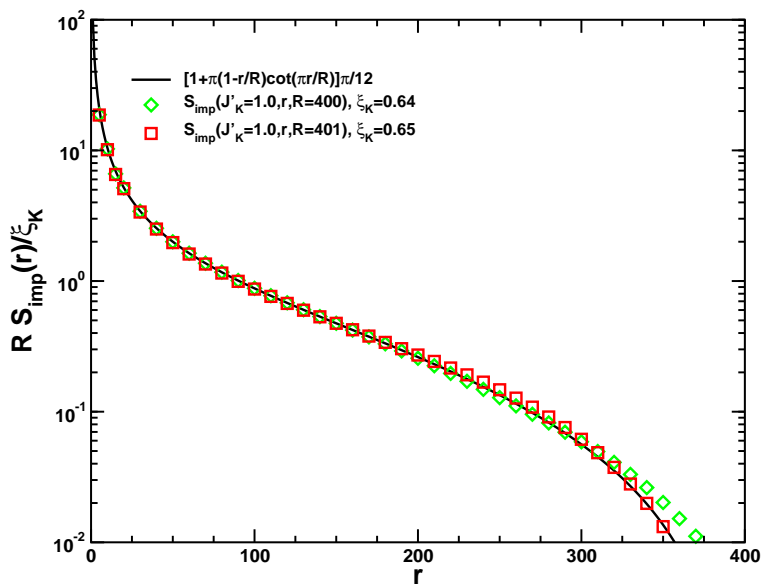}
\end{center}
\caption{$S_{imp}(J^{\prime}_K=1,r,{R})$ for ${R}=400,401$ compared to the
FLT prediction, Eq.~(\ref{FLT_FSE}).}
\label{fig:simp1FLT}
\end{figure}

The FLT expression for the uniform part of the entanglement entropy, Eq.~(%
\ref{FLT_FSE}) has been compared in detail with numerical DMRG data in Ref.~%
\cite{Sorensen06}. The resulting values for $\xi_K$ are shown in table~\ref
{tab:xiKFLT}. Excellent agreement between this expression and numerical data
for $S_{imp}(J^{\prime}_K,r,{R})$ was found for $J_K^{\prime}\lesssim 1$.
However, the fixed point impurity entanglement entropy $S_{imp}(J^{%
\prime}_K=1,r,{R})$ with $J_K^{\prime}=1$ itself should in fact be given by
this same expression for ${R}$ both even and odd. Now, with a $\xi_K=0.65$
of order ${}(1)$. In Fig.~\ref{fig:simp1FLT} we show results for $%
S_{imp}(J^{\prime}_K=1,r,{R})$ for ${R}=400,401$. In both cases do we find
good agreement with the FLT result, Eq.~(\ref{FLT_FSE}). The small
discrepancies between the numerical data and the FLT results for $r$ close
to 400 are likely due to the approximate forms used for extracting the uniform and
alternating parts of the numerical data. For $J_K^{\prime}\neq 1$ these
discrepancies are less pronounced since the numerical signal for $S_{imp}$
is larger by a factor of ${R}/\xi_K$.

It is important to note that the observed agreement implies that the fixed
point impurity entanglement entropy, $S_{imp}(J_{K}^{\prime }=1,r,{R})\sim 1/%
{R}$ for both even and odd $R$ and hence is zero in the thermodynamic
limit.
\begin{table}[tbp]
\centering
%
\begin{tabular}{|c|cccccccc|}
\hline
$J^{\prime}_K$ & 1.00 & 0.80 & 0.60 & 0.525 & 0.45 & 0.41 & 0.37 & 0.30 \\
\hline
$\xi_K$ & 0.65 & 1.97 & 5.93 & 9.84 & 17.83 & 25.65 & 38.29 & 83.79 \\
\hline
\end{tabular}
\caption{$\xi _{K}(J_{K}^{\prime })$ as determined from $S_{imp}(J_{K}^{%
\prime },r,{R})$ for ${R}=400$ using the FLT prediction, Eq.~(\ref{FLT_FSE}%
). The estimates become unreliable once $\xi _{K}$ becomes comparable to ${R}
$.}
\label{tab:xiKFLT}
\end{table}

\section{The TS- and OTS-ansatz for the Dimerized Phase, $J_2^c < J_2 \leq J/2$\label{sec:ssansatz}}

For $J_2>J_2^c$ the spin chain model, Eq.~(\ref{eq:spinch}), develops a gap and a non-zero dimerization.
The well known Majumdar-Ghosh (MG) model~\cite{MG69} ($J_2=J/2$) is part of this phase and
we therefore take the MG model as our starting point for an analysis of the entanglement entropy
in the dimerized phase. The MG point constitutes a disorder point~\cite{Schollwock96} beyond which
the short-range correlations become incommensurate
and our analysis therefore focus on the regime $J_2^c\leq J_2 < J/2$.
Due to the gap above the dimerized ground-state, it becomes possible to proceed
using variational wavefunctions for this range of parameters, yielding surprisingly precise results.
The initial assumption is that the lowest excitation in the dimerized phase is described by a single free spin
acting as domain wall or a {\it soliton}. This is an approximation since a complete description would include
states where the domain wall extends over several sites. The thin soliton states are not orthonormal.
By assuming that they are, we arrive at a further simplified ansatz which never the less proves to be surprisingly precise.
Below we detail this approach.

\subsection{The Thin Soliton (TS) Ansatz\label{sec:TS}}
For the MG model~\cite{MG69} ($J_2=J/2$) it is well known that for $\RR
$ even the singlet ground-state is two fold degenerate~\cite{MG69,SS81,Caspers82,Caspers84}.
These two ground-states corresponds to the formation of singlet states
between nearest neighbor spins either between sites $2n+1$ and $2n+2$ or $%
2n+2$ and $2n+3$. The resulting dimerization then occurs in two distinct
patterns. (See Fig.~\ref{fig:MG}.) For $\RR$ odd a very good approximation to the
ground-state is obtained by assuming that a single spin, the \textit{soliton},
is left unpaired separating regions with dimerizations in the two above
mentioned patterns. If we number the sites of the system $r=1,\ldots,%
\RR$ the number of odd sites is given by $N^o=(\RR+1)/2$. We define a state $|n\rangle$ by:
\begin{equation}
|n\rangle\equiv|\overbrace{-\ \ldots\ -}^n\ \uparrow\ -\ \ldots\ -\rangle.
\end{equation}
Here, $-$ indicates a singlet between site $r$ and $r+1$ with 
$n=0,\ldots N_d=N^o-1$ 
such singlets occurring before the soliton indicated by the $\uparrow$. 
Here, $n$ can take on the values $n=0,\ldots N_d=N^o-1$ with
$N_d$ the total number of dimers. 
Note that, singlets to the left of the soliton occur between sites $%
[2n+1,2n+2]$ and between sites $[2n+2,2n+3]$ to the right of the soliton.
We use the Marshall sign convention that $[2n+1,2n+2]=(|\uparrow\downarrow\rangle-|\downarrow\uparrow\rangle)/\sqrt{2}$
whereas $[2n+2,2n+3]=(|\downarrow\uparrow\rangle-|\uparrow\downarrow\rangle)/\sqrt{2}$.
We shall refer to these states as \textit{thin soliton} states
(TS-states) since the soliton is not ``spread" out over several sites by including
valence bonds of more than unit length. Note that, in such a dimerized state the soliton can, for $%
\RR$ odd, only be situated on \textit{odd} sites $r=1,3,5,\ldots
\RR$. We can then write an ansatz for the ground-state wavefunction
in the following manner~\cite{SS81,Caspers82,Caspers84,Sorensen98}:
\begin{equation}
|\Psi_{TS}^\Uparrow\rangle\simeq \sum_{n=0}^{N_d}\psi^{sol}_n|n\rangle.
  \label{eq:TSansatz}
\end{equation}
We shall refer to this ansatz as the thin-soliton ansatz (TS-ansatz). If
we consider $\RR=5$, we find 3 linearly independent (but not orthogonal) thin
soliton states. However, with 5 $S=1/2$ spins it is easy to see that there
are in fact 5 $S=1/2$ states. The thin soliton states do therefore not form a
complete basis for the $S=1/2$ subspace and the TS-ansatz is therefore
variational in nature for $\RR$ odd. However, for the MG-model it is known that the
TS-ansatz is very precise~\cite{Caspers82,Caspers84,Sorensen98}. Caspers et al~\cite
{Caspers82,Caspers84} improved on the TS-ansatz by including terms with longer valence
bonds in a systematic manner and showed that the resulting variational
energies were only changed slightly.

For $\RR$ \textit{even} Shastry and Sutherland~\cite{SS81} considered
excited states corresponding to 2-soliton states and studied bound-states of
solitons with relatively high energies. Subsequently, exact wave-functions
for bound soliton states were found by Caspers and Magnus~\cite{Caspers82}.
However, at low energies the solitons behave as free massive particles~\cite
{Sorensen98} with the soliton mass defined by:
\begin{equation}
\Delta_{sol}=\lim_{\RR\to\infty}E(\RR)-\frac{E(\RR%
+1)+E(\RR-1)}{2},\ \ \RR\ \mathrm{odd}.
\end{equation}
Here $E(R)$ is the ground-state energy for a system of length $R$.
Using DMRG the mass of the soliton has been estimated~\cite{Sorensen98} for
the MG model:
\begin{equation}
\Delta_{sol}^{DMRG}/J=0.1170(2).
\end{equation}
Using the TS-ansatz and periodic boundary conditions (pbc) the soliton mass
was determined to be~\cite{Caspers82,Caspers84} $\Delta_{sol}^{TS,pbc}/J=0.125$ and
including 3 and 5 spin structures in the variational calculation~\cite
{Caspers82,Caspers84} the estimate improved to $0.11701$. 
Note that the ground-state energy, $E(R)$, is an extensive quantity and the term $\propto R$,
$3JR/8$, is given exactly by the TS-ansatz. The error in the estimate of the ground-state energy
using the TS-ansatz is only $(0.125-0.11701)J\simeq 0.008J$, a small quantity independent of $R$.

It is useful to obtain
an estimate of the wave-function for a single soliton from a simple physical picture. Such an estimate can be obtained
in the following manner:
The soliton is repelled by the open ends used in the present study and we
therefore expect the thin soliton to behave as a particle in a box~\cite
{Sorensen98} with $\psi^{sol}_{-1}=\psi^{sol}_{N_d+1}=0$. In that case we
find:
\begin{equation}
\psi^{sol}_n\simeq \sqrt{\frac{2}{N_d+2}}\sin\left(\frac{\pi(2n+2)}{\RR+3}\right).  \label{eq:sinform}
\end{equation}
The $\psi^{sol}_n$ can also be determined using variationally methods as shown in
\ref{app:TSMG} and in the following we shall use such variationally determined $\psi^{sol}_n$.
However, as shown in \ref{app:TSMG} the variational estimate is only marginally better than
the above form, Eq.~(\ref{eq:sinform}).
Several other quantities can also be calculated for the MG model using the TS-ansatz, such as the on-site
magnetization, $\langle S^z_r\rangle$, and the spin-spin correlation function, $\langle \vec S_r\cdot\vec S_{r+1}\rangle$. Since these
quantities are unrelated to the main focus of the present paper, the entanglement entropy, we have included
them in \ref{app:TSMG}.

We now turn to calculating the entanglement entropy for the 
Majumdar-Ghosh model using the TS-ansatz. The two cases of $R$ even and odd
have to be considered separately and we start with $R$ odd.

\subsubsection{TS-ansatz for the Entanglement entropy for $\RR$ odd, $J^{\prime}_K=1$}
Using the TS-ansatz it is possible to obtain an explicit expression for $S(J'_K=1,r,R)$ for $R$ odd. 
We start by
separating Eq.~(\ref{eq:TSansatz}) into contributions from region $A$ ($%
1\ldots r$) and $B$ ($r+1\ldots\RR$). Without loss of generality we
initially assume that $r$ is odd. We write:
\begin{equation}
|\Psi_{TS}^\Uparrow\rangle=\sum_{i,j=0}^{3}C_{i,j}|\psi_i\rangle|\phi_j%
\rangle,
\end{equation}
with the $|\psi_i\rangle$ states in the Hilbert space for region $A$ and the
$|\phi_j\rangle$ states in the Hilbert space for region $B$. We find for $%
|\psi_i\rangle$:
\begin{eqnarray}
|\psi_1\rangle&=&\sum_{n=0}^{\frac{r-1}{2}}\psi^{sol}_n|\overbrace{-\
\ldots\ -}^n\ \uparrow\ -\ \ldots\ -\rangle  \nonumber \\
|\psi_2\rangle&=&|\overbrace{-\ -\ -\ \ldots\ -\ -\ -\ }^{\frac{r-1}{2}}\
\uparrow\rangle  \nonumber \\
|\psi_3\rangle&=&|\overbrace{-\ -\ -\ \ldots\ -\ -\ -\ }^{\frac{r-1}{2}}\
\downarrow\rangle.
\end{eqnarray}
For $|\phi_j\rangle$ we then write:
\begin{eqnarray}
|\phi_1\rangle&=&|\overbrace{-\ -\ -\ \ldots\ -\ -\ -\ }^{\frac{\RR-r%
}{2}}\rangle  \nonumber \\
|\phi_2\rangle&=&\sum_{n=0}^{\frac{\RR-r}{2}-1}\psi^{sol}_{\frac{r+1%
}{2}+n}|\downarrow\ \overbrace{-\ \ldots\ -}^n\ \uparrow\ -\ \ldots\ -\rangle
\nonumber \\
|\phi_3\rangle&=&\sum_{n=0}^{\frac{\RR-r}{2}-1}\psi^{sol}_{\frac{r+1%
}{2}+n}|\uparrow\ \overbrace{-\ \ldots\ -}^n\ \uparrow\ -\ \ldots\ -\rangle.
\nonumber \\
  \label{eq:phij}
\end{eqnarray}
With these definitions we find:
\begin{equation}
|\Psi_{TS}^\Uparrow\rangle= |\psi_1\rangle|\phi_1\rangle+\frac{1}{\sqrt{2}}%
\left[|\psi_2\rangle|\phi_2\rangle-|\psi_3\rangle|\phi_3\rangle\right],
\end{equation}
and thus:
\begin{equation}
C=\left( \begin{array}{ccc} 1 & 0 & 0\cr 0 & \frac{1}{\sqrt{2}} & 0 \cr 0 & 0 &
-\frac{1}{\sqrt{2}} \end{array}\right)
\end{equation}
Clearly, having defined the states this way, the $|\psi_i\rangle$ and $\phi_j\rangle$ are not
orthonormal. It is easy to see that $\langle\psi_3|\psi_i\rangle=0$ for $%
i=1,2$ and trivially $\langle\psi_2|\psi_2\rangle=\langle\psi_3|\psi_3%
\rangle=1$; however, using the Marshall sign convention described at the start of section~\ref{sec:TS} we find,
\begin{eqnarray}
\langle\psi_1|\psi_2\rangle&=&\sum_{n=0}^{\frac{r-1}{2}}\psi^{sol}_n(2)^{-|%
\frac{r-1}{2}-n|}=
\langle\psi_2|\psi_1\rangle
\nonumber \\
\langle\psi_1|\psi_1\rangle&=&\sum_{n,m=0}^{\frac{r-1}{2}}\psi^{sol}_n%
\psi^{sol}_m(2)^{-|n-m|}.
\end{eqnarray}
Likewise, we see that $\langle\phi_3|\phi_j\rangle=0$ for $j=1,2$ and $%
\langle\phi_1|\phi_1\rangle=1$, however,
\begin{eqnarray}
\langle\phi_1|\phi_2\rangle&=&\frac{1}{\sqrt{2}}\sum_{n=0}^{\frac{\RR%
-r}{2}-1}\psi^{sol}_{\frac{r+1}{2}+n}(2)^{-n}=
\langle\phi_2|\phi_1\rangle
\nonumber \\
\langle\phi_2|\phi_2\rangle&=&\sum_{n,m=0}^{\frac{\RR-r}{2}%
-1}\psi^{sol}_{\frac{r+1}{2}+n}\psi^{sol}_{\frac{r+1}{2}+m}(2)^{-|n-m|}
\nonumber \\
&=&\langle\phi_3|\phi_3\rangle.
\end{eqnarray}

In order obtain the reduced density matrix, $\rho$, describing region $A$
with the usual properties (i.e. $\mathrm{Tr}\rho=1$ with eigenvalues $%
\omega_i\geq0$) it is easiest to proceed by orthonormalizing the states $%
|\psi_i\rangle$. This is done in the usual manner by defining:
\begin{eqnarray}
|\tilde\psi_1\rangle&=&\frac{1}{\sqrt{\langle\psi_1|\psi_1\rangle}}%
|\psi_1\rangle  \nonumber \\
|\tilde\psi_2\rangle&=&\frac{|\psi_2\rangle-|\tilde\psi_1\rangle\langle%
\psi_2|\tilde\psi_1\rangle}{\sqrt{\langle\psi_2|\psi_2\rangle-|\langle%
\psi_2|\tilde\psi_1\rangle|^2}}.
\end{eqnarray}
With this orthonormalization the coefficient matrix $C$ then becomes:
\begin{equation}
\tilde C=\left( \begin{array}{ccc} \sqrt{\langle\psi_1|\psi_1\rangle} &
\frac{1}{\sqrt{2}}\langle\psi_2|\tilde\psi_1\rangle & 0\cr 0 &
\sqrt{\frac{1}{2}\left(\langle\psi_2|\psi_2\rangle-|\langle\psi_2|\tilde%
\psi_1\rangle|^2\right)} & 0 \cr 0 & 0 & -\frac{1}{\sqrt{2}} \end{array}%
\right)  \label{eq:Ctilde}
\end{equation}
The reduced density matrix for region $A$ can now be obtained by tracing out
region $B$:
\begin{equation}
\rho_{\Uparrow\Uparrow}(i_1,i_2)=\sum_{j_1,j_2=1}^{3}\tilde
C_{i_1,j_1}\tilde C_{i_2,j_2}\langle\phi_{j_1}|\phi_{j_2}\rangle,
\label{eq:rhoUpUp}
\end{equation}
from which the entanglement entropy is calculated employing the standard
formula $S=-\mathrm{Tr}\rho\ln\rho$. An equivalent expression for $r$ {\it even}
can be obtained by interchanging $|\psi_i\rangle$ and $|\phi_j\rangle$.

\begin{figure}[!ht]
\begin{center}
\includegraphics[width=10cm,clip]{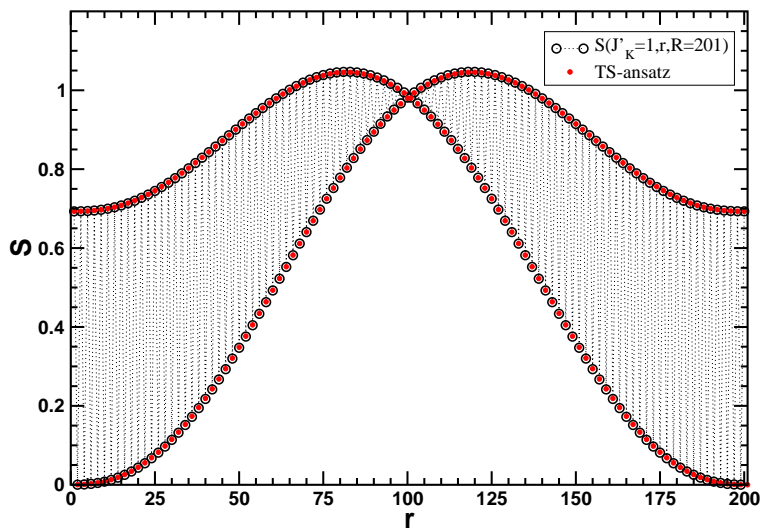}
\end{center}
\caption{DMRG results for $S(J^{\prime}_K=1,r,\RR=201)$ for the
spin-chain model with $J_2=J/2$ (MG model). The small circles represent the
theoretical result obtained using the ansatz Eq.~(\ref{eq:TSansatz}) with
variationally determined $\psi^{sol}_n$ (\ref{app:TSMG}).
}
\label{fig:TSansatz}
\end{figure}

As an example of how well the TS-ansatz, Eq.~(\ref{eq:TSansatz}), works we
show in Fig.~\ref{fig:TSansatz} the results of a calculation of $%
S(J^{\prime}_K=1,r,\RR=201)$ using this ansatz compared with DMRG
results for the same quantity. The agreement is almost perfect. 
In Fig.~\ref{fig:TSansatz}
variationally determined $\psi^{sol}_n$ (\ref{app:TSMG}) have been used.

\subsubsection{TS-ansatz for the Entanglement entropy for $\RR$ even, $J^{\prime}_K=0$}

It is now straight forward to extend our results from the previous section
to the case where $\RR$ is even and $J^{\prime}_K=0$. Our starting
point is the expression Eq.~(\ref{eq:rhoJK0}) for the reduced density
matrix. In order to calculate $\rho_{\Downarrow\Downarrow}$ and in
particular $\rho_{\Uparrow\Downarrow}$ we need an ansatz for $%
\Psi_{TS}^\Downarrow$ with all couplings
uniform and total length $R'=R-1$ {\it odd} with $r'=r-1$ denoting
the spatial coordinate. This can straightforwardly be obtained from Eq.~(\ref
{eq:TSansatz}) by inverting the soliton spin to obtain:
\begin{equation}
|\Psi_{TS}^\Downarrow\rangle\simeq \sum_{n=0}^{\frac{\RR'-1}{2}%
}\psi^{sol}_n|\overbrace{-\ \ldots\ -}^n\ \downarrow\ -\ \ldots\ -\rangle.
\label{eq:TSansatzdown}
\end{equation}
Using this ansatz we can now evaluate $\rho_{\Downarrow\Downarrow}$ as well
as $\rho_{\Uparrow\Downarrow}$.

As above, we start by separating Eq.~(\ref{eq:TSansatzdown}) into
contributions from region $A$ ($1\ldots r'$) and $B$ ($r'+1\ldots\RR'$)
again initially assuming that $r'$ is odd. We write:
\begin{equation}
|\Psi_{TS}^\Downarrow\rangle=\sum_{i,j=0}^{3}C_{i,j}|\psi_i^{\prime}\rangle|%
\phi_j^{\prime}\rangle,
\end{equation}
with the $|\psi_i^{\prime}\rangle$ states in the Hilbert space for region $A$
and the $|\phi_j^{\prime}\rangle$ states in the Hilbert space for region $B$%
. In this case it is convenient to define $|\psi_i^{\prime}\rangle$:
\begin{eqnarray}
|\psi_1^{\prime}\rangle&=&\sum_{n=0}^{\frac{r'-1}{2}}\psi^{sol}_n|\overbrace{%
-\ \ldots\ -}^n\ \downarrow\ -\ \ldots\ -\rangle  \nonumber \\
|\psi_2^{\prime}\rangle&=&|\overbrace{-\ -\ -\ \ldots\ -\ -\ -\ }^{\frac{r'-1}{2}}\ \downarrow\rangle  \nonumber \\
|\psi_3^{\prime}\rangle&=&|\overbrace{-\ -\ -\ \ldots\ -\ -\ -\ }^{\frac{r'-1}{2}}\ \uparrow\rangle.
\end{eqnarray}
For $|\phi_j^{\prime}\rangle$ we write:
\begin{eqnarray}
|\phi_1^{\prime}\rangle&=&|\overbrace{-\ -\ -\ \ldots\ -\ -\ -\ }^{\frac{%
\RR'-r'}{2}}\rangle  \nonumber \\
|\phi_2^{\prime}\rangle&=&\sum_{n=0}^{\frac{\RR'-r'}{2}-1}\psi^{sol}_{%
\frac{r'+1}{2}+n}|\uparrow\ \overbrace{-\ \ldots\ -}^n\ \downarrow\ -\
\ldots\ -\rangle  \nonumber \\
|\phi_3^{\prime}\rangle&=&\sum_{n=0}^{\frac{\RR'-r'}{2}-1}\psi^{sol}_{%
\frac{r'+1}{2}+n}|\downarrow\ \overbrace{-\ \ldots\ -}^n\ \downarrow\ -\
\ldots\ -\rangle.  \nonumber \\
  \label{eq:phipj}
\end{eqnarray}
With these definitions one finds that $\langle\psi_i|\psi_j\rangle=
\langle\psi_i^{\prime}|\psi_j^{\prime}\rangle$ and $\langle\phi_i|\phi_j%
\rangle= \langle\phi_i^{\prime}|\phi_j^{\prime}\rangle$. Hence, for this
choice of states, $\rho_{\Downarrow\Downarrow}$ is numerically identical to $%
\rho_{\Uparrow\Uparrow}$, Eq.~(\ref{eq:rhoUpUp}). However, the actual states
$|\phi_j^{\prime}\rangle$ are \textit{not} identical to the $|\phi_j\rangle$
and $\langle\phi_i|\phi^{\prime}_j\rangle$ is \textit{not} the identity
matrix. This becomes important for the evaluation of $\rho_{\Uparrow%
\Downarrow}$ which can be expressed:
\begin{equation}
\rho_{\Uparrow\Downarrow}(i_1,i_2)=\sum_{j_1,j_2=1}^{3}\tilde
C_{i_1,j_1}\tilde C_{i_2,j_2}\langle\phi_{j_1}|\phi_{j_2}^{\prime}\rangle,
\label{eq:rhoUpDown}
\end{equation}
with the coefficient matrix $\tilde C$ given as before by Eq.~(\ref
{eq:Ctilde}). It is easily seen that $\langle\phi_i|\phi_j^{\prime}\rangle=0$
except for $\langle\phi_1|\phi_1^{\prime}\rangle=\langle\phi_1|\phi_1\rangle$%
, $\langle\phi_2|\phi_1^{\prime}\rangle=\langle\phi_2|\phi_1\rangle$ and $%
\langle\phi_1|\phi_2^{\prime}\rangle=\langle\phi_1|\phi_2\rangle$. $%
\rho_{\Uparrow\Downarrow}$ can then be evaluated along with $%
\rho_{\Downarrow\Uparrow}=\rho^\dagger_{\Uparrow\Downarrow}$ and the full
density matrix, Eq.~(\ref{eq:rhoJK0}), describing $J^{\prime}_K=0$ and $%
\RR$ \textit{even} constructed. Again, by essentially
interchanging the role of $|\psi_i^{\prime}\rangle$ and
$|\phi_j^{\prime}\rangle$, it is straightforward to obtain results
for $r'=r-1$ \textit{even} as well.

With the matrices, $\rho_{\Uparrow\Uparrow}$, $\rho_{\Downarrow\Downarrow}$,
$\rho_{\Uparrow\Downarrow}$$\rho_{\Downarrow\Uparrow}$ in hand, the full
density matrix for $J^{\prime}_K=0$ and $\RR$ \textit{even} can be
constructed from Eq.~(\ref{eq:rhoJK0}) and the complete entanglement entropy
calculated.

\begin{figure}[!ht]
\begin{center}
\includegraphics[width=10cm,clip]{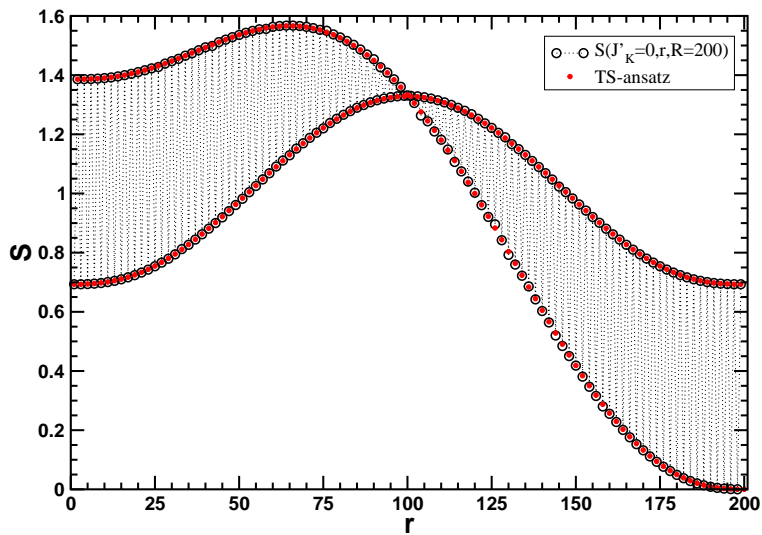}
\end{center}
\caption{DMRG results for $S(J^{\prime}_K=0,r,\RR=200)$ for the
spin-chain model with $J_2=J/2$ (MG model). The solid line represents the
theoretical result obtained using the TS-ansatz Eq.~(\ref{eq:TSansatz}) and
Eq.~(\ref{eq:TSansatzdown}) with 
variationally determined $\psi^{sol}_n$ (\ref{app:TSMG}).}
\label{fig:TSansatzJK0}
\end{figure}
As an example of how well the TS-ansatz, Eq.~(\ref{eq:TSansatz}) and Eq.~(%
\ref{eq:TSansatzdown}), work we show in Fig.~\ref{fig:TSansatzJK0} the
results of a calculation of $S(J^{\prime}_K=0,r,\RR=200)$ using this
approach compared with DMRG results for the same quantity with variationally determined $\psi^{sol}_n$ (\ref{app:TSMG}).
Excellent
agreement is observed. The small discrepancies between the DMRG results and
the TS-ansatz visible at a few $r$ values are due to complications using
spin-inversion in the DMRG calculations specific to this value of $J_2$.

We end this section by remarking that in principle the TS-ansatz could be
used for any $J_2>J^c_2$ by simply redoing the variational calculation
determining the $\psi^{sol}_n$.

\subsection{The Orthonormal Thin Soliton (OTS) Ansatz}

While relatively straight forward to use, the TS-ansatz does not
yield expressions that are intuitively easy to grasp. It therefore
seems desirable to develop a simplified picture of the underlying
physics which we will now try to do. We shall do this by assuming
that the TS states are orthonormal, thereby arriving at a simplified ansatz.
Within this simplified ansatz the ``single particle entanglement'' 
and ``impurity valence bond'' contributions to the entanglement entropy, introduced in 
in sub-section \ref{sec:SPEIVB}, are derived explicitly employing this ansatz.

\subsubsection{OTS-ansatz for the Entanglement entropy for $\RR$ odd, $J^{\prime}_K=1$}
We begin by focusing on the case of $\RR$ odd and $%
J^{\prime}_K=1$ at the MG-point ($J_2=J/2$). Our starting point for the
calculation of this quantity using the TS-ansatz was the expression $%
|\Psi^\Uparrow>=\sum_{i,j}C_{i,j}|\psi_i\rangle|\phi_j\rangle$. We now make
the simplifying assumption that all the $|\psi_i\rangle$ and $|\phi_j\rangle$
are orthonormal. We shall refer to this as the orthonormal thin soliton ansatz (OTS-ansatz).
This does not change the coefficients $C_{i,j}$: $C_{1,1}=1, C_{2,2}=-C_{3,3}=1/\sqrt{2}$ and $0$
otherwise. However, the reduced density matrix now significantly simplifies
and assuming orthonormality we find for $r$ odd:
\begin{equation}
\rho^\perp_{\Uparrow\Uparrow}= \left( \begin{array}{ccc}
\sum_{n=0}^{k-1}|\psi^{sol}_n|^2&0 & 0\cr 0&
\frac{1}{2}\sum_{n=k}^{N_d}|\psi^{sol}_{n}|^2 & 0\cr 0& 0&
\frac{1}{2}\sum_{n=k}^{N_d}|\psi^{sol}_{n}|^2 \cr \end{array}\right).
\label{eq:rhoperp}
\end{equation}
Here, $k=(r+1)/2$ and, as above, $N_d=(\RR-1)/2$. If we by $p$
denote the probability that the soliton is to the left of the point $r$ we
then see that this is simply:
\begin{equation}
\rho^\perp_{\Uparrow\Uparrow}= \left( \begin{array}{ccc} p&0 & 0\cr 0&
\frac{1-p}{2} & 0\cr 0& 0& \frac{1-p}{2} \cr \end{array}\right),\ r\
\mathrm{odd}.
\end{equation}
It immediately follows that for $r$ odd:
\begin{equation}
S(J^{\prime}_K=1,r,\RR)=-p\ln( p) -(1-p)\ln(1-p)+(1-p)\ln(2).
\label{eq:otsodd1}
\end{equation}
If we now turn to $r$ even while still considering $\RR$ odd and $%
J^{\prime}_K=1$ we instead find:
\begin{equation}
\rho^\perp_{\Uparrow\Uparrow}= \left( \begin{array}{ccc} 1-p&0 & 0\cr 0&
\frac{p}{2} & 0\cr 0& 0& \frac{p}{2} \cr \end{array}\right),\ r\ \mathrm{%
even}.
\end{equation}
This naturally follows from the fact that now $\rho^\perp_{1,1}$
describes
a situation with the soliton to the right of the point $r$ while $%
\rho^\perp_{2,2}$ and $\rho^\perp_{3,3}$ have the soliton the left. We then
see that in this case for $r$ even:
\begin{equation}
S(J^{\prime}_K=1,r,\RR)=-p\ln( p) -(1-p)\ln(1-p)+p\ln(2).
\label{eq:otsodd2}
\end{equation}
We emphasize that Eq.~(\ref{eq:otsodd1}) and (\ref{eq:otsodd2}) {\it precisely equal} the
heuristic expression Eq.~(\ref{eq:HeurSodd}) based on the SPE and IVB.

We can now explicitly obtain expressions for the uniform and alternating part
of the entanglement entropy, we find for $\RR$ odd:
\begin{eqnarray}
S_U(J^{\prime}_K=1,r,\RR)&=&-p\ln( p) -(1-p)\ln(1-p)+\frac{1}{2}%
\ln(2)  \label{eq:SUodd} \\
S_A(J^{\prime}_K=1,r,\RR) &=& (\frac{1}{2}-p)\ln(2).  \label{eq:SAodd}
\end{eqnarray}
From this we can immediately extract $S_{imp}(J^{\prime}_K=1,r,\RR)$
for $\RR$ odd since at the MG-point $S_U(J_K'=1,r,\RR)$ for $%
\RR$ even is simply $\ln(2)/2$ for any $r$. We then find:
\begin{equation}
S_{imp}(J^{\prime}_K=1,r,\RR)=-p\ln( p) -(1-p)\ln(1-p), \RR\
\mathrm{odd}.  \label{eq:sfpodd}
\end{equation}
Since this is effectively the entanglement resulting from the presence of a
single thin soliton in the ground-state we see that the $S_{imp}$ in this
case is given uniquely by the SPE.

\begin{figure}[!ht]
\begin{center}
\includegraphics[width=10cm,clip]{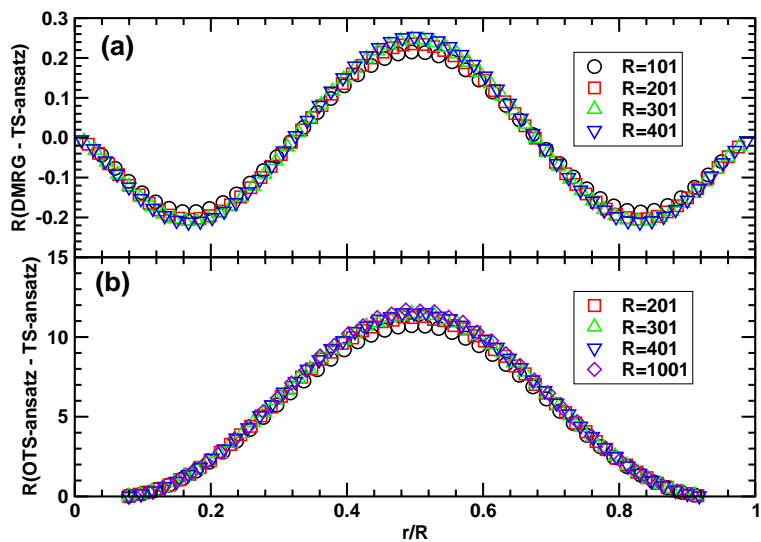}
\end{center}
\caption{Comparison of results for $S_{imp}(J_K'=1,r,\RR)$ at the MG-point $J_2=J/2$ for a range of
system sizes obtained using DMRG, TS-ansatz and the OTS-ansatz. (a) Difference between DMRG results and
TS-ansatz results scaled by $\RR$. (b) Difference between OTS and TS-ansatz results scaled by $\RR$.
For the TS-ansatz variationally determined $\psi^{sol}_n$ (\ref{app:TSMG}) have been used where as
for the OTS-ansatz the analytical result for $p(r)$ Eq.~(\ref{eq:p}) have been used.
}
\label{fig:ErrorSimp1MG}
\end{figure}
It is of considerable interest to test how well the TS- and OTS-ansatz agree at the MG-point
and to test how well either one of them agree with the DMRG results. In Fig.~\ref{fig:ErrorSimp1MG}(a) we
show results for the difference between $S_{imp}(J_K'=1,r,\RR)$ calculated with DMRG and the TS-ansatz scaled
by $\RR$. Fig.~\ref{fig:ErrorSimp1MG}(b) shows results for the difference between $S_{imp}(J_K'=1,r,\RR)$ calculated
with the OTS-ansatz, Eq.~(\ref{eq:sfpodd}), and with the TS- ansatz, again scaled by $\RR$. The data in Fig.~\ref{fig:ErrorSimp1MG} clearly show that in the
thermodynamic limit all three approaches agree. This somewhat surprising result indicates that at the MG-point
the non-orthogonality of the TS-states for the calculation of the entanglement entropy cannot play an important role.

\subsubsection{OTS-ansatz for the Entanglement entropy for $\RR$ even, $J^{\prime}_K=0$}
We now turn to the case where $J^{\prime}_K=0$ and $\RR$ is even. We
then have to consider the full $6\times 6$ density matrix. If we initially
take $r$ even we find ($\RR$ even):
\begin{equation}
\rho^\perp= \left( \begin{array}{cccccc} \frac{p}{2}&0 & 0 & -\frac{p}{2} & 0 &
0\cr 0& \frac{1-p}{4} & 0 & 0 & 0 & 0\cr 0& 0& \frac{1-p}{4} & 0 &0 &0 \cr
-\frac{p}{2} &0 &0 &\frac{p}{2}&0 & 0\cr 0& 0& 0& 0& \frac{1-p}{4} & 0\cr 0&
0& 0& 0& 0& \frac{1-p}{4} \cr \end{array}\right),\ r\ \mathrm{even}.
\label{eq:rhoodd}
\end{equation}
This follows from the observation that within the OTS ansatz
and with the states $|\phi_k\rangle$ and $|\phi'_l\rangle$ as defined
in Eqs.~(\ref{eq:phij}) and (\ref{eq:phipj}),
 $\langle\phi_k|\phi'_l\rangle=0$ unless $k=l=1$ in which case
$\langle\phi_1|\phi'_1\rangle=1$.
By diagonalizing this matrix it follows that for $r$ even ($\RR$
even):
\begin{equation}
S(J^{\prime}_K=0,r,\RR)=-p\ln( p) -(1-p)\ln(1-p)+2(1-p)\ln(2).
\label{eq:otseven1}
\end{equation}
In the same manner we write for $r$ odd ($\RR$ even):
\begin{equation}
\rho^\perp= \left( \begin{array}{cccccc} \frac{1-p}{2}&0 & 0 & 0 & 0 & 0\cr 0&
\frac{p}{4} & 0 & 0 & 0 & \frac{p}{4}\cr 0& 0& \frac{p}{4} & 0 &\frac{p}{4}
&0 \cr 0 &0 &0 &\frac{1-p}{2}&0 & 0\cr 0& 0& \frac{p}{4}& 0& \frac{p}{4} &
0\cr 0& \frac{p}{4}& 0& 0& 0& \frac{p}{4} \cr \end{array}\right),\ r\
\mathrm{odd}.
\label{eq:rhoeven}
\end{equation}
In a way analogous to Eq.~(\ref{eq:rhoodd}), we  find non-zero
off-diagonal elements. From which it follows that for $r$ odd
($\RR$ even):
\begin{equation}
S(J^{\prime}_K=0,r,\RR)=-p\ln( p) -(1-p)\ln(1-p)+\ln(2).
\label{eq:otseven2}
\end{equation}
Again we emphasize that Eq.~(\ref{eq:otseven1}) and (\ref{eq:otseven2}) {\it precisely equal} the
heuristic expression Eq.~(\ref{eq:HeurSeven}) based on the SPE and IVB.

As above, we can now explicitly obtain expressions for the uniform and
alternating part of the entanglement entropy, we find for $\RR$
even:
\begin{eqnarray}
S_U(J^{\prime}_K=0,r,\RR)&=&-p\ln( p) -(1-p)\ln(1-p)
+(\frac{3}{2}-p)\ln(2)  \label{eq:SUeven} \\
S_A(J^{\prime}_K=0,r,\RR)&=& (\frac{1}{2}-p)\ln(2).\label{eq:SAeven}
\end{eqnarray}
We note that the uniform part is simply $\mathrm{SPE}+(3/2-p)\ln(2)$. Again
we can immediately extract $S_{imp}(J^{\prime}_K=0,r,\RR)$ for $%
\RR$ even at the MG-point by using the above result for $S_U(J_K'=1,r,%
\RR)$, Eq.~(\ref{eq:SUodd}). We find:
\begin{equation}
S_{imp}(J^{\prime}_K=0,r,\RR)=(1-p)\ln(2), \RR\ \mathrm{even}%
.  \label{eq:sfpeven}
\end{equation}
We note that this expression does \textit{not} contain a contribution from
the single part entanglement (SPE), but purely a term related to the
impurity valence bond (IVB).

By numerically calculating the density matrix $\rho$ for large $\RR$ using the TS-ansatz,
we have verified that the $\rho$ indeed does follow the above approximate forms, Eqs.~(\ref{eq:rhoodd}) and (\ref{eq:rhoeven}),
for both $r$ even and odd.

It is useful to have an analytical expression for $p$, the probability of
finding the soliton in region A $(x\le r)$. Simplifying our expression for
the soliton wavefunction, Eq.~(\ref{eq:sinform}), we write:
\begin{equation}
\psi^{sol}(x)\simeq \sqrt{\frac{2}{\RR}}\sin\left(\frac{\pi x }{\RR}\right).  \label{eq:sinform2}
\end{equation}
With $p$ corresponding to the probability of finding the particle in region
A ($x\leq r$) we can then write:
\begin{equation}
p=\int_{0}^{r}|\psi^{sol}(x)|^2dx.
\end{equation}
From which we find:
\begin{equation}
p=\frac{r}{\RR}-\frac{1}{2\pi}\sin\left(2\pi r/\RR\right).
\label{eq:p}
\end{equation}
This expression agrees rather well with the probability extracted from the variationally
determined soliton wave-function. See \ref{app:TSMG}.
Since $p$ only depends on the single variable $r/\RR$ we see that
also $S_{imp}(J_K^{\prime}=0,r,\RR)$ and $S_{imp}(J_K^{\prime}=1,r,\RR)$
as obtained from the OTS-ansatz are functions of the single variable $r/\RR$.
When we refer to the OTS-ansatz we always assume that $p(r)$ has been determined
using the above analytical form, Eq.~(\ref{eq:p}).

If a similar analysis had been performed away from the MG-point, at a
different $J_2 > J_2^c$, the soliton wave function, Eq.~(\ref{eq:sinform2})
would likely change and thereby the above expression for $p$. However, it is
noteworthy that the remaining expressions such as Eqs.~(\ref{eq:sfpodd}) and
(\ref{eq:sfpeven}) only depend on $J_2$ through $p$.

\section{Numerical Results for the Fixed Point Entanglement\label{sec:numres}}
\subsection{Fixed Point Entanglement at the MG-point, $J_2=J/2$}

\begin{figure}[!ht]
\begin{center}
\includegraphics[width=10cm,clip]{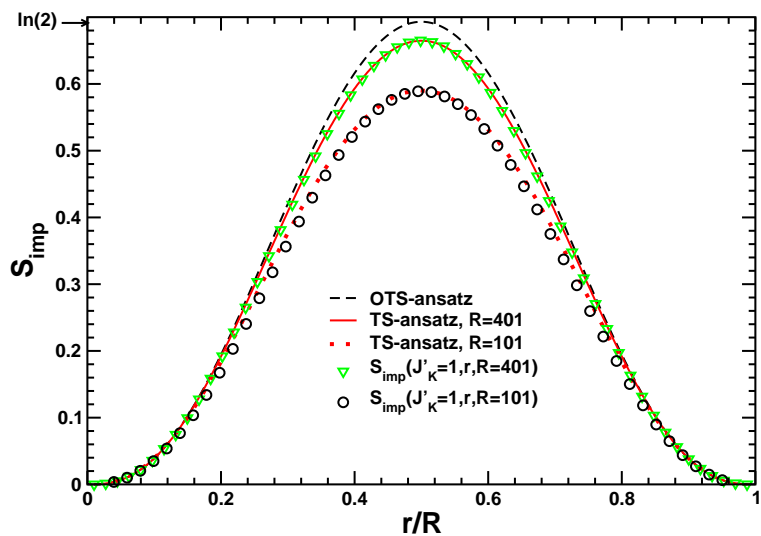}
\end{center}
\caption{DMRG results for $S_{imp}(J^{\prime}_K=1,r,\RR)$ for $%
\RR=101,401$ for the spin-chain model with $J_2=J/2$. The
dashed line represents the OTS-ansatz result Eq.~(\ref{eq:sfpodd}). For $%
\RR=101,401$ we also show the results for the TS-ansatz Eq.~(\ref
{eq:TSansatz}) as the solid and dotted lines. }
\label{fig:Simp1MG}
\end{figure}
At the MG-point ($J_2=J/2$) we can use the TS-ansatz, Eq.~(\ref{eq:TSansatz}), directly to calculate $S_{imp}(J^{\prime}_K=1,r,\RR)$
for $\RR$ odd. This is straight forward since $S(J^{\prime}_K=1,r,%
\RR)$ for $\RR$ even is simply $\ln(2)$ for $r$ odd and zero
otherwise. The result of such a calculation it the MG-point is shown in Fig.~\ref{fig:Simp1MG} (solid lines) for $\RR=101,401$ where we also
show result for the same quantity calculated using the TS- and OTS-ansatz, Eqs.~(\ref{eq:TSansatz}), with variationally determined $\psi^{sol}_n$, and (\ref{eq:sfpodd})
with $p(r)$ from Eq.~(\ref{eq:p}).
Note that, the TS-ansatz depends on $\RR$ where as the OTS-ansatz is independent of $\RR$. Excellent agreement
is observed between the theoretical and numerical results. Both the DMRG results and TS-results rapidly approach the OTS result as
implied by Fig.~\ref{fig:ErrorSimp1MG}. Clearly this fixed point impurity entanglement entropy is non-zero at
the MG-point and attains its maximum of $\ln(2)$ in the middle of the chain.

\begin{figure}[!ht]
\begin{center}
\includegraphics[width=10cm,clip]{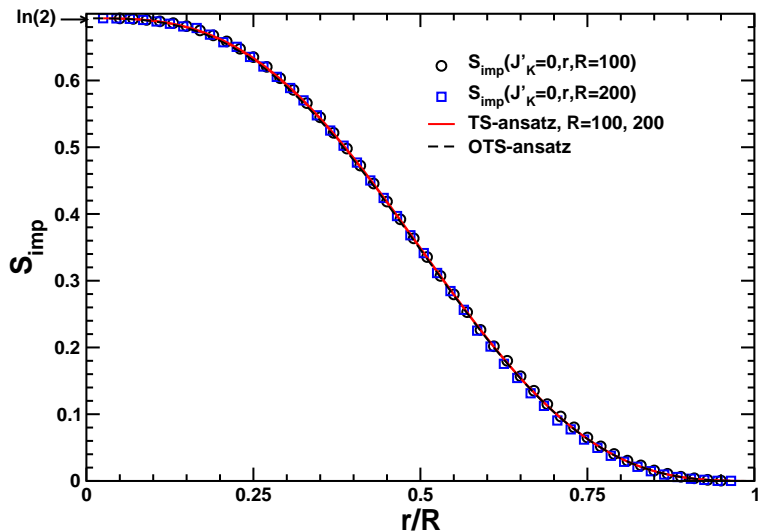}
\end{center}
\caption{DMRG results for $S_{imp}(J^{\prime}_K=0,r,\RR)$ for $%
\RR=100,200$ for the spin-chain model with $J_2=J/2$. The solid line
represents the theoretical result Eq.~(\ref{eq:sfpeven}). }
\label{fig:Simp0MG}
\end{figure}
Likewise, we can obtain $S_{imp}(J^{\prime}_K=0,r,\RR)$ for $\RR$ even 
at the MG-point using the TS-ansatz with variationally determined $\psi^{sol}_n$. The results are
indistinguishable from the OTS-ansatz, Eq.~(\ref{eq:sfpeven}), with $p(r)$ from Eq.~(\ref{eq:p}).
Our results obtained at the MG-point are shown in Fig.~\ref{fig:Simp0MG} for system
sizes $\RR=100,200$. Almost no variation with $\RR$ is observed and the numerical DMRG results display
almost complete agreement with the TS- and OTS-ansatz. The TS- and OTS-ansatz yields results that are
almost identical and the difference between the two is not visible in Fig.~\ref{fig:Simp0MG}. The lack
of finite-size effects is presumably related to the fact that $S_{imp}(J'_K=0,r,\RR)$  does not contain
a contribution from the single particle entanglement entropy (SPE). $S_{imp}(J'_K=0,r/\RR)$ display
a distinct cross-over between $\ln(2)$ and zero and clearly remains non-zero as $\RR\to\infty$.

\subsection{Fixed Point Entanglement in the Dimerized Phase, $J_2^c\leq J_2 < J/2$}
\begin{figure}[!ht]
\begin{center}
\includegraphics[width=10cm,clip]{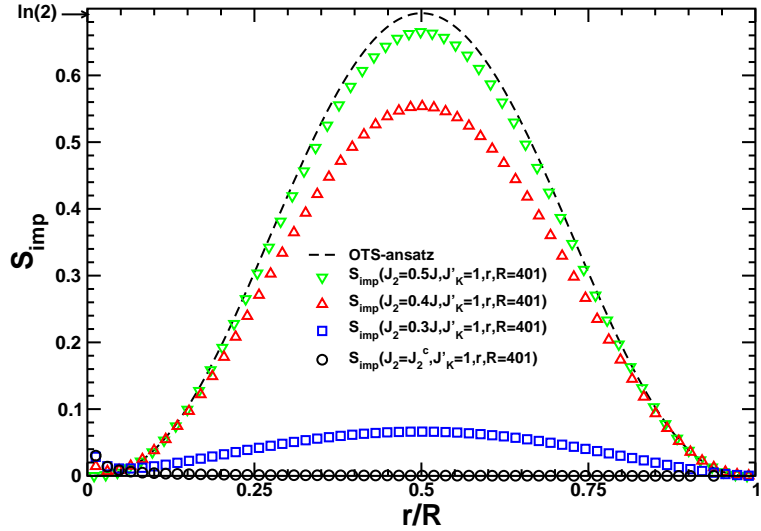}
\end{center}
\caption{DMRG results for $S_{imp}(J^{\prime}_K=1,r,\RR)$ for $%
\RR=401$ for the spin-chain model with $J_2=J_2^c,0.3J,0.4J,0.5J$. The
dashed line represents the OTS result Eq.~(\ref{eq:sfpodd}), valid at the MG-point ($J_2=J/2$). }
\label{fig:Simp1Scal}
\end{figure}
We now turn to a discussion of the variation of the fixed point entanglement entropies $S_{imp}(J_K'=1,r/\RR)$
and $S_{imp}(J'_K=0,r/\RR)$ with $J_2$ as we decrease $J_2$ towards the critical point $J_2^c$.
In Fig.~\ref{fig:Simp1Scal} we show DMRG results for $S_{imp}(J_K'=1,r/\RR)$ for $J_2=J^c_2,0.3J,0.4J,0.5J$.
In all cases with $\RR=401$. The dashed line represents the OTS-result, Eq.~(\ref{eq:sfpodd}), with $p(r)$ from Eq.~(\ref{eq:p}), valid at the MG-point.
It is seen that $S_{imp}(J_K'=1,r/\RR)$ quickly diminishes as $J_2$ is decreased towards $J^c_2$. At $J_2^c$ we have argued in
section~\ref{sec:FLT} that $S_{imp}(J_K'=1,r/\RR)$ approaches zero in the thermodynamic limit. 
In fact, from the structure of the OTS-ansatz it is clear that this approximation
must break down at the critical point since almost any form for the soliton wavefunction, $\psi^{sol}$, would imply
a non-zero $p$ and thus a non-zero result in Eq.~(\ref{eq:sfpodd}). We also note that at the critical point
the soliton becomes massless.

\begin{figure}[!ht]
\begin{center}
\includegraphics[width=10cm,clip]{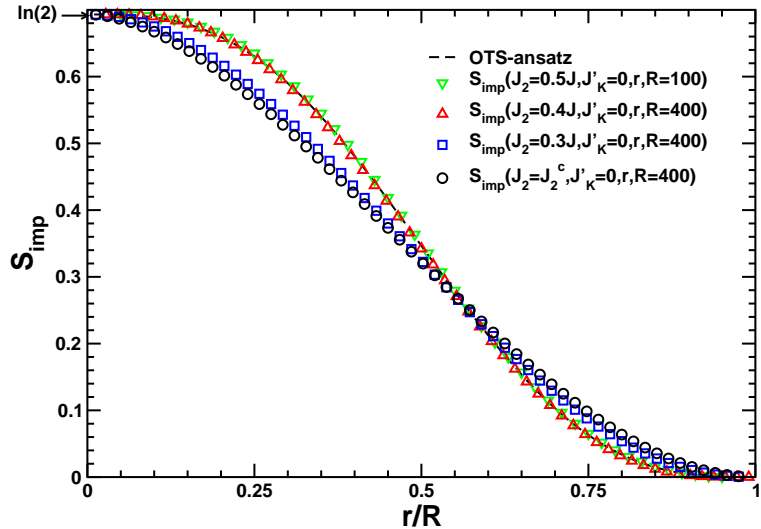}
\end{center}
\caption{DMRG results for $S_{imp}(J^{\prime}_K=0,r,\RR)$ for $%
\RR=400$ for the spin-chain model with $J_2= J_2^c,0.3J,0.4J,0.5J$. The
dashed line represents the OTS result Eq.~(\ref{eq:sfpeven}). }
\label{fig:Simp0Scal}
\end{figure}
We now focus on $S_{imp}(J_K'=0,r/\RR)$ for which we
show DMRG results for $S_{imp}(J_K'=1,r/\RR)$ with $J_2=J^c_2,0.3J,0.4J,0.5J$ in Fig.~\ref{fig:Simp0Scal}.
The dashed line represents the OTS-result Eq.~(\ref{eq:sfpeven}), with $p(r)$ from Eq.~(\ref{eq:p}), valid at the MG-point.
Although  some variation of $S_{imp}(J_K'=0,r/\RR)$ with $J_2$ is seen, this variation is rather small
and this fixed point impurity entanglement entropy clearly remains non-zero at the critical point, $J_2^c$.
There are some finite size effects but at $J_2^c$ these affects are rather small. Perhaps surprisingly, the OTS result
Eq.~(\ref{eq:sfpeven}) is not dramatically different from the DMRG results even at the critical point $J_2^c$.
We speculate that this is due to the fact that $S_{imp}(J_K'=0,r/\RR)$ does not contain a contribution from the single particle entanglement
but can be described entirely in terms of the impurity valence bond picture. $S_{imp}(J_K'=1,r/\RR)$, on the contrary, is completely
described by the single particle entanglement.

\subsection{Fixed Point Entanglement in the Gapless Heisenberg Phase, $J_2 < J_2^c$}
\begin{figure}[!ht]
\begin{center}
\includegraphics[width=10cm,clip]{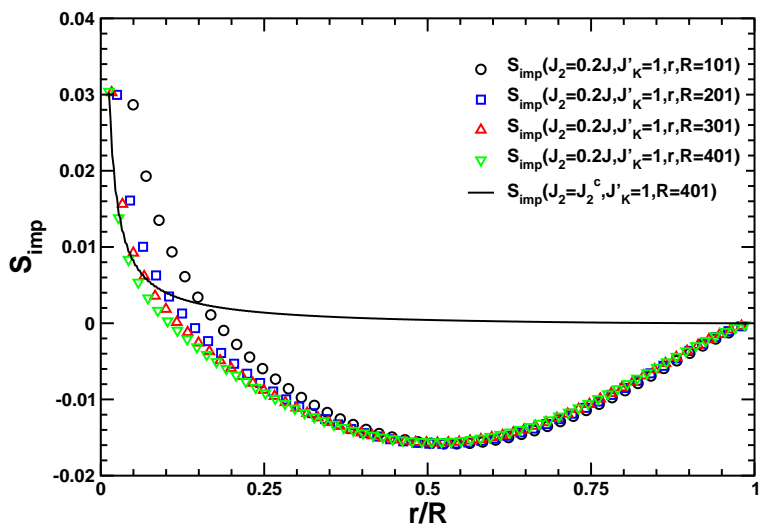}
\end{center}
\caption{DMRG results for $S_{imp}(J_2=0.2J,J^{\prime}_K=1,r,\RR)$ for $%
\RR=101,201,301,401$ for the spin-chain model with $J_2=0.2J$.
The solid
line is DMRG results for $S_{imp}(J_2=J_2^c,J^{\prime}_K=1,r,\RR=401)$ calculated at the
critical point, $J_2=J_2^c$.
}
\label{fig:Simp1J202}
\end{figure}
Finally we turn to the gapless Heisenberg phase where $J_2< J_2^c$. Our results for $S_{imp}(J_2=0.2J,J_K'=1,r/\RR)$ are shown
in Fig.~\ref{fig:Simp1J202} for a series of system sizes, $\RR=101,201,301$ and $401$ all calculated at $J_K'=0.2J$.
For comparison we have included results for $S_{imp}(J_2=J_2^c,J_K'=1,r/\RR)$ at the critical point. It is seen that $S_{imp}(J_2=0.2J,J_K'=1,r/\RR)$
is rather small and {\it negative} for this value of $J_2$ with relatively little variation with the system size $\RR$.
In comparison $S_{imp}(J_2=J_2^c,J_K'=1,r/\RR)$ remains positive. A likely explanation for this is that the single particle
entanglement, largely characterizing $S_{imp}(J'_K=1, r/\RR)$, is very sensitive to $J_2$ and no longer is a sensible quantity
for $J_2<J_2^c$ due to the gapless nature of this phase.

\begin{figure}[!ht]
\begin{center}
\includegraphics[width=10cm,clip]{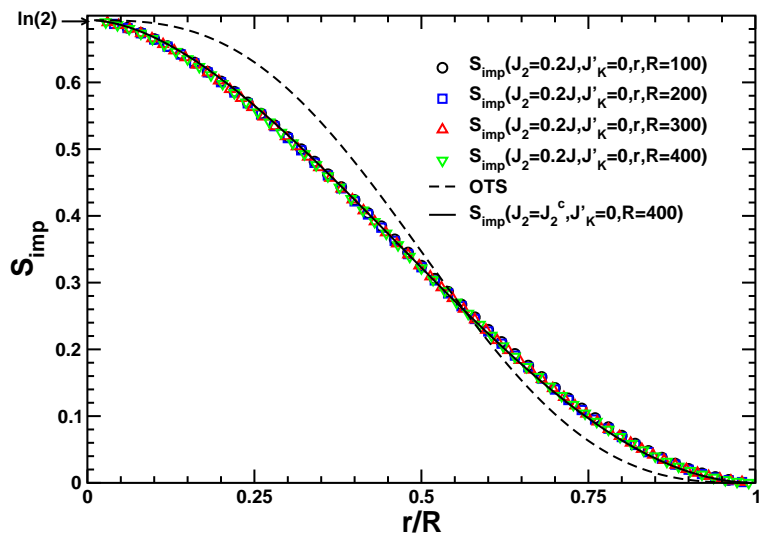}
\end{center}
\caption{DMRG results for $S_{imp}(J_2=0.2J,J^{\prime}_K=0,r,\RR)$ for $%
\RR=100,200,300,400$ for the spin-chain model with $J_2= 0.2J$. The
dashed line represents the OTS result Eq.~(\ref{eq:sfpeven}).
The solid
line are DMRG results for $S_{imp}(J_2=J_2^c,J^{\prime}_K=0,r,\RR=400)$ calculated at the
critical point, $J_2=J_2^c$.
}
\label{fig:Simp0J202}
\end{figure}
Results for $S_{imp}(J_2=0.2J,J^{\prime}_K=0,r,\RR)$ for $\RR=100,200,300,400$ are shown in Fig.~\ref{fig:Simp0J202}
for $J_2=0.2J$. We also show results for $S_{imp}(J^{\prime}_K=0,r,\RR=400)$ calculated at the
critical point, $J_2=J_2^c$. In this case the results at $J_2=0.2J$ are very similar to the results
at the critical point $J_2=J_2^c$ and it seems possible that the results coincide in the thermodynamic limit.
We speculate that the fixed point entanglement found at the critical point is representative for all of
the gapless Heisenberg phase $J_2< J_2^c$.

\section{The alternating part, $S_A$\label{sec:sa}}
\subsection{General discussion}

Entanglement entropy in spin chains also has an interesting staggered part 
as we pointed out in \cite{Laflorencie06}.  For an open chain with no impurity, 
this was shown to decay away from the boundary with a power law $1/2$, the 
same power-law exhibited by the dimerization. The operator of dimension 1/2, representing the 
dimerization, is related by a chiral SU(2) transformation to the staggered spin 
density, $\vec n$, reviewed in \ref{app:field_theory}.
 It has no counterpart in a non-interacting fermion system. (Or, 
more correctly, the counterpart, while it exists, must always come together 
with the operator $\cos\phi_c$ from the charge sector as reviewed 
in \ref{app:field_theory}).) This alternating entropy for 
the open chain with no impurity decays with a different exponent than what 
occurs for an open chain of free fermions \cite{Laflorencie06}. 
If we now include a weak coupling of 
an impurity spin to the end of the chain, we expect that the associated 
change in the alternating part of the entanglement entropy will continue 
to be different that of the free fermion Kondo model. This is in 
striking contrast to the uniform part which we have argued to be 
the same (at long length scales) for free fermion and spin chain 
Kondo models. In the rest of this section we discuss the behavior of the 
dimerization and examine the behavior of the alternating part of the 
entanglement entropy for both critical and dimerized spin chains. 

\subsection{The Alternating Part of the Energy Density, $E_A$\label{sec:ea}}
We start by deriving a field theory epxression for the alternating part of the energy density, $E_A$,
that we shall find sheds some light on the alternating part of the entanglement entropy.
The energy density for XXZ antiferromagnetic spin chains:
\begin{equation}
\left\langle h_{r}\right\rangle =\left\langle (S_{r}^{+}S_{r+1}^{-}+S_{r}^{-}S_{r+1}^{+})/2+\Delta S_{r}^{z}S_{r+1}^{z}\right\rangle 
\end{equation}
is uniform in periodic chains.
On the other hand, an open end breaks translational invariance and
there will be a slowly decaying alternating term or
''dimerization'' in the energy density
\begin{equation}
\left\langle h_{r}\right\rangle =E_{U}(r)+(-1)^{r}E_{A}(r),
\end{equation}
where $E_{A}(r)$ becomes nonzero near the boundary and decays
slowly away from it. We can calculate $E_{A}(r)$ by Abelian
bosonization modified by open boundary conditions~\cite{Tsai00}. In the critical region
$\left| \Delta \right| \leq 1$, the low energy effective
Hamiltonian is just a free massless relativistic boson.

The staggered part of $h_{r}\sim (-1)^{r+1}(\psi _{R}^{\dagger
}\psi
_{L}-\psi _{L}^{\dagger }\psi _{R})\sim (-1)^{r+1}\sin (\sqrt{4\pi K}\phi )$%
. Here we follow the notation of Ref.~\cite{Eggert92},
but define the Luttinger parameter as 
$K=\pi /(2(\pi -\cos ^{-1}\Delta ))$ so that $K=1$ for an $XY$ spin chain and $K=1/2$ for the Heisenberg model.
In a system with finite ${\RR}$ and open boundary conditions,
\begin{equation}
E_{A}(r,{\RR})\propto \left\langle \sin (\sqrt{4\pi K}\phi
)\right\rangle \propto \frac{1}{[\frac{2{\RR}}{\pi }\sin (\frac{\pi
r}{{\RR}})]^{K}}. \label{EA}
\end{equation}
This is our basic result for $E_A$, from which it follows that $E_A(r,\RR)=f(r/\RR)/\sqrt{\RR}$
for the Heisenberg model but $E_A(r,\RR)=f(r/\RR)/\RR$ for an $XY$ spin chain, 
corresponding to free fermions. 

At the Heisenberg point, $\Delta =1$, Eq. (\ref{EA}) will have
some logarithmical corrections due to the presence of a marginally
irrelevant coupling constant, $g$, in the low energy Hamiltonian, Eq. (\ref{margint}), that we
now try to take into account.  This interaction is reviewed in \ref{app:field_theory}. 
 Ignoring boundaries, the staggered energy density $E_{A}\sim \sin (\sqrt{2\pi }\phi )$ 
has the anomalous dimension
\begin{equation}
\gamma (g )=1/2-3g/4 .  \label{AD}
\end{equation}
With a boundary, the renormalization group equation for $E_{A}$ is the naive 
one, involving the anomalous dimension, $\gamma (g)$ in the usual way.
\begin{equation}
\lbrack \partial /\partial (\ln r)+\beta (g )\partial
/\partial g +\gamma (g )]E_{A}(r,r/\RR )=0.  \label{RG}
\end{equation}
(Here the partial derivative with respect to $r$ is taken with $r/\RR$ held fixed.) 

This result may require some justification since, in general, the presence 
of a boundary can strongly affect the scaling behavior.  It is crucial 
here that we are considering $E_A$ far from the boundary compared to 
the ultraviolet cut-off (i.e. $r\gg 1$). The boundary condition dictates 
that we should regard the right moving factor in the operator $e^{i\sqrt{2\pi}\phi (r)}$, 
which occurs here as a left moving operator at the reflected point $-r$:
\begin{equation} e^{i\sqrt{2\pi}\phi (r)}=e^{i\sqrt{2\pi}(\phi_L(r)+\phi_R(r))}
\to e^{i\sqrt{2\pi}(\phi_L(r)-\phi_L(-r))}.\end{equation}
To calculate the anomalous dimension of this bi-local operator 
we can consider the operator product expansion (OPE) with 
the marginal interaction (reviewed in \ref{app:field_theory}). 
This marginal interaction, $\cos \sqrt{8\pi}\phi$ also 
becomes bi-local in the presence of the boundary.  The 
OPE of these two bilocal operators, 
$e^{i\sqrt{2\pi}(\phi_L(r)-\phi_L(-r))}$ and 
$e^{-i\sqrt{8\pi}(\phi_L(r)-\phi_L(-r))}$
is the produce of the OPE's of the operators at $\pm r$ separately. 
Fortunately, this gives exactly the same result as without 
the boundary.  Therefore we expect the naive RG equation to apply. 

The general solution of Eq. (%
\ref{RG}) is
\begin{equation}
E_{A}(r,r/\RR )=F[g (r),r/\RR ]\exp \{-\int_{r_{0}}^{r}d(\ln r^{\prime
})\gamma [g (r^{\prime })]\},
\end{equation}
where $F$ is an arbitrary function of $g (r)$ and $r/\RR$. 
\begin{figure}[ht]
\begin{center}
\includegraphics[width=10cm,clip]{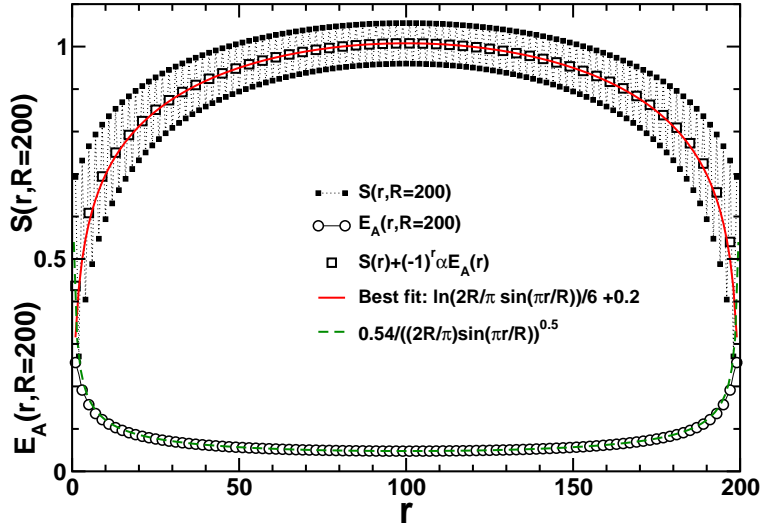}
\end{center}
\caption{Entanglement entropy $S(r)$ (black squares) and alternating part of the                                                                                 
energy density $E_A(r)$ (open circles) computed with DMRG at $J_{2}^{c}$ for                                                                                     
$R=200$ sites. The dashed line is Eq.~(\ref{EA}). The uniform part of $S(r)$, obtained by taking $S(r)+(-1)^r\alpha                                                                                 
E_A(r)$ with $\alpha=1.001699$, is represented by open squares. The best fit,                                                                                    
shown by a red curve, is indicated on the plot.} 
\label{fig:Dim}
\end{figure}

In the weak coupling limit, we may evaluate the scaling function $F(g(r),r/\RR )$ 
at $g(r)=0$ and with the
second order
beta function, $\beta =-g^{2},$ we obtain:
$E_{A}(r,r/\RR )=F(r/\RR )/[\sqrt{r}(\ln \left| r\right| )^{3/4}]$.
One can push this a bit further following a similar calculation in Ref.~%
\cite{Barzykin99, Ian98}. Provided with the beta function up to
third order
\begin{equation}
\beta (g)=g^{2}-(1/2)g^{3},  \label{beta}
\end{equation}
the effective coupling solved from Eq. (\ref{beta}) is
\begin{equation}
\frac{1}{g (r)}-\frac{1}{g_{0}}=\{\ln (r/r_{0})+\frac{1}{2}\ln [\ln (r/r_{0})]\},
\end{equation}
and expanding $F(g(r),r/\RR )$ in powers of $g (r),$ we can improve
the solution as
\begin{eqnarray}
E_{A}(r,r/\RR )&=&\frac{F(r/\RR )}{\sqrt{r}[\ln (r/a_{1})+\frac{1}{2}\ln \ln (r/a_{1})]}
\nonumber \\
&&\cdot \left\{ 1+\frac{a_{2}}{[\ln (r/a_{1})]^{2}}\right\} ,  \label{LogEA}
\end{eqnarray}
where a term proportional to $1/[\ln (r/a_1)]$
which could have occurred inside the curly brackets can always be adsorbed by redefining
$a_{1}$. Note that since $r/\RR$ is being held fixed here, as we take 
$r\to \infty$, it follows that we can always replace $r$ by $\RR$ 
inside the logarithms in Eq. (\ref{LogEA}) by rescaling $a_1$ by $\RR /r$.

As reviewed in \ref{app:field_theory}, at the critical value of $J_2$, 
the marginal coupling constant vanishes and all logarithmic corrections vanish. 
Hence, we can check Eq.~(\ref{EA}) directly using our DMRG results if we work at the critical
point $J_2^c$. This is shown in Fig.~\ref{fig:Dim} where results for $E_A(r,R=200)$ for a uniform ($J'_K=1)$ system with $R=200$
are plotted along with Eq.~(\ref{EA}). The agreement is quite good. In Ref.~\cite{Laflorencie06} it was argued
that the alternating part of the entanglement entropy is given by $S_A=-\alpha E_A$. With $\alpha=1.001699$ we
show $S(r,R)+(-1)^r\alpha E_A(r,R)$ in Fig.~\ref{fig:Dim} and excellent agreement with this quantity and the uniform
part of $S(r,R)$ is observed. Fitting this uniform part to Eq.~(\ref{eq:sobc}) and using the fact that~\cite{Eggert92} $\ln g=-(1/4)\ln 2$
we estimate $s_1\simeq 0.746$.

\subsection{Scaling of $S^A_{imp}$ at $J_2^c$}
\begin{figure}[!ht]
\hfill\includegraphics[width=14cm,clip]{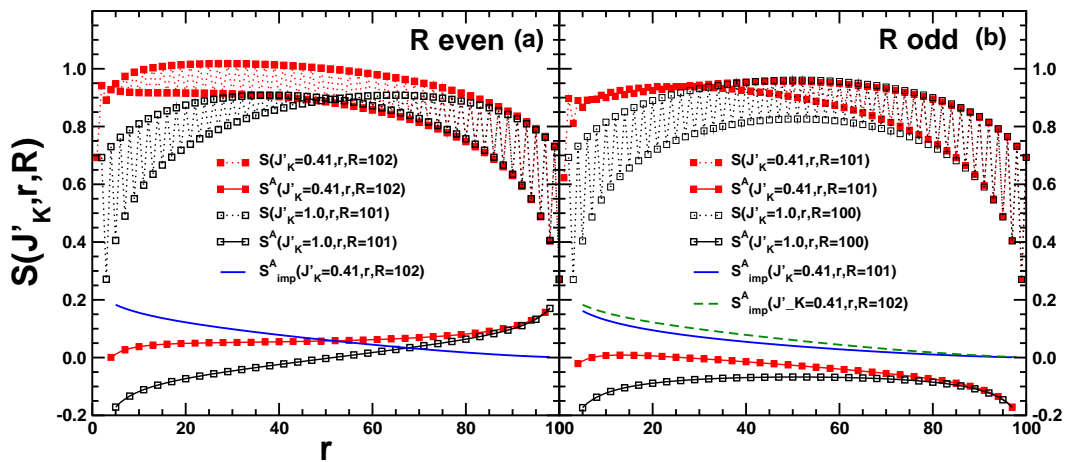}
\caption{(a) DMRG results for the total entanglement entropy, $S(J'_K,r,R)$ for a 102 site spin chain at $J^c_2$,
  with a $J'_K=0.41$ Kondo impurity ($\blacksquare$) along with $S(J'_K=1,r-1,R-1)$ ($\square$).
    For both cases is the extracted alternating part shown along with the resulting $S^A_{imp}(J'_K=0.41,r,R=102)$ for $R$ even.
(b) DMRG results for the total entanglement entropy, $S(J'_K,r,R)$ for a 101 site spin chain at $J^c_2$,
  with a $J'_K=0.41$ Kondo impurity ($\blacksquare$) along with $S(J'_K=1,r-1,R-1)$ ($\square$).
    For both cases is the extracted alternating part shown along with the resulting $S^A_{imp}(J'_K=0.41,r,R=101)$ for $R$ odd.
    For comparison we also show $S^A_{imp}(J'_K=0.41,r,R=102)$ for $R$ even from panel (a) (dashed line).
}
\label{fig:RawASimp}
\end{figure}
Our fundamental definition of $S_{imp}$, Eq.~(\ref{Simpdef}), focused only on the uniform part
of the entanglement entropy. It is also possible to define the alternating part of the impurity
entanglement entropy following Eq.~(\ref{Simpdef}):
\begin{equation}
S^A_{imp}(J'_K,r,{\RR})\equiv S_A(J'_K,r,{\RR})-S_{A}(1,r-1,{\RR}-1).\label{ASimpdef}
  \end{equation}
As before, we have subtracted $S_A$ when the impurity is absent,
in which case both $r$ and $\RR$ are reduced by one and
the coupling at the end of this reduced chain, linking site $2$ to
$3$ and $4$, has unit strength. Applying this definition to numerical data involves some subtleties.
First of all $S^A$ is only defined up to an overall sign. Secondly, when calculating $S_{A}(1,r-1,{\RR}-1)$
we define this as $-S_{A}(1,r',{\RR}')$ with $R'=R-1$ since the shift from $r$ to $r-1$ implies a sign change in the
alternating part. For convenience we have therefore always exploited this degree of freedom to use a sign convention
that makes the resulting $S^A_{imp}$ positive in all cases. In Fig.~\ref{fig:RawASimp} we show data for the
total entanglement entropy along with the extracted alternating parts and the resulting $S^A_{imp}$ for both
$R=102$ even and $R=101$ odd. The initial data are the same as shown in Fig.~\ref{fig:RawSimp} with $\xi_K=25.65$.
As was the case for $S_{imp}$ we do not observe any special features in $S^A_{imp}(r)$ for fixed $R, J_K'$ associated with the length
scale $\xi_K$ and in all cases $S^A_{imp}$ decays monotonically with $r$.

\begin{figure}[!ht]
\begin{center}
\includegraphics[width=15cm,clip]{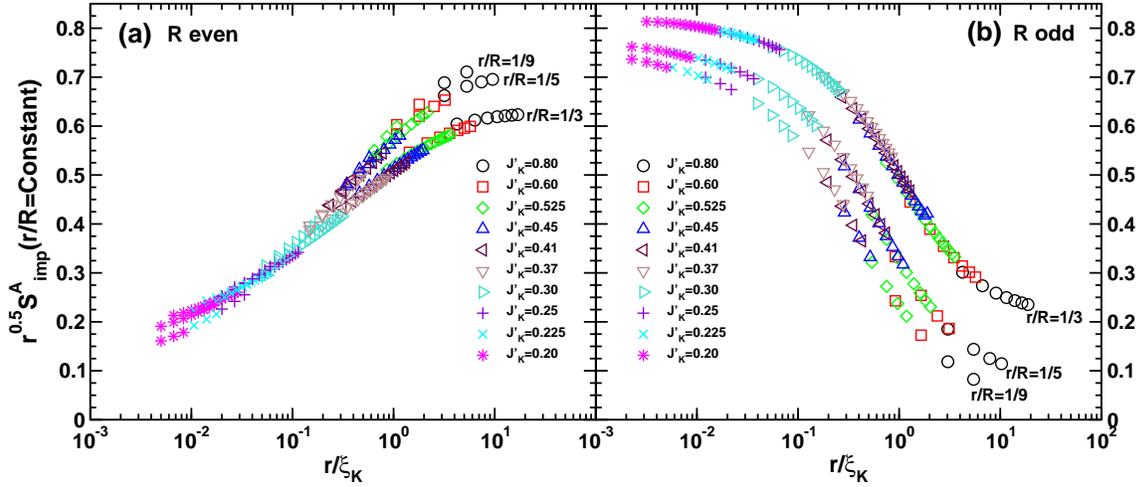}
\end{center}
\caption{$\sqrt{r}S^A_{imp}$ for fixed $r/R$ and a range of couplings $J'_K$ at $J_2^c$. (a) for $R$ even and (b) for $R$ odd.
The values of $\xi_K$ used are
obtained from the scaling of $S_{imp}$ and are listed in table~\ref{tab:xiKScal}.
DMRG results with $m=256$ states. }
\label{fig:ASimprovr}
\end{figure}
We now turn to a discussion of a possible scaling form for $S^A_{imp}$.
In Ref.~\cite{Laflorencie06} it was shown that for $J_K'=1$ the alternating part of
the entanglement entropy, $S^A$, is proportional to the alternating part in the energy, $E_A$
and it was shown that $E_A(r)=f(r/R)/\sqrt(r)$ for some scaling function $f$. See also the
detailed derivation in subsection~\ref{sec:ea}. A first guess for a scaling for $S_{imp}^A$ would
then simply follow a generalization of the above formula to the case $J_K'\neq1$. Naively, this would
imply that $\sqrt(r)S^A_{imp}$ should be a scaling function, $f(r/R,r/\xi_K)$.
Our results for $\sqrt(r)S^A_{imp}$ for fixed $r/\RR$ are shown in Fig.~\ref{fig:ASimprovr} for a range
of $J'_K$ and $\RR$. The values for $\xi_K$ used to attempt the scaling are the ones previously
determined from the scaling of $S_{imp}$ for fixed $r/\RR$ at $J_2^c$, listed in table~\ref{tab:xiKScal}.
Clearly the results for $\sqrt{r}S^A_{imp}$ follow the expected scaling form. We expect that the scaling
would have been better had we allowed the $\xi_K$ to vary instead of using the data from table~\ref{tab:xiKScal}.

\subsection{Alternating Part of the Fixed Point Entanglement Entropy}
\begin{figure}[!ht]
\begin{center}
\includegraphics[width=10cm,clip]{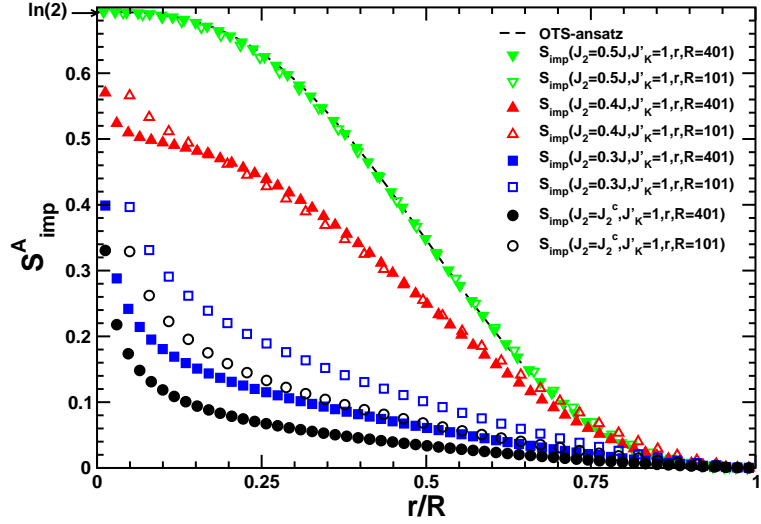}
\end{center}
\caption{DMRG results for the alternating part of the fixed point entanglement, $S^A_{imp}(J_2,J'_K=1,r/R)$,
shown for a range of values of $J_2=J_2^c,0.3J,0.4J,0.5J$, in each case for 2 different values
  of $R=101,401$.
  The dashed line indicates the result for the OTS-ansatz, Eq.~(\ref{eq:SAimpodd}).
}
\label{fig:ASimp1Scal}
\end{figure}
Following our definition of the alternating part of the impurity
entanglement entropy, $S^A_{imp}$, given in Eq.~(\ref{ASimpdef}),
we can also analyze the alternating part of the fixed point
entropy by studying $S^A_{imp}(J'_K=0)$ and $S^A_{imp}(J'_K=1)$.
If we first focus on $S^A_{imp}(J'_K=0)$ it is clearly zero when
$R$ is odd as was the case for $S_{imp}(J'_K=0)$. For $R$ even, we
see from Eq.~(\ref{eq:SAeven}) and Eq.~(\ref{eq:SAodd}) that at
the MG-point ($J_2=J/2$) the OTS-ansatz predicts that also in this
case is $S^A_{imp}(J'_K=0)$ zero. Numerical DMRG results confirms
this and shows that also for $J_2<J/2$ is $S^A_{imp}(J'_K=0)$
negligible. We have verified that this also holds at the critical
point $J_2=J_2^c$ as well as in the gapless Heisenberg phase at
$J_2=0.2J$.

The interesting fixed point entropy is then $S^A_{imp}(J'_K=1)$. As for $S_{imp}(J'_K=1)$, we see that the difference between
$R$ even and $R$ odd amounts to a change of sign of the resulting fixed point entanglement entropy. We therefore only
consider $R$ odd. At the MG-point we expect the OTS-ansatz to yield very precise results. If we combine Eq.~(\ref{eq:SAodd}) with the
fact that for $R$ even $S^A(J'_K=1)=\ln(2)/2$ for any $r$, we find the OTS result:
\begin{equation}
S^A_{imp}(J'_K=1)=(1-p)\ln(2),\ R\ {\rm odd}.\label{eq:SAimpodd}
\end{equation}
Where $p$ is given in Eq.~(\ref{eq:p}).
As we mentioned in the previous section there is a slight subtlety here since, in order to calculate $S^A_{imp}(J'K=1)$ for $R$ odd,  we need
$S_{A}(1,r-1,{\RR}-1)$ for $R-1$ even. Due to the shift, $r\to r-1$, we have defined this as $-\ln(2)/2$
as opposed to the $\ln(2)/2$ we just quoted. This sign convention renders the resulting fixed point entanglement
entropy positive for all $r/R$. Choosing the other possible sign convention would have resulted in a trivial shift
in the fixed point entanglement entropy of $\ln(2)$, yielding $-p\ln(2)$.

Our numerical DMRG results for $S^A_{imp}(J'_K=1)$ for $R$ odd as a function of $r/R$ are shown in Fig.~\ref{fig:ASimp1Scal} for a range
of second nearest neighbor couplings $J_2=J_2^c,0.3J,0.4J,0.5J$ throughout the dimerized phase $J_2^c\leq J_2\leq J/2$.
In each case we show results for two system sizes $R=101$ and $401$. At the MG-point we see that finite size effects are minimal
and the results for $R=101$ and $401$ agree very well. In both cases the agreement with the result from the OTS-ansatz, Eq.~(\ref{eq:SAimpodd}), shown
as the dashed line, is almost perfect. As $J_2$ is decreased from the value of $J/2$ pronounced finite-size effects develop and the fixed point
entanglement entropy clearly tends toward zero with increasing $R$. This is also true at the critical point, $J_2=J_2^c$. For clarity we
have not shown results for $J_2=0.2J$ in the gapless phase where $S^A_{imp}(J'_K=1,r/R)$ for both $R=101$ and $401$ is smaller than at the critical
point.

\section{Conclusions~\label{sec:conclusion}}
The presence of impurities will clearly affect the entanglement and we have here defined
the impurity contribution to the entanglement entropy, $S_{imp}$. 
Using the equivalence of the electronic Kondo model and the $J_1-J_2$ spin chain model
in the regime $J_2<J_2^c$ we have shown numerical evidence that $S_{imp}$ follows a scaling
form $S_{imp}(r/\xi_K,r/R)$ demonstrating the presence of the length scale $\xi_K$ associated
with screening of the impurity. 
We have provided rather
strong arguments in favor of this scaling and analytical results based on a Fermi liquid approach
valid for $r\gg\xi_K$. 
The D-dimensional Kondo model has been shown to yield the same
entanglement as the 1-D Kondo model which in turn is equivalent to the entanglement in the
$J_1-J_2$ spin chain. 
The single-site entanglement obtained by only considering the entanglement
of the impurity spin has been shown to display weak scaling violations. For $J_2>J_2^c$ in the 
dimerized phase we have shown that almost exact
calculations are possible using the variational TS-ansatz. Simplifying the TS-ansatz we have
shown that the contributions arising from impurity valence bonds (IVB) and single particle entanglement (SPE)
naturally arise at the MG-point. Finally, we have argued that at finite temperatures the entanglement entropy
for a sub-system $A$ of size $r$, $S(T)$, will approach the thermal entropy, $S_{th}(T)$, for $T\gg v/r$.
In light of the generality of our model we expect our results to be quite
widely applicable. The study of quantum impurity entanglement as it occurs in more complex models with 
many body ground-states displaying non-trivial order would clearly be of considerable interest.

%
\ack
We are grateful to J. Cardy for interesting discussions. This research was
supported by NSERC (all authors), the CIAR (IA) CFI (ESS) and SHARCNET
(ESS). Numerical simulations have been performed on the WestGrid network
and the SCHARCNET facility at McMaster University.

\appendix

\section{Review of field theory results on the Kondo effect and spin chains}
\label{app:field_theory}
The usual Kondo model describes electrons moving in 3 dimensions with 
a short range exchange interaction with a single S=1/2 impurity spin.  No 
other electron-electron interactions are taken into account.  Although 
similar conclusions can be arrived at more generally, the 
simplest case is where a free electron quadratic dispersion 
relation is assumed, and the Kondo 
interaction is taken to be a $\delta$-function, implying spherical symmetry.
The Hamiltonian is given in Eq. (\ref{H3D}). 

Kondo model calculations are traditionally carried out by reducing
the 3$D$ fermionic model of Eq.~(\ref{H3D}) to an equivalent
1$D$ fermionic model. This can be done by expanding $\psi
(\vec{r})$:
\begin{equation}
\psi (\vec{r})=(1/r)\sum_{l,m}Y_{l,m}(\hat{r})\psi _{l,m}(r),
\end{equation}
where the $Y_{l,m}(\hat{r})$ are the usual spherical harmonics,
depending only on the direction of $\vec{r}$ and the $\psi
_{l,m}(r)$ are 1$D$ quantum fields. The spherically symmetric $H$ of
Eq.~(\ref {H3D}) reduces to a sum of commuting terms:
$H=\sum_{l,m}H_{l,m}$. Due to the $\delta$-function interaction, 
all harmonics are non-interacting except for the $s$-wave.
The higher harmonics contain only the centrifugal
potential energy $V_{l}(r)=l(l+1)/2r^{2}$.
 A low energy form of the $l=0$ Hamiltonian is obtained by linearizing the dispersion relation
 around the Fermi surface, 
$\epsilon (k)=k^{2}/2m-\epsilon _{F}$, yielding a 1$D$
Dirac fermion theory on the half-line, $r>0$, coupled to the impurity spin
at $r=0$.  We introduce left and right movers by:
\be \psi_{l=0} (r)\approx e^{ik_Fr}\psi_R(r)+e^{-ik_Fr}\psi_L(r),\ee
where $\psi_{L/R}(r)$ vary slowly and $k_F$ is the Fermi momentum. 
\begin{eqnarray}
H_{1D} &\approx &(iv/2\pi )\int_{0}^{R}dr[\psi _{L}^{\dagger
}(d/dr)\psi _{L}-\psi _{R}^{\dagger }(d/dr)\psi _{R}]  \nonumber\\
&&+v\lambda_{K}\psi _{L}^{\dagger }(0)(\vec{\sigma}/2)\psi _{L}(0)\cdot \vec{S}.
\label{H1D}
\end{eqnarray}
Note that we have adopted an unconventional normalization for these 1D fermion fields:
\be \{\psi_{L/R}^\dagger (r),\psi_{L/R}(r')\}=2\pi \delta (r-r').\ee
Here $\lambda _{K}\propto J_K$ and the left and right movers obey a
boundary condition: $\psi _{L}(0)=\psi _{R}(0)$ \cite{Affleck90,Affleck91a,Affleck91b}. Up to some
trivial geometric factors of $r$, all the physics of Eq.~(\ref{H3D})
can then be obtained from this 1$D$ fermion model.  If the 3D system 
is originally defined inside a sphere of radius $\RR$, then 
the 1D system exists on a line of length $\RR$.

It is now very convenient to bosonize this 1D model. This 
allows for the introduction of separate bosonic 
fields representing the spin and charge degrees of 
freedom of the conduction electrons.  The non-interacting 
free electron Hamiltonian can be written as a sum of 
decoupled spin and charge terms:
\be  H_{0}=H_{s0}+H_{c0}.\ee
The spin part can be written in terms of the spin current operators:
\be \vec J_{L/R}\equiv \psi^\dagger_{L/R}{\vec \sigma \over 2}\psi_{L/R},\ee
as:
\be H_{s0}=(v/2\pi )\int_0^{\RR}dr (1/3)[\vec J_L\cdot \vec J_L+\vec J_R\cdot \vec J_R].\ee
 The free boundary condition on the fermion fields at $r=0$ implies a 
boundary condition in the continuum limit: $\psi_L(0)=-\psi_R(0)$ which 
in turn implies $\vec J_L(0)=\vec J_R(0)$. 
Since the Kondo interaction only involves $\vec J_L$ it can 
be written entirely in terms of the spin boson field. 
\be H_K=v\lambda_K\vec J_L(0)\cdot \vec S.\ee
(Actually, there are irrelevant operators which couple 
spin and charge sectors together, but these can be 
ignored in the low energy effective theory.) 

It is actually possible to continue to the negative $r$ axis, defining:
\be \psi_L(-r)\equiv -\psi_R(r), \ \  (0<r<\RR).\ee
This in turn implies that $\vec J_R(r)=\vec J_L(-r)$. 
The free spin Hamiltonian then becomes:
\be H_{s0}=(v/6\pi )\int_{-\RR}^{\RR} dr \vec J_L\cdot \vec J_L.\label{H0s}\ee
Imposing convenient boundary conditions on left and right fields at $r=\RR$, 
we obtain the periodic boundary conditions in the left-moving formulation:
$\vec J_L(-\RR )=\vec J_L(\RR )$. 

Now consider the $J_1$-$J_2$ spin chain model of Eq. (\ref{eq:spinch}). 
The connection
between spin chains and the Kondo model was pointed out in~\cite{Eggert92,Rommer00},
using field theory arguments
and in~\cite{Frahm97} the Hamiltonian of Eq. (\ref{eq:spinch}) with $J_2=0$ was, among others,
solved exactly by Bethe ansatz noting connections with Kondo physics.
The low energy theory can again be described by bosonization.  
One approach is to bosonize the weakly coupled Hubbard model at half-filling 
and then to extrapolate to strong coupling. We again 
obtain spin and charge bosons.  The Hubbard interaction gives 
a gap to the charge boson field, which can thus be eliminated 
from the low energy theory. The spin operators at site $j$ 
can be represented:
\be \vec S_j\approx [\vec J_L+\vec J_R]+(-1)^j\hbox{constant}\cos \phi_c \cdot \vec n.\ee
Here $\phi_c$ is the charge boson and $\vec n$ is the antiferromagnetic order parameter 
and can be written entirely in terms of the spin boson field. The Hubbard 
interaction makes $<\cos \phi_c >\neq 0$ so we may replace this 
factor by a constant:
\be \vec S_j\approx [\vec J_L+\vec J_R]+(-1)^j\hbox{constant}\cdot \vec n.\ee
The low energy Hamiltonian is simply the spin part of the non-interacting 
electron Hamiltonian, $H_{s0}$, up to irrelevant operators. In particular, 
the spin current operators have the same Green's functions as in 
the non-interacting model. The field $\vec n$ can also be 
expressed in terms of the non-interacting spin bosons.  
It is represented in terms of exponentials of bosons 
(in the usual abelian bosonization scheme).  

Now suppose that the chain has a free end at $r=0$.  The boundary 
conditions on the fermion fields can be translated into 
boundary conditions on the spin boson field.  These can be seen 
to imply:
\be \vec n(0)\propto \vec J_L(0)=\vec J_R(0).\label{nbc}\ee

Now suppose that we weakly couple one additional spin to the spin chain, 
with a coupling constant $J_k'\ll 1$. We may write the low 
energy effective Hamiltonian using bosonization. If the extra 
spin couples at the end of the chain then the resulting interaction 
term is particularly simple:
\be H_K\propto \vec J_L(0)\cdot \vec S.\ee
The coupling constant here is proportional to $J_K'$ but 
the constant of proportionality is non-trivial involving 
the proportionality constant in Eq. (\ref{nbc}). It 
can be extracted by studying the end-to-end spin correlation 
function in the chain without the weakly coupled spin \cite{KEAC}, 
yielding:
\be J_K'\approx 1.3807 \lambda_K.\label{1.38}\ee
Thus the same low energy effective Hamiltonian is 
obtained for the spin chain with a weakly coupled spin at 
the end as for the usual free fermion Kondo model.  

Two caveats should be made here.  First of all, 
if the weakly coupled spin couples to a point far 
from the end of the spin chain then a much different 
low energy effective Hamiltonian occurs. Secondly, 
even when the weakly coupled spin is at the end, 
there are subtle differences from the usual 
Kondo model due to differences in the irrelevant operators. 
In the free fermion model all irrelevant operators 
have dimensions of at least 3 and are strictly irrelevant. 
However, the spin chain model has a notorious 
marginally irrelevant operator in its low energy effective Hamiltonian.
\be \delta H = -(gv/2\pi )\vec J_L\cdot \vec J_R.\label{margint}\ee 
This leads to logarithmic corrections to essentially all low 
energy properties which are difficult to calculate in detail 
and cause problems with fitting numerical data.  Fortunately, 
there is a way of avoiding this difficulty.    As we vary 
$J_2$, this marginal coupling constant, $g$ also 
varies.  For small $J_2$, $g>0$, and is marginally irrelevant. 
A critical point occurs at $J_2=J_c\approx .2412$. For 
larger $J_2$ the system goes into a gapped spontaneously 
dimerized phase. Right at the critical point 
$g =0$ and all logarithmic corrections vanish.  
Much better agreement between numerical simulations 
and field theory predictions can be obtained at this 
point.  All remaining irrelevant operators 
are strictly irrelevant, leading only to corrections 
which vanish with power-laws of the energy (or inverse length) 
scale, not logarithms. In a separate paper\cite{KEAC}, 
we study the $J_2=0$ Kondo spin chain model in more detail. 
Here we will focus on the $J_2=J_c$ case.  Note 
that this is actually much closer to the free fermion 
version of the Kondo model.  We note that Eq. (\ref{1.38}) was 
obtained at $J_2=J_c$. 

At low energies and long length scales, the effective Kondo coupling becomes
large, and the effective Hamiltonian flows to the strong coupling fixed
point. In the electron version of the Kondo model we may think of this fixed
point as one where the impurity spin is ``screened'', i.e. it forms a
singlet with a conduction electron. The remaining electrons behave, at low
energies and long length scales, as if they were non-interacting, except
that they obey a modified boundary condition reflecting the fact that they
cannot break up the singlet by occupying the same orbital as the screening
electron. This modified boundary condition corresponds to a $\pi /2$ phase
shift. Correspondingly in the spin chain Kondo model, the impurity spin gets
``adsorbed into the chain'' and no longer behaves like a paramagnetic spin
at low energies and long distances. The leading corrections to this low
energy long distance picture are described by lowest order perturbation
theory in the leading irrelevant operator at the strong coupling fixed
point. This is an interaction between the remaining conduction electrons,
near the screened impurity. (It doesn't involve the impurity itself since it
is screened and doesn't appear in the low energy effective Hamiltonian.)
This leading irrelevant operator is $\vec{J}_{L}(0)\cdot \vec{J}_{L}(0)$\cite
{Affleck91a,Affleck91b}. From (\ref{H0s}), we see that this is proportional
to the spin part of the free electron energy density. It is the entire
energy density in the low energy effective Hamiltonian for the spin chain.
The energy density has dimensions of (energy)/(length) so the corresponding
coupling constant in the effective Hamiltonian must have dimensions of
length. On general scaling grounds we expect it to be proportional to $\xi
_{K}$. The precise constant of proportionality simple corresponds to giving
a precise definition of what we mean by $\xi _{K}$. We adopt the convention:
\begin{equation}
H_{int}=-(\pi \xi _{K}){\cal H}_{s,L}(0).  \label{Hintcon}
\end{equation}
Here the subscripts $s$ and $L$ are a reminder that this is the
spin only part of the energy density for left movers. For the purpose of doing first order
perturbation theory in $H_{int}$ for quantities like the
susceptibility, specific heat or ground state energy, which are
translationally invariant in $0^{th}$ order, we may replace~\cite{Affleck90} $%
H_{int}$ by:
\begin{equation}
H_{int}\to -[\pi \xi _{K}/(2L)]\int_{-R}^{R}{\cal H}_{s,L}(x).
\end{equation}
This is equivalent to a length dependent reduction of the velocity:
\begin{equation}
v\to v[1-\pi \xi _{K}/(2R)].  \label{vshift}
\end{equation}
This then implies that the susceptibility, which is $R/(2\pi v)$ in the
absence of the Kondo impurity becomes:
\begin{eqnarray}
\chi \to R/\{(2\pi v)[1-\pi \xi _{K}/(2R)]\}&\approx& R/(2\pi v)
+\xi _{K}/(4v)\nonumber\\
&=&R/(2\pi v)+1/(4T_{K}).
\end{eqnarray}
Thus the zero temperature impurity susceptibility is $1/(4T_{K})$. It is this
form of the impurity susceptibility, simply related to the high temperature,
free spin behavior, $1/(4T)$, which motivates the definition of $\xi _{K}$
(and hence $T_{K}=v_{F}/\xi _{K}$) implied by (\ref{Hintcon}). We note that
this interaction $H_{int}$ is present even in the absence of an impurity,
for free fermions but then the coupling constant is of order a lattice
constant. Similarly it is also present for the spin chain with no impurity
(i.e. $J_{K}^{\prime }=1$) with a coupling constant of order a lattice
constant. The effect of a weak Kondo coupling is to make this coupling
constant large. We emphasize that this precise choice of definition of $%
T_{K} $ has no physical consequences. The power of Fermi liquid theory is to
predict not only the form of low energy quantities but also ratios of
various low energy quantities such as impurity susceptibility, impurity
specific heat, resistivity, etc., corresponding to various generalized
Wilson ratios. In the limit $\xi _{K}\ll r$, we can also calculate $S_{imp}$
using the FLT interaction of Eq.~\ref{Hintcon} in the lowest order
perturbation theory.

\section{3D to 1D reduction and entanglement entropy}
\label{app:3D-1D}
In this appendix we prove that the impurity entanglement 
entropy is the same for the $D$ and one-dimensional Kondo model and consequently also for
the spin chain model.  We also present a new derivation of the free fermion entanglement entropy
in D-dimensions.

With a spherically symmetric dispersion relation and Kondo interaction, 
as in Eq. (\ref{H3D}), the Hamiltonian is 
a sum of terms acting on subspaces of different $l$, $m$ quantum numbers for single electrons. 
That is, the kinetic energy term, for each electron separates into such a sum and so 
does the Kondo interaction.  If the Kondo interaction is a $\delta$-function then 
only the s-wave term is non-zero.  If it is longer range there are small larger $l$ terms. 
However, according to the usually RG picture of the multi-channel Kondo model the largest 
term grows under renormalization and the others shrink, so that in the low energy effective 
Hamiltonian we can ignore the higher harmonics. 

This form of $H$ implies that the ground state wave-function is a product of factors, 
one from each channel. This follows because the single-particle wave-functions all 
have definite $l$, $m$ quantum numbers.  We may define electron creation operators 
$c^\dagger_{l,m,r}$ which create an electron at  distance $r$ from the origin in 
the $l$, $m$ channel, and write the ground state as a product over $l$ and $m$ of 
factors for each $l$, $m$.  Ignoring the Kondo interaction (as we can basically do 
for $l>0$) each factor just corresponds to the product of creation operators 
corresponding to a filled Fermi sea. (Each channel is filled up to the Fermi energy.) 
Including the Kondo interaction this factorization is still valid but the wave-function 
for channels with a Kondo interaction are non-trivial. So, of course, the pure 
density matrix $|\psi ><\psi |$ is also a product of factors for each $l$, $m$. 

Now consider tracing over the region outside a sphere of radius $r$. It is important 
here that we choose a sphere, not some other shape, so as to preserve the 
rotational symmetry. This should leave a reduced density matrix which is also 
a product over $l$ and $m$.  This may be especially clear if we introduce an ultra-violet 
cut-off by only allowing the particles to sit on the surfaces of spheres at various 
distances from the origin. The reduced density matrix is obtained by tracing over 
all the spherical surfaces further from the origin than $r$. So we can write:
\be \rho (r)= \prod \rho_{l,m}(r).\ee 
Here each reduced density matrix, $\rho_{l,m}$ acts on a different sector of the Hilbert 
space. So we can write:
\be \ln \rho (r) = \sum_{l,m}\ln \rho_{l,m}(r).\ee
Actually, this notation implies that each term is a product of a non-trivial operator 
in one channel and the identity operator in all the others. When we calculate tr$\rho \ln \rho$, 
we get a sum of terms like tr$\rho \ln \rho_{l,m}$. Because we can write a complete set of 
states as a set of products in each channel, the trace of an operator 
written in product form reduces to a product of traces.  So:
\be \Tr \rho \ln \rho_{l,m}=[\prod_{(l',m')\neq (l,m)}\Tr \rho_{l',m'}]\Tr (\rho_{l,m}\ln \rho_{l,m}).\ee
Since tr$\rho_{l',m'}=1$, we get:
\be \Tr [\rho \ln \rho ] =\sum_{l,m} \Tr [\rho_{l,m}\ln \rho_{l,m}].\ee

Now if we compare the zero Kondo coupling to finite Kondo coupling case, only the s-wave channel 
changes.  (More correctly, for a finite range Kondo interaction there is also some change 
in the higher $l$ channels.  However, this is presumably just a short distance effect and 
drops off much more quickly than the s-wave part.  i.e. like $a/r$ where $r$ is a cut off scale, 
instead of $\xi_K/r$ for the s-wave.) So the impurity entanglement entropy for the 3D system is the 
same as for 1D. Exploiting the equivalence of the 1D Kondo model and the spin chain model the entanglement
entropy is the same also for this model.

We remark that the well-known $r^2\ln r$ form of the entanglement entropy \cite{Wolf06,Gioev06,Barthel06} for 
free fermions (at $R\to \infty$) can apparently be recovered from the 
decomposition into angular momentum channels.  Note that the Hamiltonian 
for all $l>0$ contains  a centrifugal potential:
\be H_{l,m}=-{1\over 2}{d^2\over dr^2}+{l(l+1)\over 4r^2}.\ee
This centrifugal potential will not be too important when the 
size, $r$, of region $A$ is sufficiently large and $l$ is not too large. 
Thus we expect to recover essentially the usual 1D result 
$S=(1/6)\ln r$ for these values of $l$ and $m$. However, 
for large enough $l$, we expect the entanglement entropy 
to be reduced, and eventually to vanish at large $l$ for 
any fixed $r$.  This can be seen from the fact that 
the density of electrons in region $A$ with 
angular momentum $l$ vanishes at large $l$.  If 
the region $A$ is essentially empty there can't be 
any entanglement. We may estimate the order of 
magnitude of $l$ at which the sphere of size $r$ 
becomes empty by simply comparing the centrifugal potential 
to the Fermi energy,
\be E_F\approx {l_{max}(l_{max}+1)\over 4r^2}.\ee
This gives
\be l_{max}\approx 2r\sqrt{E_F}.\ee
Thus we estimate the free fermion entanglement entropy in 3D as:
\be S(r)\approx (1/6)\ln r \sum_{l=0}^{l_{max}}(2l+1)
\approx {E_Fr^2\over 3}\ln r.\ee
(The prefactor of $E_F\propto n^{-2/3}$ where $n$ is 
the electron density is only expected to be correct 
in order of magnitude. )  This 
argument generalizes to any dimension $D$, giving:
\be S\propto r^{D-1}\ln r.\ee

\section{Separating Staggered and Uniform Parts~\label{app:7p}}
In this appendix we derive the 7-point formula used for extracting the uniform
and alternating parts of the entanglement entropy from the numerical data.

We focus on functions, $f$, defined on a set of discrete lattice sites $i$,
with a uniform part, $u$, in addition to a staggered part, $s$:
\begin{equation}
f(i)=u(i)+(-1)^is(i).
\end{equation}
We assume that both $s$ and $u$ are slowly varying. Often $s$ and $u$ are
extracted by using the simplest possible 2-point approximation
\begin{eqnarray}
u(i)&=&\frac{f(i)+f(i+1)}{2}  \nonumber \\
s(i)&=&(-1)^i\frac{f(i)-f(i+1)}{2},
\end{eqnarray}
by effectively assuming that $u(i)\simeq u(i+1)$ and $s(i)\simeq
s(i+1)$. This somewhat crude approximation has many drawbacks and
is insufficient for the present study. We therefore focus on
higher order n-point approximations by approximating $u$ and $s$
locally by polynomials. If one is interested in developing an
n-point formula that is symmetric around the point of interest it
is not possible to use the same degree of polynomial for both $u$
and $s$. For the present study the most frequently occurring case
is $u$ varying more rapidly than $s$. We have then found it
sufficient to develop a 7-point approximation by assuming that
$u(i)\simeq a i^3+bi^2+c i + d$ and $s(i)=e i^2+f i +g$, arriving
at the equations:
\begin{eqnarray}
f(i-3)&\simeq&-27a+9b-3c+d-(9e-3f+g)  \nonumber \\
f(i-2)&\simeq&-8a+4b-2c+d+(4e-2f+g)  \nonumber \\
f(i-1)&\simeq&-a+b-c+d-(e-f+g)  \nonumber \\
f(i)&\simeq&d+g  \nonumber \\
f(i+1)&\simeq&a+b+c+d-(e+f+g)  \nonumber \\
f(i+2)&\simeq&8a+4b+2c+d+(4e+2f+g)  \nonumber \\
f(i+3)&\simeq&27a+9b+3c+d-(9e+3f+g)  \nonumber \\
\end{eqnarray}
Solving these equations for $d,g$ we immediately get:
\begin{eqnarray}
u(i)&=& -\frac{15}{496}f(i-3) -\frac{1}{248}f(i-2) +\frac{71}{248}f(i-1)
\nonumber \\
&+&\frac{1}{2}f(i) +\frac{137}{496}f(i+1) +\frac{1}{248}f(i+2) -\frac{1}{31}%
f(i+3)  \nonumber \\
s(i)&=&(-1)^i\big[ \frac{15}{496}f(i-3) +\frac{1}{248}f(i-2) -\frac{71}{248}%
f(i-1)  \nonumber \\
&+&\frac{1}{2}f(i) -\frac{137}{496}f(i+1) -\frac{1}{248}f(i+2) +\frac{1}{31}%
f(i+3)\big]  \nonumber \\
&=&(-1)^i(f(i)-u(i))
\end{eqnarray}

\section{Detailed Proof of limited $SU(2)$ invariance of $S$}
\label{app:SU2}
In this appendix we show that the von Neumann entropy
for a system in doublet state ($\RR$ odd) is $SU(2)$ invariant.
For $\RR$ even the ground-state is a non-degenerate singlet an $S$ is manifestly
$SU(2)$ invariant.

Consider the case of $\RR$ odd, where the ground states form a spin
doublet. The most general linear combination of the doublet of ground states
can be written as a unitary transformation of the spin-up ground state:
\begin{equation}
|\psi > = U|\uparrow >.
\end{equation}
Here we can choose
\begin{equation}
U=\exp [i(a_xS^x_T+a_yS^y_T)]\end{equation}
so that
\begin{eqnarray}
 U|\uparrow >
&=& [\cos |\vec a/2|\mathcal{I}+2\hat a \cdot \vec S_T \sin |\vec
a/2|]|\uparrow >  \nonumber \\
&=&\cos |\vec a/2||\uparrow > +{\frac{(a_x-ia_y)}{|\vec a|}}\sin |\vec
a/2||\downarrow >,
\end{eqnarray}
with $\mathcal{I}$ the identity matrix and
\begin{equation}
\vec S_T\equiv \sum_{j=1}^{\RR}\vec S_j,
\end{equation}
and we have used the fact that the states $|\uparrow >$, $|\downarrow >$ transform
like an S=1/2 doublet under SU(2) to treat the Taylor expansion of the
exponential.
Now we observe that:
\begin{equation}
U = U_A\otimes U_B,
\end{equation}
where
\begin{equation}
U_A=\exp [i\vec a \cdot \sum_{j=1}^{r}\vec S_j], \ \ U_B=\exp [i\vec a \cdot
\sum_{j=r+1}^{\RR}\vec S_j].
\end{equation}
Thus:
\begin{equation}
|\psi > = U_A\otimes U_B|\uparrow >.
\end{equation}
The pure density matrix can then be written:
\begin{equation}
\rho = U_A\otimes U_B|\uparrow ><\uparrow |U_B^\dagger \otimes U_A^\dagger .
\end{equation}
It follows that the reduced density matrix is:
\begin{equation}
\rho_A^U = U_A\rho_AU^\dagger ,  \label{id}
\end{equation}
where $\rho_A$ is the reduced density matrix obtained from the pure state $%
|\uparrow >$ and $\rho_A^U$ is the reduced density matrix for the unitarily
transformed state $U|\uparrow >$. To prove Eq. (\ref{id}), consider
decomposing the pure state $|\uparrow >$ into a sum of a product of complete
bases of states in regions $A$ and $B$:
\begin{equation}
|\uparrow > = \sum_{ab}C_{ab}|a>\otimes |b>.
\end{equation}
Then the pure state density matrix is:
\begin{equation}
\rho_U=\sum_{aba^{\prime}b^{\prime}}C_{ab}C^*_{a^{\prime}b^{%
\prime}}U_A|a>U_B|b><b^{\prime}|U_B^\dagger <a^{\prime}|U_A^\dagger
\end{equation}
To obtain the reduced density matrix we perform the partial trace over region $B$:
\begin{equation}
\rho_A^U=\sum_{aba^{\prime}b^{\prime}n}C_{ab}C^*_{a^{\prime}b^{%
\prime}}U_A|a><n|U_B|b>
<b^{\prime}|U_B^\dagger |n><a^{\prime}|U_A^\dagger
\end{equation}
Here $\sum_n$ is a sum over a complete set of states in region $B$. If we now
use the following identity:
\begin{equation}
\sum_n <n|U_B|b><b^{\prime}|U_B^\dagger |n>=\hbox{tr}U_B|b><b^{%
\prime}|U_B^\dagger
= \hbox{tr}|b><b^{\prime}|,
\end{equation}
we see that:
\begin{equation}
\rho_A^U=U_A\sum_{aba^{\prime}b^{\prime}}C_{ab}C^*_{a^{\prime}b^{%
\prime}}|a><a^{\prime}|\hbox{tr}(|b><b^{\prime}|)U_A^\dagger
= U_A\rho_AU^\dagger
\end{equation}
Eq. (\ref{id}) implies that $\rho_A^U$ and $\rho_A$ have the same
eigenvalues, and hence the same entanglement entropy. 

\section{Connection between $S$ and the thermodynamic entropy\label{app:FiniteT}}
In this appendix we give arguments and provide numerical data showing that quite generally,
for sufficiently high $T$, the entanglement entropy will approach the thermodynamic entropy.
\begin{figure}[!ht]
\begin{center}
\includegraphics[width=10cm,clip]{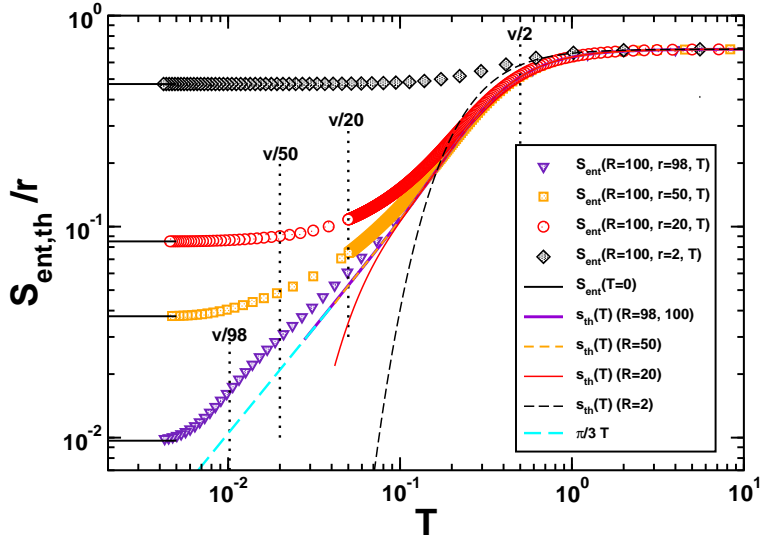}
\end{center}
\caption{The finite temperature entanglement entropy per unit length of the subsystem for an $XX$ spin chain shown along with
  the thermal entropy for system sizes of $r=2,20,50,98$.
    For the entanglement entropy the total system size was $R=100$. All entropies are plotted per unit length.
}
\label{fig:SEvsST}
\end{figure}

As usual, let us consider a subsystem, $A$ of linear extent $r$ with the entire
system, $A+B$, of size $R$. Let us also consider the subsystem 
$A$ decoupled from $B$. Following Gibbs, it is well known that the thermal
entropy for subsystem $A$ (decoupled from $B$)
 can be expressed as $S_{th}=-\Tr [\rho \ln \rho ]$
where $\rho = e^{-H_A/T}/Z$ is the thermal density matrix and $H_A$ is the
hamiltonian describing $A$. 
This mixed state is expected to occur if we start with our system 
$A$ weakly coupled to an infinite reservoir, and then trace 
over the reservoir. Before performing the trace, we could regard 
the system as being in a pure state of system plus reservoir. 
On the other hand, the finite $T$ entanglement entropy, 
$S=-tr[\rho_A\ln \rho_A]$ is defined 
by beginning with the system $A+B$ in the mixed state 
corresponding to the thermal density matrix $e^{-(H_A+H_B)/T}/Z$ 
and then tracing over region $B$ to obtain the reduced density 
matrix $\rho_A(T)$. Again, we could arrive 
at this thermal density matrix by beginning with sytem $A+B$ weakly 
coupled to an infinite reservoir and then tracing over the reservoir. 
While $S_{th}(T)$ is, by construction, independent of the coupling to $B$, 
the entanglement
entropy $S(T)$ can clearly depend on it.
We shall argue that for  $T\gg v/r$ the coupling to $B$ can also be neglected when
calculating $S(T)$ and the entanglement entropy $S$ will then approach $S_{th}$. 
Such a connection was previously noted in~\cite{Cardy04} and \cite{Korepin04}.
\begin{figure}[!ht]
\begin{center}
\includegraphics[width=10cm,clip]{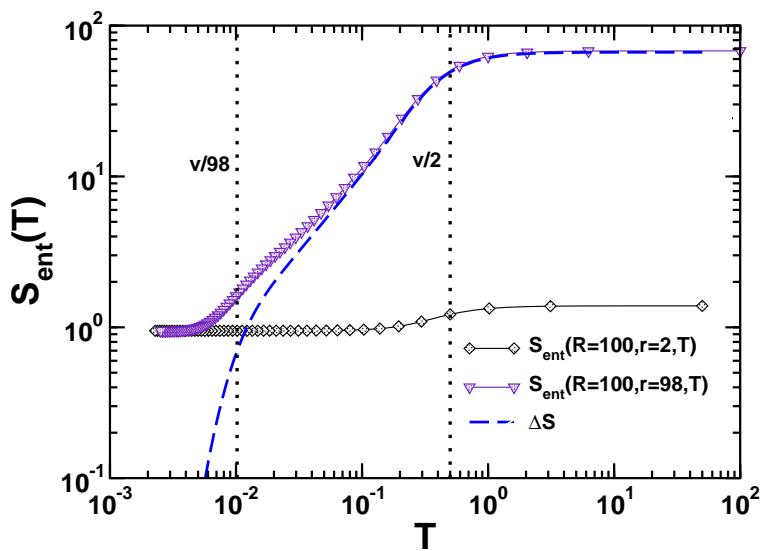}
\end{center}
\caption{The finite temperature entanglement entropy for subsystems $r=2,98$ for a $R=100$ site $XX$ spin chain.
  At $T=0$ the two entanglement entropies are identical and the dashed line indicates the difference between
    the two entropies, quickly approaching 0 as $T\to 0$.
}
\label{fig:SAmSB}
\end{figure}

Essentially, we want to argue that we may regard region $B$ approximately as a kind of
additional reservoir for region $A$ when $T$ is sufficiently large. Normally a
reservoir should be sufficiently weakly coupled to the system so as not to
disturb its energy eigenvalues or eigenstates. This is not true in the case of
subsystem $B$ which is strongly coupled to subsystem $A$. However, this strong
coupling only exists at the boundary (at the point $r$). For most
eigenfunctions we can neglect the perturbation due to this coupling.  The only
important exception to this statement are the low energy states, with
wave-lengths of order the system size and energies of O($v/r$). We can expect
these to be strongly perturbed by the coupling to subsystem $B$ and
consequently expect $S(T)$ to be strongly modified from the thermal entropy,
$S_{th}(T)$ at low temperatures.  However, when $T\gg v/r$, we expect these
states to make a negligible contribution to $S(T)$ since the density of
states is much larger at higher energies.  Therefore we expect $S(T)$ to
approach $S_{th}(T)$ when $T\gg v/r$.  This argument suggests that this
should occur regardless of $R$.  We also expect this correspondance to hold 
regardless of boundary conditions.  The thermal entropy for region $A$ 
becomes independent of whether it is calculated with pbc or obc when $T\gg v/r$. 
Similarly the entanglement entropy becomes independent of whether region $A$ is 
part of a system, $A+B$ which obeys pbc or obc at $T\gg v/r$.

In order to provide numerical evidence that $S(T)$ indeed does approach $S_{th}$, the thermal entropy, we have performed exact
calculations for both quantities on an $XX$ spin chain of length $R=100$. The 
finite $T$ entanglement entropy is calculated for a subsystem, $A$, of size $r$ within a total
system $A+B$ of size $R$ obeying periodic boundary conditions.  We have 
also calculated the thermal entropy, $S_{th}(T)$, for a system of size $r$ also with periodic boundary conditions.  We then 
compare $S(T)/r$ to $S_{th}(T)/r$. 
Our results are shown in Fig.~\ref{fig:SEvsST}
where both entropies are plotted per unit length of the subsystem. Four different sub-system sizes of $r=2,20,50,98$ are
considered and in all cases do we observe excellent agreement with the thermal entropy at sufficiently high temperature.
In Fig.~\ref{fig:SAmSB} is shown the finite temperature entanglement entropy for a $R=100$ site $XX$ spin chain for two
different subsystem sizes of $r=2,98$. At $T=0$ the two entanglement entropies are identical and the difference between
the two is shown as the solid line, quickly approaching 0 as $T\to 0$. 

\section{Solitons in the Majumdar-Ghosh Model\label{app:TSMG}}
In this appendix we focus on various properties of the MG model that also can be calculated with
very high precision using the TS-ansatz as described in section~\ref{sec:TS}. We begin by a determination
of $\psi^{sol}_n$.

\subsection{Determination of $\psi^{sol}_n$}
It is important to note that even though the thin soliton states $%
|n\rangle $ are linearly independent they are not orthogonal. The matrix of
overlaps, $B$, is given by:
\begin{equation}
B_{n,m}=\langle n|m\rangle=2^{-|n-m|}.
\end{equation}
If we now consider the action of the Hamiltonian Eq.~(\ref{eq:spinch}) {\it within} the
subspace spanned by the TS-states and
define $H^{\prime}=H+3J\RR/8\ I$ ($I$ is the identity matrix) it can be shown that~\cite{Caspers82,Caspers84}:
\begin{eqnarray}
H^{\prime}|0\rangle &\approx& \frac{J}{4}\left[2|0\rangle-|1\rangle\right], \ \ n=0
\nonumber \\
H^{\prime}|n\rangle &\approx& \frac{J}{4}\left[\frac{5}{2}|n\rangle-|n-1%
\rangle-|n+1\rangle\right], \ \ n\neq0,N_d  \nonumber \\
H^{\prime}|N_d\rangle &\approx& \frac{J}{4}\left[2|N_d\rangle-|N_d-1\rangle%
\right], \ \ n=N_d.  \label{eq:tb}
\end{eqnarray}
Note that $H'$ applied to a TS-state will in general also generate a contribution that is not in the
TS-subspace. Here we have dropped the part of such ``fat soliton" states that is orthogonal to the
TS-states since we have restricted $H'$ to the TS-subspace.
We rewrite, Eq.~(\ref{eq:tb}) in the following way:
\begin{equation}
H'|n\rangle=\frac{J}{4}h'_{nm}|m\rangle,
\end{equation}
where $h'_{nm}$ is a symmetric matrix with $h'_{00}=h'_{N_dN_d}=2$ and $h'_{nn}=5/2$ for $ n\neq 0,N_d$
and $h'_{n,n+1}=h'_{n+1,n}=-1$ for $n=1,\ldots, N_d-1$.
Using the full hamiltonian $H$ in Eq~(\ref{eq:spinch}) we can then write:
\begin{equation}
\frac{\langle\Psi_{TS}^\Uparrow|H|\Psi_{TS}^\Uparrow\rangle}{\langle\Psi_{TS}^\Uparrow|\Psi_{TS}^\Uparrow\rangle}= \frac{%
\sum_{m,n=0}^{N_d} \psi_m^{sol}A_{m,n}\psi^{sol}_n}{\sum_{m,n}%
\psi^{sol}_mB_{m,n}\psi^{sol}_n}-\frac{3}{8}J\RR,  \label{eq:ErgVar}
\end{equation}
where, perhaps surprisingly, it turns out that the $A_{m,n}$ is simply given in matrix notation by $%
A\equiv h' B J/4 = 3J I/8$. We can now determine the $%
\psi_n^{sol}$ in a variational manner by minimizing the energy,
Eq.~(\ref {eq:ErgVar}). This is straightforward to do using the
method of Lagrange multipliers and the constraint
$\sum_{m,n}\psi^{sol}_mB_{m,n}\psi^{sol}_n=1$, coming from
normalizing the wave-function. Due to the simple form of the
matrix $A$ and the fact that the matrix $B$ has only positive
definite eigenvalues, it is then seen that the optimal values for
the $\psi^{sol}_n$ is given by the eigenvector of $B$
corresponding to the largest eigenvalue, $\lambda_{max}$.
Furthermore, since $\langle H'\rangle$ coincides with our
definition of the soliton mass, we see that for open boundary
conditions (obc) $\Delta_{sol}^{TS,obc}$ is given by:
\begin{equation}
\Delta_{sol}^{TS,obc}=\frac{3J}{8\lambda_{max}}.
\end{equation}
For $\RR=201,401$ we find $\Delta_{sol}^{TS,obc}/J=0.12523, 0.12506$%
, respectively, in excellent agreement with the result for periodic boundary
conditions.
\begin{figure}[!ht]
\begin{center}
\includegraphics[width=10cm,clip]{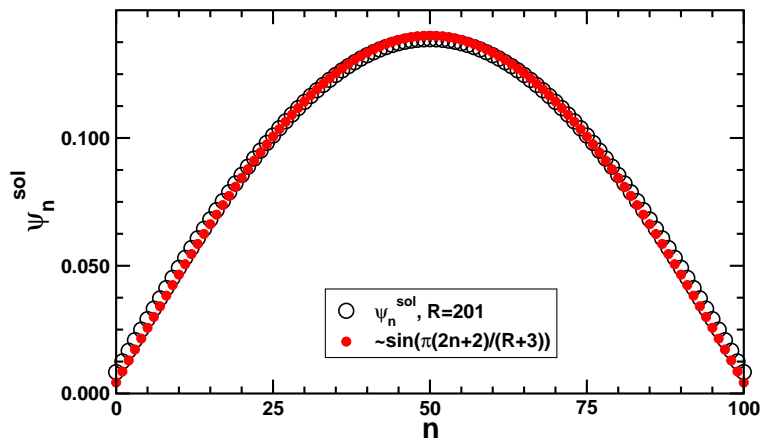}
\end{center}
\caption{Variationally determined $\psi^{sol}_n$ (open circles) for $%
\RR=201$ shown with the the free particle form given by Eq.~(\ref
{eq:sinform}). }
\label{fig:fn}
\end{figure}

We note that the same $\psi^{sol}_n$ can be found by assuming that $\langle
n|m\rangle = \delta_{n,m}$ and then solving the tight binding model
resulting from Eq.~(\ref{eq:tb}). In matrix form this tight binding model will simply
be given by the matrix $h'$. However, since $h' B$ is proportional to $I$ an eigenstate of
$h'$ is also an eigenstate of $B^{-1}$ ($B^{-1}$ exists since $B$ has positive definite eigenvalues)
and then also of $B$.
Hence, finding the eigenvector with {\it lowest} eigenvalue of $h'$
will also find the eigenvector with the {\it largest} eigenvalue of $B$ and hence the optimal $\psi^{sol}_n$.
This somewhat surprising observation,
implies that at the MG-point the
non-orthogonality of the thin soliton states cannot play an important role as
we shall discuss in more detail later.

As an illustration we show in Fig.~\ref{fig:fn} the variationally determined
$\psi^{sol}_n$ for $\RR=201$
along with the free particle form, Eq.~(\ref{eq:sinform}). Clearly the
agreement is relatively good even for this relatively small system size. 

\subsection{$S^z_r$ for $\RR$ odd ($J^{\prime}_K=1$)}
We now turn to a discussion of $\langle S^z_r\rangle$ which also can be calculated
employing the TS-ansatz, showing several surprising features.
\begin{figure}[!ht]
\begin{center}
\includegraphics[width=10cm,clip]{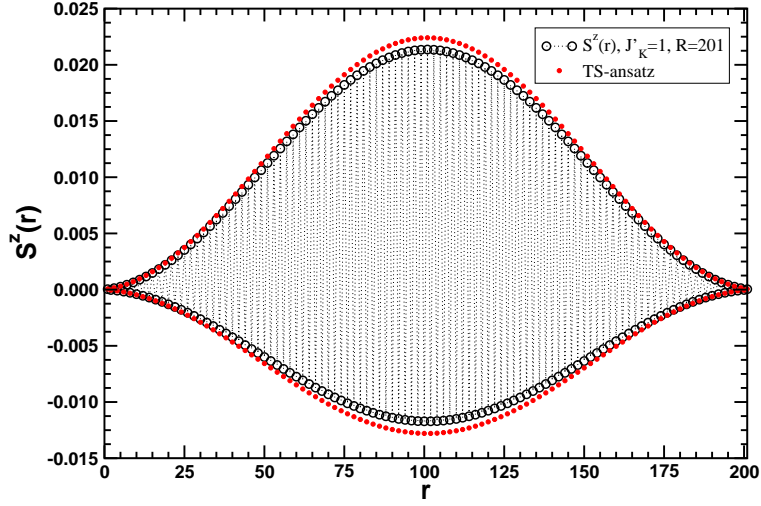}
\end{center}
\caption{$\langle S^z_r\rangle$ at the MG point ($J_2=J/2)$ calculated using
the TS-ansatz, Eq.~(\ref{eq:TSansatz}) and variationally determined $%
\psi^{sol}_n$ for $\RR=201$ (solid circles). DMRG results for $%
\langle S^z_r\rangle$ at the MG point keeping $m=256$ states (open circles).
}
\label{fig:Szj}
\end{figure}

The TS-ansatz, Eq.~(\ref{eq:TSansatz}), is constructed using the single
soliton states which do not include states where the soliton is on an
\textit{even} site when $\RR$ is odd. It would then seem natural to
assume that if we calculate $\langle S^z_r\rangle$ using the TS-ansatz it
would then be zero for $r$ even. However, due to the non-orthogonality of the
thin soliton states this turns out {\it not} to be the case. In fact we find that:
\begin{equation}
\langle\Psi_{TS}^\Uparrow|S^z_r|\Psi_{TS}^\Uparrow\rangle= \frac{%
\sum_{m,n=0}^{N_d} \psi_m^{sol}D_{m,n}\psi^{sol}_n}{\sum_{m,n}%
\psi^{sol}_mB_{m,n}\psi^{sol}_n},  \label{eq:Szj}
\end{equation}
with, for $r$ odd:
\begin{equation}
D_{n,m}=\left\{\begin{array}{cc}
\frac{1}{2} 2^{-|n-m|} &\ n\le \frac{r-1}{2}, m\ge \frac{r-1}{2}\\
\frac{1}{2} 2^{-|n-m|} &\ m\le \frac{r-1}{2}, n\ge \frac{r-1}{2}\\
0 &\ {\rm otherwise}
\end{array}\right. ,
\end{equation}
and for $r$ even
\begin{equation}
D_{n,m}=\left\{\begin{array}{cc} -\frac{1}{2}2^{-|n-m|} &\ n < \frac{r}{2}, m\ge
\frac{r}{2}\\ -\frac{1}{2}2^{-|n-m|} &\ m < \frac{r}{2}, n\ge \frac{r}{2}\\
0 &\ {\rm otherwise} \end{array}\right..
\end{equation}

In Fig.~\ref{fig:Szj} we show results for $\langle S^z_r\rangle$ at the
MG-point ($J_2=J/2)$ obtained using the TS-ansatz as well as numerically
using DMRG methods. Fairly good agreement is observed. Notably, the
TS-ansatz clearly yields a non-zero value for $\langle S^z_r\rangle$ when $r$
is \textit{even}. However, longer valence bonds, not accounted for in the
TS-ansatz, clearly contribute for this value of $\RR$.

We also stress that, if we use
the OTS-ansatz to describe $\langle S^z_r\rangle$ we simply find $\langle S^z_r\rangle=|\psi^{sol}_{(r-1)/2}|^2$
for $r$ odd, 0 for $r$ even. The non-zero result for $\langle S^z_r\rangle$ for even $r$ obtained with
the TS-ansatz (See Fig.~\ref{fig:Szj}) is then purely a result of the non-orthogonality of the TS-states.

\subsection{Dimerization for $\RR$-odd from the TS-ansatz}
\begin{figure}[!ht]
\begin{center}
\includegraphics[width=10cm,clip]{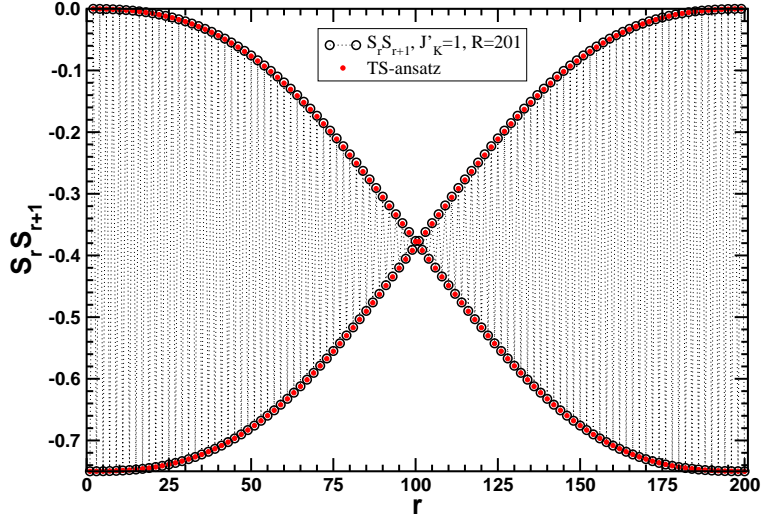}
\end{center}
\caption{$\langle \vec S_r\cdot\vec S_{r+1}\rangle$ at the MG point ($%
J_2=J/2)$ calculated using the TS-ansatz, Eq.~(\ref{eq:TSansatz}) and
variationally determined $\psi^{sol}_n$ for $\RR=201$ (solid
circles). DMRG results for $\langle \vec S_r\vec S_{r+1}\rangle$ at the MG
point keeping $m=256$ states (open circles). }
\label{fig:SSj}
\end{figure}
Finally, at the MG-point, $J_2=J/2$, it is also possible to obtain very precise results for
the dimerization using the TS-ansatz. Most measures of the dimerization are
based on the spin correlation function.
We therefore consider calculations of $\langle\vec S_r\cdot\vec S_{r+1}\rangle$
using the TS-ansatz. With $r=2n+1$, $n=0\ldots N_d=(\RR-1)/2$ and $%
B_{n,m}=2^{-|n-m|}$ we find that:
\begin{equation}
\langle\Psi_{TS}^\Uparrow|\vec S_r\cdot\vec
S_{r+1}|\Psi_{TS}^\Uparrow\rangle= \frac{\sum_{m,n=0}^{N_d}
\psi_m^{sol}F_{m,n}\psi^{sol}_n}{\sum_{m,n}\psi^{sol}_mB_{m,n}\psi^{sol}_n},
\label{eq:SSj}
\end{equation}
with, for $r$ odd:
\begin{equation}
F_{n,m}=\left\{\begin{array}{cc} -\frac{3}{4}2^{-|n-m|} &\ n > \frac{r-1}{2}\ {\rm or}\
m> \frac{r-1}{2}\\ 0 &\ {\rm otherwise} \end{array}\right.,
\end{equation}
and for $r$ even
\begin{equation}
F_{n,m}=\left\{\begin{array}{cc} -\frac{3}{4}2^{-|n-m|} &\ n < \frac{r}{2}\ {\rm or}\
m< \frac{r}{2}\\ 0 &\ {\rm otherwise} \end{array}\right..
\end{equation}

In Fig.~\ref{fig:SSj} we show results for $\langle \vec S_r\cdot\vec
S_{r+1}\rangle$ at the MG-point ($J_2=J/2)$ obtained using the TS-ansatz as
well as numerically using the DMRG methods. Almost perfect agreement is
observed.

\bibliographystyle{unsrt}
\bibliography{qie}

\begin{thebibliography}{100}

\bibitem{Bennett00}
C.~H. Bennett and D.~P. DiVincenzo.
\newblock {\em Nature}, 404:247, 2000.

\bibitem{Wilczek94}
C.~Holzhey, F.~Larsen, and F.~Wilczek.
\newblock {\em Nucl. Phys. B}, 424:443, 1994.

\bibitem{Osborne02a}
T.~J. Osborne and M.~A. Nielsen.
\newblock {\em Phys. Rev. A}, 66:032110, 2002.

\bibitem{Osterloh02}
A.~Osterloh, L.~Amico, G.~Falci, , and R.~Fazio.
\newblock {\em Nature}, 416:608, 2002.

\bibitem{Vidal03}
G.~Vidal, J.~I. Latorre, E.~Rico, and A.~Kitaev.
\newblock {\em Phys. Rev. Lett.}, 90:227902, 2003.

\bibitem{Wei05}
T.-C. Wei, D.~Das amd S.~Mukhopadyay, S.~Vishveshwara, and P.~M. Goldbart.
\newblock {\em Phys. Rev. A}, 71:060305, 2005.

\bibitem{Kopp06a}
A.~Kopp, X.~Jia, and S.~Chakravarty.
\newblock Ann. Phys., 2006.

\bibitem{Neumann27}
J.~von Neumann.
\newblock {\em G{\"o}tt. Nachr.}, 273, 1927.

\bibitem{Wehrl78}
A.~Wehrl.
\newblock {\em Rev. Mod. Phys.}, 50:221, 1978.

\bibitem{Hill97}
S.~Hill and W.~K. Wooters.
\newblock {\em Phys. Rev. Lett.}, 78:5022, 1997.

\bibitem{Wooters98}
W.~K. Wooters.
\newblock {\em Phys. Rev. Lett.}, 80:2245, 1998.

\bibitem{Bennett96b}
C.~H. Bennett, D.~P. DiVincenzo, J.~A. Smolin, and W.~K. Wooters.
\newblock {\em Phys. Rev. A}, 54:3824, 1996.

\bibitem{Bennett96prl}
C.~H. Bennett, G.~Brassard, S.~Popescu, B.~Schumacher, J.~A. Smolin, and W.~K.
  Wooters.
\newblock {\em Phys. Rev. Lett.}, 76:722, 1996.

\bibitem{Vedral97}
V.~Vedral, M.~B. Plenio, M.~A. Rippin, and P.~L. Knight.
\newblock {\em Phys. Rev. Lett.}, 78:2275, 1997.

\bibitem{Vedral98}
V.~Vedral and M.~B. Plenio.
\newblock {\em Phys. Rev. A}, 57:1619, 1998.

\bibitem{Vedral02}
V.~Vedral.
\newblock {\em Rev. Mod. Phys.}, 74:197, 2002.

\bibitem{MHorodecki01}
M.~Horodecki.
\newblock {\em Quant. Inf. Comp.}, 1:3, 2001.

\bibitem{Osborne02b}
T.~J. Osborne and M.~A. Nielsen.
\newblock {\em Quant. Inf. Proc.}, 1:45, 2002.

\bibitem{Vidal04}
G.~Vidal.
\newblock {\em Phys. Rev. Lett.}, 93:040502, 2004.

\bibitem{Verstraete04}
F.~Verstraete and J.~I. Cirac.
\newblock cond-mat/0407066, 2004.

\bibitem{MPS}
F.~Verstaete and J.~I. Cirac.
\newblock {\em Phys. Rev. B}, 73:094423, 2006.

\bibitem{White92}
S.~R. White.
\newblock {\em Phys. Rev. Lett.}, 69:2863, 1992.

\bibitem{Schol05}
U.~Schollw{\"o}ck.
\newblock {\em Rev. Mod. Phys.}, 77:259, 2005.

\bibitem{Bombelli86}
L.~Bombelli, R.~K. Koul, J.~Lee, and R.~D. Sorkin.
\newblock {\em Phys.\ Rev.\ D}, 34:373, 1986.

\bibitem{Srednicki93}
M.~Srednicki.
\newblock {\em Phys.\ Rev.\ Lett.}, 71:666, 1993.

\bibitem{Cardy04}
P.~Calabrese and J.~Cardy.
\newblock {\em J. Stat. Mech.}, page 06002, 2004.

\bibitem{Kitaev06}
A.~Kitaev and J.~Preskill.
\newblock {\em Phys.\ Rev.\ Lett.}, 96:110404, 2006.

\bibitem{Levin06}
M.~Levin and X.~G. Wen.
\newblock {\em Phys.\ Rev.\ Lett.}, 96:110405, 2006.

\bibitem{Fendley06}
P.~Fendley, M.~P.~A. Fisher, and C.~Nayak.
\newblock cond-mat/0609072, 2006.

\bibitem{Furukawa06}
S.~Furukawa and G.~Misguich.
\newblock cond-mat/0612227, 2006.

\bibitem{Ryu06}
S.~Ryu and T.~Takayanagi.
\newblock {\em Phys.\ Rev.\ Lett.}, 96:181602, 2006.

\bibitem{Ryu07}
S.~Ryu and T.~Takayanagi.
\newblock hep-th/0605073, 2006.

\bibitem{Ghosh03}
S.~Ghosh, T.~F. Rosenbaum, G.~Aeppli, and S.~N. Coppersmith.
\newblock {\em Nature}, 425:48, 2003.

\bibitem{Vedral04}
V.~Vedral.
\newblock {\em New J. Phys.}, 6:102, 2004.

\bibitem{Brukner06}
C.~Brukner, V.~Vedral, and A.~Zeilinger.
\newblock {\em Phys. Rev. A}, 73:012110, 2006.

\bibitem{Horodecki96}
M.~Horodecki, P.~Horodecki, and R.~Horodecki.
\newblock {\em Phys. Lett. A}, 223:1, 1996.

\bibitem{Terhal00}
B.~M. Terhal.
\newblock {\em Phys. Lett. A}, 271:319, 2000.

\bibitem{Lewenstein00}
M.~Lewenstein, B.~Krauss, J.~I. Cirac, and P.~Horodecki.
\newblock {\em Phys. Rev. A}, 62:052310, 2000.

\bibitem{Jordan04}
A.~N. Jordan and M.~B{\"u}ttiker.
\newblock {\em Phys. Rev. Lett.}, 92:247901, 2004.

\bibitem{Toth05}
G.~T\'oth.
\newblock {\em Phys. Rev. A}, 72:010301, 2005.

\bibitem{Wu05}
L.-A. Wu, S.~Bandyopadhyay, M.~S. Sarandy, and D.~A. Lidar.
\newblock {\em Phys. Rev. A}, 72:032309, 2005.

\bibitem{Wiesniak05}
M.~Wie\'sniak, V.~Vedral, and C.~Brukner.
\newblock {\em New J. Phys.}, 8:258, 2005.

\bibitem{Anders06}
J.~Anders, D.~Kazlikowski, C.~Lunkes, T.~Ohshima, and V.~Vedral.
\newblock {\em New J. Phys.}, 8:140, 2006.

\bibitem{Rappoport06}
T.~G. Rappoport, L.~Ghivelder, J.~C. Fernandes, R.~B. Guimar{\~a}es, and M.~A.
  Continentino.
\newblock cond-mat/0608403, 2006.

\bibitem{Vertesi06}
T.~V\'ertesi and E.~Bene.
\newblock {\em Phys. Rev. B}, 73:134404, 2006.

\bibitem{Bose05}
I.~Bose and A.~Tribedi.
\newblock {\em Phys. Rev. A}, 72:022314, 2005.

\bibitem{Zhou05}
H.-Q. Zhou, T.~Barthel, J.~Fjaerestad, and U.~Schollw{\"o}ck.
\newblock cond-mat/0511732, 2005.

\bibitem{Levine04}
G.~C. Levine.
\newblock {\em Phys. Rev. Lett.}, 93:226402, 2004.

\bibitem{Wang04}
X.~Wang.
\newblock {\em Phys. Rev. E}, 69:066118, 2004.

\bibitem{Peschel05}
I.~Peschel.
\newblock {\em J. Phys. A}, 38:4327, 2005.

\bibitem{Fan06}
H.~Fan, V.~Korepin, V.~Roychowdhury, and C.~Hadley a~S.~Bose.
\newblock cond-mat/0605133, 2006.

\bibitem{Zhao06}
J.~Zhao, I.~Peschel I, and X.Q. Wang.
\newblock {\em Phys. Rev. B}, 73:024417, 2006.

\bibitem{Cho06}
S.~Y. Cho and R.~H. McKenzie.
\newblock {\em Phys. Rev. A}, 73:012109, 2006.

\bibitem{Affleckg}
I.~Affleck and A.W.W. Ludwig.
\newblock {\em Phys.\ Rev.\ Lett.}, 67:161, 1991.

\bibitem{KEAC}
N.~Laflorencie, E.~S.~S\o rensen, and I.~Affleck.
\newblock in preparation, 2007.

\bibitem{Eggert96}
S.~Eggert.
\newblock {\em Phys. Rev. B}, 54:15590, 1996.

\bibitem{Haldane82}
F.~D.~M. Haldane.
\newblock {\em Phys.\ Rev.\ Lett.}, 25:4925, 1982.

\bibitem{MG69}
C.~K. Majumdar and D.~K. Ghosh.
\newblock {\em J.~Phys.\ C}, 3:911, 1969.

\bibitem{Loss98}
D.~Loss and D.~P. DiVincenzo.
\newblock {\em Phys. Rev. A}, 57:120, 1998.

\bibitem{Simon06}
C.~Simon, Y.-M. Niquet, X.~Caillet, J.~Eymery, J.-P. Poizat, and J.-M. Gerard.
\newblock quant-phys/0609030, 2006.

\bibitem{Eriksson04}
M.~A. Eriksson, M.~Friesen, S.~N. Coppersmith, R.~Joynt, L.~J. Klein,
  K.~Slinker, C.~Tahan, P.~M. Mooney, J.~O. Chu, and S.~J. Koester.
\newblock {\em Quant. Inf. Proc.}, 3:133, 2004.

\bibitem{Costi03}
T.~Costi and R.~H. McKenzie.
\newblock {\em Phys. Rev. A}, 68:034301, 2003.

\bibitem{Caldeira}
A.~O. Caldeira and A.~J. Leggett.
\newblock {\em Ann. Phys.}, 149:374, 1983.

\bibitem{Kopp06b}
A.~Kopp and K.~Le Hur.
\newblock cond-mat/0612095, 2004.

\bibitem{Lloyd03}
S.~Lloyd.
\newblock {\em Phys. Rev. Lett.}, 90:167902, 2003.

\bibitem{Bose03}
S.~Bose.
\newblock {\em Phys. Rev. Lett.}, 91:207901, 2003.

\bibitem{Plenio05}
M.~B. Plenio and F.~L. Semi{\~a}o.
\newblock {\em New J. Phys.}, 7:73, 2005.

\bibitem{Christandl04}
M.~Christandl, N.~Datta, A.~Ekert, and A.~J. Landahl.
\newblock {\em Phys. Rev. Lett.}, 92:187902, 2004.

\bibitem{Burgarth05a}
D.~Burgarth and S.~Bose.
\newblock {\em New J. Phys.}, 7:135, 2005.

\bibitem{Burgarth05b}
D.~Burgarth, V.~Giovannetti, and S.~Bose.
\newblock {\em J. Phys. A}, 38:6793, 2005.

\bibitem{Wojcik05}
A.~W\'ojcik, T.~Luczak, P.~Kurzy\'nski, A.~Grudka, T.~Gdala, and M.~Bednarska.
\newblock {\em Phys. Rev. A}, 72:034303, 2005.

\bibitem{Zhang05}
J.~Zhang, G.~L. Long, W.~Zhang, Z.~Deng, W.~Liu, and Z.~Lu.
\newblock {\em Phys. Rev. A}, 72:012331, 2004.

\bibitem{Karbach05}
P.~Karbach and J.~Stolze.
\newblock {\em Phys. Rev. A}, 72:030301, 2005.

\bibitem{Fitzsimons06}
J.~Fitzsimons and J.~Twamley.
\newblock {\em Phys. Rev. Lett.}, 97:090502, 2006.

\bibitem{Sorensen06}
E.~S.~S\o rensen, M.-S. Chang, N.~Laflorencie, and I.~Affleck.
\newblock {\em J. Stat. Mech.}, page L01001, 2007.

\bibitem{Nozieres}
P.~Nozi\`{e}res.
\newblock {\em J. Low Temp. Phys.}, 17:31, 1974.

\bibitem{Affleck90}
I.~Affleck.
\newblock {\em Nucl. Phys. B}, 336:517, 1990.

\bibitem{Affleck91a}
I.~Affleck and A.~W.~W. Ludwig.
\newblock {\em Nucl. Phys. B}, 352:849, 1991.

\bibitem{Affleck91b}
I.~Affleck and A.~W.~W. Ludwig.
\newblock {\em Nucl. Phys. B}, 360:641, 1991.

\bibitem{Arnesen01}
M.~C. Arnesen, S.~Bose, and V.~Vedral.
\newblock {\em Phys. Rev. Lett.}, 87:017901, 2001.

\bibitem{Gunlycke01}
D.~Gunlycke, V.~M. Kendon, V.~Vedral, and S.~Bose.
\newblock {\em Phys. Rev. A}, 64:042302, 2001.

\bibitem{Latorre04}
J.~I. Latorre, E.~Rico, and G.~Vidal.
\newblock {\em Quant. Inf. Comp.}, 4:48, 2004.

\bibitem{Verstraete04b}
F.~Verstraete, M.~A. Martin-Delgado, and J.~I. Cirac.
\newblock {\em Phys. Rev. Lett.}, 92:087201, 2004.

\bibitem{Fan04}
H.~Fan, V.~Korepin, and V.~Roychowdhury.
\newblock {\em Phys. Rev. Lett.}, 93:227203, 2004.

\bibitem{Peschel04}
I.~Peschel.
\newblock {\em J. Stat. Mech.}, page P12005, 2004.

\bibitem{Subrahmanyam04}
V.~Subrahmanyam.
\newblock {\em Phys. Rev. A}, 69:022311, 2004.

\bibitem{Laflorencie06}
N.~Laflorencie, E.~S. S{\o}rensen, M.-S. Chang, and I.~Affleck.
\newblock {\em Phys. Rev. Lett.}, 96:100603, 2006.

\bibitem{Amico04}
L.~Amico, A.~Osterloh, F.~Plastina, R.~Fazio, and G.~M. Palma.
\newblock {\em Phys. Rev. A}, 69:022304, 2004.

\bibitem{Cardy05}
P.~Calabrese and J.~Cardy.
\newblock {\em J. Stat. Mech.}, page 04010, 2004.

\bibitem{dechiara06}
G.~De Chiara, S.~Montangero, P.~Calabrese, and R.~Fazio.
\newblock {\em J. Stat. Mech.}, page P03001, 2006.

\bibitem{SS81}
B.~S. Shastry and B.~Sutherland.
\newblock {\em Phys. Rev. Lett.}, 47:964, 1981.

\bibitem{Jin04}
B.-Q. Jin and V.~E. Korepin.
\newblock {\em J. Stat. Phys.}, 116:79, 2004.

\bibitem{Sorensen98b}
E.~S{\o}rensen.
\newblock {\em J. Phys. Cond. Matt.}, 10:10655, 1998.

\bibitem{Refael04}
G.~Refael and J.~E. Moore.
\newblock {\em Phys. Rev. Lett.}, 93:260602, 2004.

\bibitem{alet07}
F.~Alet, S.~Capponi, N.~Laflorencie, and M.~Mambrini.
\newblock cond-mat/0703027, 2007.

\bibitem{Sorensen96}
E.~S. S{\o}rensen and I.~Affleck.
\newblock {\em Phys. Rev. B}, 53:9153, 1996.

\bibitem{Barzykin98}
V.~Barzykin and I.~Affleck.
\newblock {\em Phys. Rev. B}, 57:432, 1998.

\bibitem{Barzykin99}
V.~Barzykin and I.~Affleck.
\newblock {\em J. Phys. A}, 32:867, 1999.

\bibitem{Abrikosov70}
A.~A. Abrikosov and A.~A. Migdal.
\newblock {\em J. Low Temp. Phys.}, 3:519, 1970.

\bibitem{Frahm97}
H.~Frahm and A.A. Zvyagin.
\newblock {\em J. Cond. Matt.}, 9:9939, 1997.

\bibitem{IntegralTable}
I.~S. Gradshteyn and I.~M. Ryzhik.
\newblock {\em Table of Integrals, Series, and Products}.
\newblock Academic Press, 2000.

\bibitem{Schollwock96}
U.~Schollw{\"o}ck, Th. Jolic{\oe}ur, and T.~Garel.
\newblock {\em Phys.\ Rev.\ B}, 53:3304, 1996.

\bibitem{Caspers82}
W.~J. Caspers and W.~Magnus.
\newblock {\em Phys. Lett. A}, 88A:103, 1982.

\bibitem{Caspers84}
W.~J. Caspers, K.~M. Emmett, and W.~Magnus.
\newblock {\em J.~Phys. A}, 17:2687, 1984.

\bibitem{Sorensen98}
E.~S. S{\o }rensen, I.~Affleck, D.~Augier, and D.~Poilblanc.
\newblock {\em Phys. Rev. B}, 58:14701, 1998.

\bibitem{Tsai00}
S.~W. Tsai and J.~B. Marston.
\newblock {\em Phys. Rev. B}, 62:5546, 2000.

\bibitem{Eggert92}
S.~Eggert and I.~Affleck.
\newblock {\em Phys. Rev. B}, 46:10866, 1992.

\bibitem{Ian98}
I.~Affleck.
\newblock {\em J. Phys. A}, 31:4573, 1998.

\bibitem{Rommer00}
S.~Rommer and S.~Eggert.
\newblock {\em Phys.\ Rev.\ B}, 62:4370, 2000.

\bibitem{Wolf06}
M.~M. Wolf.
\newblock {\em Phys. Rev. Lett.}, 96:010404, 2006.

\bibitem{Gioev06}
D.~Gioev and I.~Klich.
\newblock {\em Phys. Rev. Lett.}, 96:100503, 2006.

\bibitem{Barthel06}
T.~Barthel and U.~Schollw{\"o}ck M.-C.~Chung.
\newblock cond-mat/0602077, 2006.

\bibitem{Korepin04}
V.~E. Korepin.
\newblock {\em Phys. Rev. Lett.}, 92:096402, 2004.

\end{thebibliography}
\end{document}